\documentclass[twocolumn,secnumarabic,amssymb, nobibnotes, aps, prd]{revtex4-1}

\usepackage{graphicx, mathrsfs, subfig, amsmath, amssymb, natbib, units, color}
\usepackage[figuresright]{rotating}

\newcommand\pd[2]{\frac{\partial#1}{\partial#2}}
\newcommand\vc[1]{\boldsymbol{#1}}
\newcommand\dv[2]{\frac{\ud #1}{\ud #2}}
\newcommand\Dv[2]{\frac{\textrm{D} #1}{\textrm{D} #2}}
\newcommand\ud{\textrm{d}}

\newcommand\Atw{\mathscr{A}}
\newcommand\Atwlab{\mathscr{A}_\textrm{gss}}
\newcommand\Atwid{\mathscr{A}_\textrm{ideal}}
\newcommand\Atwapp{\mathscr{A}_\textrm{app}}
\newcommand{\ageq}{\,\textnormal{\raisebox{-2pt}{${\stackrel{>}{\scriptstyle\sim}}$}}\,}
\newcommand{\aleq}{\,\textnormal{\raisebox{-2pt}{${\stackrel{<}{\scriptstyle\sim}}$}}\,}

\makeatletter
\newcommand\Rmnum[1]{\expandafter\@slowromancap\romannumeral #1@}
\makeatother

\begin{document}

\title{The Rotating Rayleigh-Taylor Instability} 

\author{M.~M.~Scase$^1$}
\email{matthew.scase@nottingham.ac.uk}
\author{K.~A.~Baldwin$^2$}
\author{R.~J.~A.~Hill$^3$}
\affiliation{$^1$School of Mathematical Sciences, University of Nottingham, Nottingham NG7 2RD, UK\\
$^2$Faculty of Engineering, University of Nottingham, Nottingham NG7 2RD, UK\\
$^3$School of Physics and Astronomy, University of Nottingham, Nottingham NG7 2RD, UK}
\date{February 29, 2016} 
\maketitle

The effect of rotation upon the classical two-layer Rayleigh-Taylor instability is considered theoretically and compared with previous experimental results.  In particular we consider a two-layer system with an axis of rotation that is perpendicular to the interface between the layers.  In general we find that a wave mode's growth rate may be reduced by rotation.  We further show that in some cases, unstable axisymmetric wave modes may be stabilized by rotating the system above a critical rotation rate associated with the mode's wavelength, the Atwood number and the flow's aspect ratio.  We compare our theory with experiments conducted in a magnetic field using `heavy' diamagnetic and `light' paramagnetic fluids and present comparisons between the theoretical predictions and experimental observations.

\section{\label{sec:intro}Introduction}
Understanding of the Rayleigh-Taylor instability has increased progressively since Lord Rayleigh's \citep{rayleigh83} initial work and the investigations of \citet{taylor50} and \citet{lewis50}.  The motivation for research into this fundamental problem has changed over time, from the original interests of Taylor and Lewis to the energy supply and astrophysical aspects of more recent work.  The now familiar structure of the Rayleigh-Taylor instability has been observed from small scales in, for example, inertial confinement fusion problems \citep[see e.g.,][]{freeman77}, to extremely large scales, such as the crab nebula \citep[see, e.g.,][]{wangchevalier} where pulsar winds accelerate through dense supernova remnants.  
In many cases of practical interest, it would be desirable to have some further control over the instability after the setting of the initial density profiles. One possibility is to rotate the system; the often stabilising effect of rotation on flow is well-known \citep[see e.g.,][]{fultz62}.  \citet{tao} investigated whether rotation may be used to influence the Rayleigh-Taylor instability at the surface of an inertial confinement fusion target by considering instability at an interface {\em parallel} to the axis of rotation. In inertial confinement fusion, the Rayleigh-Taylor instability reduces the efficiency of fusion during both the acceleration phase, between the ablator and the fuel, and during the deceleration phase, between the hot and cold fuel regions \citep[see e.g.,][]{lindl}. The efficiency is reduced due to the increased interfacial surface area between the two layers in each case. The work of \citet{tao} suggested that the instability may be suppressed around the equatorial region of a spherical rotating target. 

\begin{figure}
\begin{center}
\includegraphics[width = 240pt]{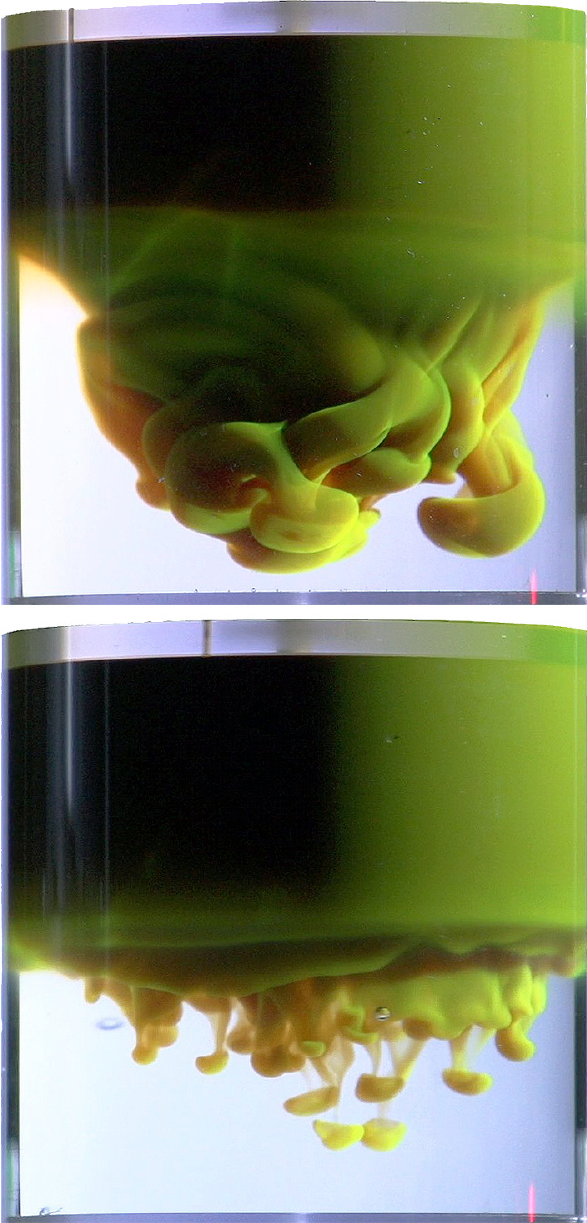}
\end{center}
\caption{\label{fig:quali} The upper image, taken of our experiments, is of the Rayleigh-Taylor instability developing in a non-rotating system.  The instability develops in time, forming large vortices that transport the `denser' (green) fluid downwards.  The lower image is of the same fluids but here the system is rotating.  The effect of the rotation can be seen to restrict the size of the vortices that form and inhibit the bulk vertical transport of fluid.  The times shown are 1.1\,s and 1.2\,s after initiation in the upper and lower images respectively, the fluids are as described in \S \ref{sec:exp}.  The tank diameter is 90\,mm, and the rotation rate in the lower image was $2.52$\,rad\,s$^{-1}$.}
\end{figure}

In a previous paper \cite{scirep} we reported results of experiments to study the development of the Rayleigh-Taylor instability in a two-layer fluid system with axis of rotation perpendicular to the layers.  The presence of rotation introduces a restoring force on fluid elements moving perpendicular to the axis of rotation: the Coriolis force. This fictitious force, which appears in a rotating reference frame, acts to restore a fluid element, traveling in a direction perpendicular to the axis of rotation, to its original position, following a curved path. The presence of the Coriolis force therefore allows the fluid to support inertial wave motions, the rotational counterpart to the internal gravity waves supported by a density stratification.  As will be shown, the Coriolis force acts to inhibit large-scale overturning motions at the unstable interface and is consequently important in changing the character of the developing Rayleigh-Taylor instability as the rate of rotation is increased. The effect is shown qualitatively in Fig.~\ref{fig:quali}. It can be seen that the large-scale overturning motion required to form large vortices (top) is restricted in the presence of rotation (bottom).  

In this paper we present a theoretical study of the Rayleigh-Taylor instability under the influence of rotation and we review our previous experimental results in light of this theory.
\citet{miles59, miles64} considered the effects of rotation on infinitesimal free-surface waves on a body of water, remarking on Fultz's \citep{fultz62} observation that the parabolic nature of the free-surface is important and cannot be neglected as previous authors had \citep[see, e.g.,][]{lamb32}
\begin{quote}
`The planar [horizontal hydrostatic interface] approximation is necessarily inconsistent for axisymmetric gravity waves in the sense that both the rotation induced shift \ldots~and the free-surface slope are of the same order of magnitude.'
\end{quote}
We develop the theory of \citet{miles59, miles64} to allow for a two-layer fluid system that may have either a stable or unstable interface.  We find in the limit of high, stable density difference that we recover Miles' \citep{miles64} result, and in the limit of an unstable density difference with no rotation we recover the classical Rayleigh-Taylor model \citep{taylor50}.  In the special limit of semi-infinite fluid layers and a strictly horizontal interface we recover the model of \citet{chandra}.  For axisymmetric waves we are able to find a critical rotation rate above which a given wave mode behaves as an oscillating standing wave, but below which exhibits Rayleigh-Taylor growth.  In general, non-axisymmetric waves cannot be stabilized indefinitely but we are able to say for a given mode whether the growth rate is reduced or increased by rotation and find that there is a strong dependence on the aspect ratio of the layers.

Previous experimental investigations of the classical, non-rotating instability have used a variety of methods, each with their own associated drawbacks.  The main techniques of the last half century include using: compressed gas to accelerate slugs of fluid vertically downwards at rates considerably higher than gravity \citep[e.g.,][]{lewis50, nevmerzhitsky94}, rocketry to rapidly accelerate gravitationally stable stratifications vertically downwards \citep[e.g.,][]{read84}, linear electric motors or other methods to reverse the apparent direction of gravity \citep[e.g.,][]{dimonte96,wilkinsonjacobs}, and more recently, using barrier removal techniques to allow a dense layer of fluid to impinge on lower layers \cite[e.g.,][]{dalziel, linden94, jacobsdalziel}. Other techniques include using a splitter plate to separate dense horizontal flows above from light horizontal flows below and then using the downstream distance from the end of the splitter plate as a proxy for time after release \citep[e.g.,][]{sniderandrews}.  Recent studies have made use of magnetic fields to induce the Rayleigh-Taylor instability in a two-dimensional system \citep[see, e.g.,][]{carlesetal, huangetal} or rotating magnetic fields with a view to controlling the initial conditions of the Rayleigh-Taylor instability in a ferrofluid \citep[e.g.,][]{rannacherengel, poehlmannetal}.  Our previous experiments made use of the magnetic field of a superconducting solenoid magnet to apply magnetic body forces to a rotating two-fluid system \citep{scirep}.  The gradient magnetic field attracts the light paramagnetic fluid in the upper layer toward the solenoid, and repels the desnse diamagnetic fluid in the lower layer, with a force proportional to the magnetic field strength and its gradient. Above a particular magnetic field strength and field gradient, determined by the relative magnetic susceptibilities and densities of the two fluids, the paramagnetic and diamagnetic body forces acting on the fluids overcome the gravitational stability of the system, inducing the onset of Rayleigh-Taylor instability, and the paramagnetic fluid exchanges places with the diamagnetic fluid (see Supplementary Information).  We compare these experimental findings with the theory presented here.

The structure of the paper is as follows: in \S \ref{sec:modeling} we develop an inviscid theory based on the previous theories of Rayleigh-Taylor instability due to \citet{taylor50} and the modeling of surface oscillations on rotating bodies of fluid due to \citet{lamb32} and \citet{miles59, miles64}.  In \S \ref{sec:mag} we develop the theory presented in \S \ref{sec:modeling} to allow for magnetic initiation of the rotating Rayleigh-Taylor instability on linearly magnetizable fluids such as the para- and diamagnetic fluids used in the experiments.  In \S \ref{sec:exp} we
review the results from our earlier experiments in light of the theory presented here.  Finally in \S \ref{sec:conc} we discuss our results and draw our conclusions.

\section{\label{sec:modeling}Modeling}

\subsection{\label{sec:growth}Growth of the instability}

\begin{figure}
\setlength{\unitlength}{1pt}
\begin{center}
\begin{picture}(215,250)(0,10)
\put(0,10){\includegraphics[height = 250pt]{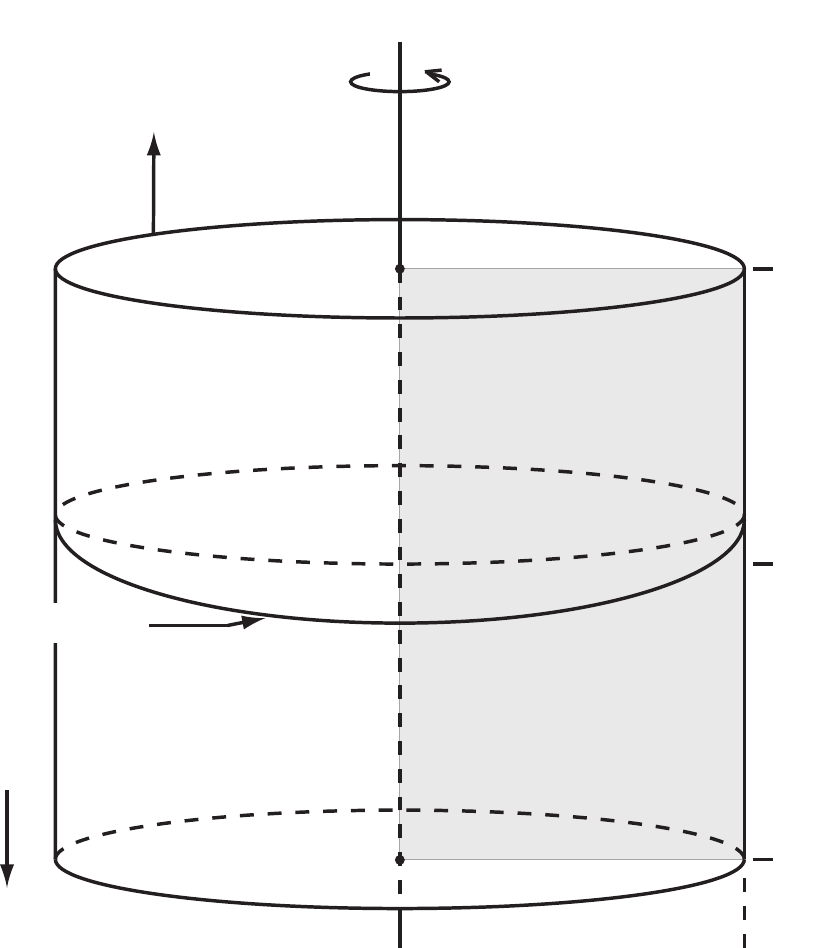}}
\put(-8,38){$g$}
\put(0,94){$z = z_0(r)$}
\put(124,230){$\Omega$}
\put(28,210){$g_1$}
\put(48,68){$\vc{u}_2$, $\rho_2$}
\put(48,148){$\vc{u}_1$, $\rho_1$}
\put(103,1){$0$}
\put(194,2){$a$}
\put(144,146){$\mathscr{D}_1$}
\put(144,62){$\mathscr{D}_2$}
\put(207,31){$-d$}
\put(206,109){$0$}
\put(206,186){$d$}
\end{picture}
\end{center}
\caption{Two layers of incompressible fluid of density $\rho_1$ and $\rho_2$ occupy a cylindrical tank of radius $a$ that is being accelerated \citep[see][]{taylor50} at a rate $g_1$.  When the tank is not rotating we take the interface between the fluids to be at $z=0$ (coordinates moving with the tank), the base of the tank at $z=-d$ and the lid of the tank at $z=d$.  The tank is spun up to have a constant angular velocity $\Omega$ about the $z$-axis.  The isobar describing the interface is given by $z = z_0(r)$ where $z_0(r) = \Omega^2(r^2 - \frac{1}{2}a^2)/(2g)$ and $p=p_0$ on $z=z_0(r)$.  The meridional plane is split into two domains, $\mathscr{D}_1$ and $\mathscr{D}_2$ representing the upper and lower layers respectively (shaded gray).}
\label{fig:schematic}
\end{figure}
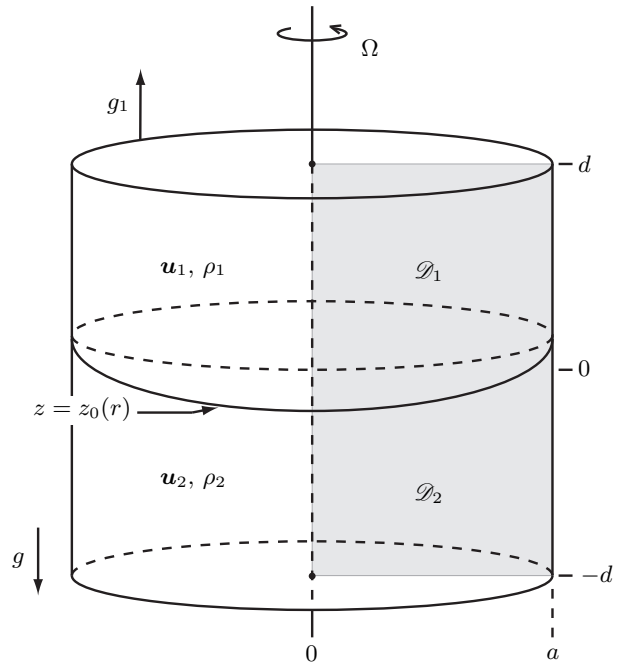

We begin by considering a two-layer rotating fluid as shown in Fig.~\ref{fig:schematic}.  The upper layer is denoted by a subscript 1 and the lower layer by a subscript 2.  We assume cylindrical polar coordinates with unit vectors $\vc{e_r}$, $\vc{e}_\theta$, and $\vc{e}_z$ in the radial, azimuthal, and vertical directions respectively and take the rotation to be described by the pseudovector $\vc{\Omega} = \Omega\,\vc{e}_z$.  The radius of the cylinder is $a$, and the lid and base of the cylinder are at $z=\pm d$.  Ignoring the effects of viscosity, we write the rotating Euler equation for the fluid in each layer as
\begin{equation} \label{eq:Eul_Cont2L}
\Dv{\vc{u}_j}{t} = -\frac{1}{\rho_j}\nabla p_j + \vc{g}^* 
-\vc{\Omega}\times\left(\vc{\Omega}\times\vc{x}\right) - 2\,\vc{\Omega}\times\vc{u}_j,
\end{equation}
for $j=1,2$, where $\vc{g}^* = -(g+g_1)\vc{e}_z$ and $\vc{u}_j$ and $\vc{x}$ are velocity and position vectors respectively, in the rotating frame.  For simplicity we drop the $g_1$ notation and will write $\vc{g}^*=-g\vc{e}_z$, with the understanding that $g$ may not be equal to the acceleration due to gravity, and may change sign as a result of external bulk acceleration of the system.  
When the fluid system is spun up into a hydrostatic regime (in the rotating, non-inertial reference frame) then $\vc{u}_j\equiv0$ and
\begin{equation} \label{eq:hydro_press}
p_j = p_0 - \rho_j\left\{gz - \frac{\Omega^2}{2}(r^2-{\textstyle \frac{1}{2}}a^2)\right\},\quad j = 1,2,
\end{equation}
where $p_0$ is a constant reference pressure equal to the pressure at the interface when the system is not rotating.  We take $z = z_0(r)$ to be the position of the interface between the two fluid layers.  In the absence of viscosity, requiring the stress to be continuous across the interface is equivalent to requiring continuity of pressure across the interface.  Hence we may write $p_1=p_2$ on $z=z_0(r)$, and it follows that the interface is an isobar on which $p_j=p_0$ and has profile given by
\begin{equation} \label{eq:interface}
z_0(r) = \frac{\Omega^2 (r^2-\frac{1}{2}a^2)}{2g}.
\end{equation}
The shape and position of the interface are independent of the densities of the fluid in the upper and lower layers.
Hence, whilst the value of $p_0$ and the stability of the interface may change according as to whether $\rho_1<\rho_2$ or {\it vice-versa}, the profile remains the familiar `concave' paraboloid such as may be observed at the free surface of a vigorously stirred beverage.  

Following \citet{taylor50} we investigate the development of the Rayleigh-Taylor instability under rotation by considering the development of a perturbation to the interface.  The strength of a stratification can be characterized by an Atwood number, defined here as $\Atw = (\rho_2 - \rho_1)/(\rho_2 + \rho_1)$.  Using this definition we have that for a stable stratification $\Atw>0$ and for an unstable stratification $\Atw<0$ [\textsc{n.b.}, in experimental investigations of the Rayleigh-Taylor instability, many authors, dealing only with unstable flows, define the Atwood number with opposite sign].  The amplitude of the perturbation and the velocity and pressure deviation from the hydrostatic are all assumed to be small.  We describe the fluid velocity and pressure perturbations in terms of a scalar potential, unifying the approaches of \citet{taylor50}, in modeling the non-rotating Rayleigh-Taylor instability, and \citet{miles59,miles64}, in modeling surface waves on a rotating fluid.  \citet{taylor50} used a standard velocity potential and \citet{miles64} used an `acceleration potential' of the kind proposed by \citet{poincare85}.  Here we make use of the `generalized potential' described by \citet{hart81}.  Specifically, for an interface perturbation
\begin{equation} \label{eq:intdef2L}
z = z_0(r) + \epsilon\, \zeta(r, \theta, t),
\end{equation}
where $\epsilon|\zeta|\ll d$, we take the velocity perturbation to the hydrostatic background to be
\begin{multline} \label{eq:udef2L}
\vc{u}_j = \epsilon \Bigg\{\!\left(1 + \frac{1}{4\Omega^2}\pd{^2}{t^2}\right)\!\nabla\phi_j - \frac{1}{2\Omega}\pd{}{t}\left(\vc{e}_z\!\times\!\nabla\phi_j\right) 
\\
+ \vc{e}_z\!\times\!\left(\vc{e}_z\!\times\!\nabla\phi_j\right)\Bigg\},
\end{multline}
for $j=1,2$, and the pressure to be
\begin{equation} \label{eq:pdef2L}
p_j = p_0 - \rho_j g\left[z - z_0(r)\right] - \epsilon\rho_j\left\{\pd{\phi_j}{t} + \frac{1}{4\Omega^2}\pd{^3\phi_j}{t^3}\right\},
\end{equation}
for  $j=1,2$.

Substitution of \eqref{eq:udef2L} and \eqref{eq:pdef2L} into \eqref{eq:Eul_Cont2L} shows that the rotating Euler equation is satisfied at leading order by the order 1 hydrostatic pressure terms and at order $\epsilon$ by the generalized potential $\phi$.  (We note that both the present formulation, and that of \citet{miles59,miles64}, necessarily imply a swirl component to the flow as soon as the radial velocity is non-zero.)  By further assuming that the fluid in each layer is incompressible, i.e., $\nabla\cdot\vc{u}_j=0$, we obtain the governing wave equation for each fluid layer
\begin{equation}\label{eq:potgov2L}
\left\{\partial_t^2\nabla^2 + 4\Omega^2\partial_z^2\right\}\phi_j = 0, \qquad j=1,2.
\end{equation}
Solutions to this type of wave equation in the context of inertial waves and internal gravity waves are well-known \citep[see, e.g.,][and references therein]{lighthill78}.

We seek to solve the governing equation \eqref{eq:potgov2L} together with the following boundary conditions: that there is no flow through the tank walls
\begin{equation} \label{eq:bcs1}
\left.\begin{array}{ll}
\vc{u}\cdot\vc{e}_r = 0,~\textnormal{on}~r=a,\\
\vc{u}\cdot\vc{e}_z=0,~\textnormal{on}~z=\pm d;
\end{array}\right\}
\end{equation}
the velocity on the axis of rotation, $r=0$, is sufficiently regular, specifically that 
\begin{equation} \label{eq:regularity}
r\partial \phi_j^2/\partial r \to 0 ~\textrm{as}~ r \to 0,
\end{equation}
(this condition allows for finite fluid velocities across the axis of rotation); and finally, we also require continuity of stress across the interface.  In the absence of viscosity we therefore require 
\begin{equation} \label{eq:pcont1}
p\,\big|^+_-=0,~\textnormal{across}~z = z_0 + \epsilon \zeta.
\end{equation}
Since $\zeta$ is unknown we require the kinematic condition that the interface moves with the local fluid velocity to close the system:
\begin{equation} \label{eq:kin1}
\Dv{}{t}(z_0 + \epsilon \zeta) = \vc{u}\cdot\vc{e}_z,~\textnormal{on}~z = z_0 + \epsilon \zeta.
\end{equation}

Following \citet{taylor50} and \citet{miles64} we adopt a variational formulation and seek normal mode solutions of the form
\begin{equation} \label{eq:stw}
\phi = \hat\phi(r,z)\exp\{\textrm{i}\left(\omega t + m\theta\right)\},
~
\zeta = \hat\zeta(r)\exp\{\textrm{i}\left(\omega t + m\theta\right)\},
\end{equation}
where $m\in\mathbb{N}_0$ is an azimuthal wavenumber.  Substitution into \eqref{eq:potgov2L} yields the governing equation
\begin{equation} \label{eq:potgov3}
\frac{1}{r}\pd{}{r}\left(r\pd{\hat\phi_j}{r}\right) - \frac{m^2}{r^2}\hat\phi_j + \left(1-\mu^2\right)\pd{^2\hat\phi_j}{z^2}=0,
~ j = 1,2,
\end{equation}
where we adopt Miles' \citep{miles64} notation by defining $\mu = 2\Omega/\omega$. 
The boundary conditions \eqref{eq:bcs1} and \eqref{eq:regularity} become
\begin{equation} \label{eq:bcs1a}
\left.\begin{array}{ll}
r\partial\hat\phi_j^2/\partial r \to 0  & \textnormal{as}~r \to 0, \\
r\partial\hat\phi_j/\partial r+\mu m \hat\phi_j = 0, &\textnormal{on}~r=a,\\
\partial\hat\phi_j/\partial z=0, &\textnormal{on}~z=\pm d,
\end{array}\right\}
\end{equation}
where the plus or minus is taken according to whether $j=1$ or $2$ respectively.
The condition of pressure continuity across the interface \eqref{eq:pcont1} yields at order $\epsilon$
\begin{equation} \label{eq:pcont2}
\textrm{i}\,\omega\mu^2\hat\zeta = \frac{2\Omega^2}{g}\left(1 - \frac{1}{\mu^2}\right)\left(\frac{1+\Atw }{\Atw }\hat\phi_2 -\frac{1-\Atw }{\Atw }\hat\phi_1\right),
\end{equation}
on $z = z_0$.  The kinematic condition \eqref{eq:kin1} at order $\epsilon$ can be written as
\begin{equation} \label{eq:kin2}
\textrm{i}\,\omega\mu^2\hat\zeta = z_0'\left(\pd{\hat\phi_j}{r} + \frac{\mu m}{r}\hat\phi_j\right) - \left(1-\mu^2\right)\pd{\hat\phi_j}{z},
~ j=1,2,
\end{equation}
on $z = z_0$ for each layer, where $z_0' \equiv \textrm{d}z_0/\textrm{d}r$.

The variational functional $\Phi[\hat\phi_1, \hat\phi_2]$ is defined by multiplying the governing equation \eqref{eq:potgov3} by $\rho_j\hat\phi_j$ and integrating over the domain $\mathscr{D} = \mathscr{D}_1\cup\mathscr{D}_2 =  [0,a]\times[-d,d]$ (see Fig.~\ref{fig:schematic}) so that 
\begin{equation} \label{eq:ig}
\Phi = \int_\mathscr{D}\rho\hat\phi\left\{\frac{1}{r}\pd{}{r}\left(r\pd{\hat\phi}{r}\right) - \frac{m^2}{r^2}\hat\phi +  (1-\mu^2)\pd{^2\hat\phi}{z^2}\right\}\textrm{d}A.
\end{equation}
Following the method outlined in \citet{miles64} we write the integral \eqref{eq:ig} in conservative form giving
\begin{multline} \label{eq:meth1}
\Phi = \int_\mathscr{D}\rho\left[\frac{1}{r}\pd{}{r}\!\left(r\hat\phi\pd{\hat\phi}{r}\right) + (1-\mu^2)\pd{}{z}\!\left(\hat\phi\pd{\hat\phi}{z}\right)\right]\textrm{d}A
 \\
-\int_\mathscr{D}\rho\left[\left(\pd{\hat\phi}{r}\right)^2 +\frac{m^2}{r^2}\hat\phi^2+ (1-\mu^2)\left(\pd{\hat\phi}{z}\right)^2\right]\textrm{d}A.
\end{multline}
We consider the first integral in \eqref{eq:meth1} and integrate over $\mathscr{D}_1$ and $\mathscr{D}_2$ separately.  Defining $I_1$ to be the integral over $\mathscr{D}_1$ and $I_2$ to be the integral over $\mathscr{D}_2$, we have
\begin{multline}
I_1 = \int_{z_0(0)}^{z_0(a)}\int_0^{r_0(z)} \frac{\rho_1}{r}\pd{}{r}\left(r\hat\phi_1\pd{\hat\phi_1}{r}\right)\,r\textrm{d}r\textrm{d}z 
\\ + \int_{z_0(a)}^d\int_0^a
 \frac{\rho_1}{r}\pd{}{r}\left(r\hat\phi_1\pd{\hat\phi_1}{r}\right)\,r\textrm{d}r\textrm{d}z \\
 + \int_0^a\int_{z_0(r)}^d\rho_1(1-\mu^2)\pd{}{z}\left(\hat\phi_1\pd{\hat\phi_1}{z}\right)r\textrm{d}z\textrm{d}r,
\end{multline}
where $r_0(z)$ is the well-defined inverse of $z_0(r)$.  Integrating and enforcing the boundary conditions $\partial\hat\phi_1/\partial z|_{z=d}=0$, $(r \partial\hat\phi_1/\partial r+\mu m\hat\phi_1)|_{r=a}=0$,  and $r\partial \hat\phi_1^2/\partial r\to 0$ as $r\to0$ implies
\begin{multline} \label{eq:theabove}
I_1= \rho_1\int_{z_0(0)}^{z_0(a)} \left. r \hat\phi_1\pd{\hat\phi_1}{r}\right|_{r=r_0(z)}\textrm{d}z \\
- \rho_1 \mu m\int_{z_0(a)}^d \left.  \hat\phi_1^2 \right|_{r=a}\textrm{d}z  \\
- \rho_1(1-\mu^2)\int_0^a \left.\hat\phi_1\pd{\hat\phi_1}{z}\right|_{z=z_0(r)} r\textrm{d}r.
\end{multline}
Transforming the first term in \eqref{eq:theabove} by making the substitution $z = z_0(r)$ gives the result
\begin{subequations} \label{eq:I1,2}
\begin{multline} 
I_1 = -\rho_1\mu m\int_{z_0(a)}^d \left. \hat\phi_1^2 \right|_{r=a}\textrm{d}z \\ + \rho_1 \int_0^a\hat\phi_1\left.\left\{z_0'\pd{\hat\phi_1}{r} - 
(1-\mu^2)\pd{\hat\phi_1}{z}\right\}\right|_{z=z_0(r)} r\,\textrm{d}r.
\end{multline}
Following a similar procedure, we may also show
\begin{multline}
I_2 = -\rho_2\mu m\int_{-d}^{z_0(a)} \left. \hat\phi_2^2 \right|_{r=a}\textrm{d}z \\ - \rho_2\int_0^a\hat\phi_2\left.\left\{z_0'\pd{\hat\phi_2}{r} - 
(1-\mu^2)\pd{\hat\phi_2}{z}\right\}\right|_{z=z_0(r)} r\,\textrm{d}r.
\end{multline}
\end{subequations}
Eliminating the interface perturbation, $\zeta$, from the pressure continuity condition \eqref{eq:pcont2} and the kinematic condition \eqref{eq:kin2} we see that
\begin{multline} \label{eq:newbcs}
z_0'\pd{\hat\phi_j}{r} - (1-\mu^2)\pd{\hat\phi_j}{z} = - z_0'\frac{\mu m}{r} \hat\phi_j \\ +
\frac{2\Omega^2}{g}\left(1 - \frac{1}{\mu^2}\right)\left(\frac{1+\Atw }{\Atw }\hat\phi_2 -\frac{1-\Atw }{\Atw }\hat\phi_1\right),
\end{multline}
for $j=1$, $2$ on $z=z_0(r)$.  Thus, we may rewrite (\ref{eq:I1,2}a,b) as 
\begin{subequations} \label{eq:intsv2}
\begin{multline} 
I_1 = -\rho_1\mu m\int_{z_0(a)}^d \left.\hat\phi_1^2 \right|_{r=a}\textrm{d}z 
 \\+ \int_0^a\rho_1\hat\phi_1\left.\left\{\frac{2\Omega^2}{g}\left(1 - \frac{1}{\mu^2}\right)\left(\frac{1+\Atw }{\Atw }\hat\phi_2 -\frac{1-\Atw }{\Atw }\hat\phi_1\right) 
 \right.\right. \\ \left.\left.
 - z_0' \frac{\mu m}{r} \hat\phi_1\right\}\right|_{z=z_0(r)} r\textrm{d}r,
\end{multline}
\begin{multline}
I_2 = -\rho_2\mu m\int_{-d}^{z_0(a)} \left.\hat\phi_2^2 \right|_{r=a}\textrm{d}z 
\\
- \int_0^a\rho_2\hat\phi_2\left.\left\{\frac{2\Omega^2}{g}\left(1 - \frac{1}{\mu^2}\right) \left(\frac{1+\Atw }{\Atw }\hat\phi_2 -\frac{1-\Atw }{\Atw }\hat\phi_1\right)
 \right.\right. \\ \left.\left.
  - z_0' \frac{\mu m}{r} \hat\phi_2\right\}\right|_{z=z_0(r)}  r\textrm{d}r.
\end{multline}
\end{subequations}
Substituting \eqref{eq:intsv2} into \eqref{eq:meth1} we have that
\begin{multline} \label{eq:new1}
\Phi[\phi_1, \phi_2] =  -\rho_1\mu m\int_{z_0(a)}^d \left.\hat\phi_1^2 \right|_{r=a}\textrm{d}z 
 \\
 + \int_0^a\rho_1\hat\phi_1\left.\left\{\frac{2\Omega^2}{g}\left(1 - \frac{1}{\mu^2}\right)\left(\frac{1+\Atw }{\Atw }\hat\phi_2 -\frac{1-\Atw }{\Atw }\hat\phi_1\right) 
 \right.\right. \\ \left.\left.
 - z_0' \frac{\mu m}{r} \hat\phi_1\right\}\right|_{z=z_0(r)} r\textrm{d}r
 -\rho_2\mu m\int_{-d}^{z_0(a)} \left.\hat\phi_2^2 \right|_{r=a}\textrm{d}z 
\\
- \int_0^a\rho_2\hat\phi_2\left.\left\{\frac{2\Omega^2}{g}\left(1 - \frac{1}{\mu^2}\right) \left(\frac{1+\Atw }{\Atw }\hat\phi_2 -\frac{1-\Atw }{\Atw }\hat\phi_1\right)
 \right.\right. \\ \left.\left.
  - z_0' \frac{\mu m}{r} \hat\phi_2\right\}\right|_{z=z_0(r)}  r\textrm{d}r
  \\
  -\int_{\mathscr{D}_1}\rho_1\left[\left(\pd{\hat\phi_1}{r}\right)^2 +\frac{m^2}{r^2}\hat\phi_1^2+ (1-\mu^2)\left(\pd{\hat\phi_1}{z}\right)^2\right]\textrm{d}A\\
   -\int_{\mathscr{D}_2}\rho_2\left[\left(\pd{\hat\phi_2}{r}\right)^2 +\frac{m^2}{r^2}\hat\phi_2^2+ (1-\mu^2)\left(\pd{\hat\phi_2}{z}\right)^2\right]\textrm{d}A.
\end{multline}
Taking the functional derivative of $\Phi$ with respect to, for example, $\hat\phi_1$, where $\delta_1\Phi \equiv \Phi[\hat\phi_1 + \delta\hat\phi_1, \hat\phi_2] - \Phi[\hat\phi_1,\hat\phi_2]$ yields, after some manipulation,
\begin{multline}
\delta_1\Phi =  2\rho_1\int_{\mathscr{D}_1} 
\left\{\frac{1}{r}\pd{}{r}\left(r\pd{\hat\phi_1}{r}\right)  \right.\\\left.
- \frac{m^2}{r^2}\hat\phi_1 + \left(1-\mu^2\right)\pd{^2\hat\phi_1}{z^2}\right\}\delta\hat\phi_1\textrm{d}A
\\
-2\rho_1\int_{z_0(a)}^d \left.\left\{\mu m \hat\phi_1 + r\pd{\hat\phi_1}{r}\right\}\delta\hat\phi_1\right|_{r=a}\textrm{d}z
\\
+2\rho_1\int_0^a\left.\left\{\frac{2\Omega^2}{g}\left(1 - \frac{1}{\mu^2}\right)\left(\frac{1+\mathscr{A}}{\mathscr{A}}\hat\phi_2 - \frac{1-\mathscr{A}}{\mathscr{A}}\hat\phi_1\right)
\right.\right. \\ \left.\left.
- z_0'\frac{\mu m}{r}\hat\phi_1 - \left[z_0'\pd{\hat\phi_1}{r} - (1-\mu^2)\pd{\hat\phi_1}{z}\right]\right\}\delta\hat\phi_1\right|_{z = z_0(r)}r\textrm{d}r.
\end{multline}
So we see that the functional $\Phi$ is stationary with respect to first-order variations of $\hat\phi_1$ about the solution of the governing equation \eqref{eq:potgov3} in $\mathscr{D}_1$, the boundary condition \eqref{eq:newbcs} for $j=1$ at the interface $z=z_0(r)$ and at the no-radial flow condition at $r=a$ on the boundary of $\mathscr{D}_1$.  Similarly, $\Phi$ is stationary with respect to first-order variations of $\hat\phi_2$ about the solution of the governing equation \eqref{eq:potgov3} in $\mathscr{D}_2$, the boundary condition \eqref{eq:newbcs} for $j=2$ at the interface $z=z_0(r)$ and the no-radial flow condition at $r=a$ on the boundary of $\mathscr{D}_2$.  (The Euler-Lagrange equation for $\Phi$ as expressed in \eqref{eq:ig} is the governing equation \eqref{eq:potgov3} multiplied by $2\rho$.)  Following \citet{miles64} we pose trial solutions that satisfy the governing equation \eqref{eq:potgov3}, the regularity condition at $r=0$ and the boundary conditions on $r=a$ and $z=\pm d$ exactly, and invoke the variational principle only in respect to the final boundary condition on $z=z_0(r)$.

If $\hat\phi$ is an exact solution of the governing equation \eqref{eq:potgov3}, it follows from the definition of $\Phi$ that $\Phi(\hat\phi)=0$.  Therefore, if $\hat\phi$ is a solution of \eqref{eq:potgov3}, it follows from \eqref{eq:meth1} and \eqref{eq:I1,2} that
\begin{multline} \label{eq:areaint}
\int_\mathscr{D}\rho\left[\left(\pd{\hat\phi}{r}\right)^2 +\frac{m^2}{r^2}\hat\phi^2+ (1-\mu^2)\left(\pd{\hat\phi}{z}\right)^2\right]\textrm{d}A = \\
-\rho_1\mu m\int_{z_0(a)}^d \left. \hat\phi_1^2 \right|_{r=a}\textrm{d}z 
-\rho_2\mu m\int_{-d}^{z_0(a)} \left. \hat\phi_2^2 \right|_{r=a}\textrm{d}z \\
+ \int_0^a\rho_1\hat\phi_1\left.\left\{z_0'\pd{\hat\phi_1}{r} - (1-\mu^2)\pd{\hat\phi_1}{z}\right\}\right|_{z=z_0(r)} r\textrm{d}r \\
- \int_0^a\rho_2\hat\phi_2\left.\left\{z_0'\pd{\hat\phi_2}{r} - (1-\mu^2)\pd{\hat\phi_2}{z}\right\}\right|_{z=z_0(r)} r\textrm{d}r.
\end{multline}
Substituting  \eqref{eq:areaint} into \eqref{eq:new1} we therefore have, after simplification
\begin{multline} \label{eq:Phi}
\Phi \propto \int_0^a\left\{\omega^2\left[\frac{1+\Atw }{\Atw }\hat\phi_2 - \frac{1-\Atw }{\Atw }\hat\phi_1\right]^2
\right.\\ \left.
+\left[\left(\frac{gz_0'}{\Omega^2 r}\right)\frac{\Omega^2}{1-\mu^2}\left(r\pd{}{r} + 2\mu m\right) - g\pd{}{z}\right]  \right. \\ \left.
\left[\frac{1+\Atw }{\Atw }\hat\phi_2^2 - \frac{1-\Atw }{\Atw }\hat\phi_1^2\right] \right\}\Bigg|_{z=z_0(r)}r\,\textrm{d}r.
\end{multline}
The constant of proportionality is $(\rho_2-\rho_1)(1-\mu^2)/4g$, but as interest is focussed upon stationary values of $\Phi$, it will be disregarded.  We may further simplify \eqref{eq:Phi} by noting that, for $z_0$ as defined by \eqref{eq:interface}, the factor $gz_0'/(\Omega^2 r)=1$.  The expression in \eqref{eq:Phi} is the two-layer equivalent of the functional given in (3.2) of \citet{miles64} and it can be seen that Miles' expression is recovered in the limit $\Atw =1$ (the stable single layer limit).  The cross term in the first term of the integrand is crucial in coupling the behavior of the two fluid layers.

Again, following \citet{miles64}, we seek to construct a series solution based on trial solutions of the form
\begin{equation} \label{eq:eigsol}
\hat\phi_{jn}(r,z) = \mathcal{J}_m\!\left(\frac{k_n r}{a}\right)
\cosh\!\left(\frac{k_n}{a}\frac{[z\mp d]}{\sqrt{1-\mu^2}}
\right),
\end{equation}
for $n = 1, 2, \ldots$, where $\mathcal{J}_m$ is a Bessel function of the first kind and we take the minus or plus sign in \eqref{eq:eigsol} according to whether $j=1$ or $2$ respectively.   The trial solutions \eqref{eq:eigsol} satisfy both the governing equation \eqref{eq:potgov3} and the no-vertical-flow boundary conditions at $z=\pm d$.  The radial no-flow condition at $r=a$ sets the possible modes of solution and so in general we sum over the countable number of solutions, $k_n$, of
\begin{equation} \label{eq:keq}
k \mathcal{J}_{m+1}(k) = m(1+\mu)\mathcal{J}_m(k),
\end{equation}
which follows from substituting \eqref{eq:eigsol} into \eqref{eq:bcs1a} and setting $r=a$.  (The ratio $k_n/a$ may be regarded as the radial wavenumber associated with the  $n^\textnormal{th}$ mode.)   We assume that as the number of terms in the series increases we approach a full solution.  Thus, we approximate $\hat\phi_1$ and $\hat\phi_2$ by
\begin{equation} \label{eq:modesum}
\hat\phi_j\approx\hat\phi_j^{(N)} = \sum_{n=1}^N c_{jn}\, \hat\phi_{jn},
~ j=1,2, ~\textnormal{for some}~N\geqslant1.
\end{equation}
We adopt a variational approach applied to \eqref{eq:Phi} in order to find the coefficients $c_{jn}$ such that our solution satisfies \eqref{eq:newbcs} on $z=z_0$, the remaining unsatisfied condition.  Specifically, by seeking stationary values of the functional $\Phi$, by taking the partial derivatives $\partial\Phi/\partial c_{jn}$, $j=1,2$, $n=1,\ldots,N$, we may construct $2N$ linear equations in the $2N$ coefficients.  The eigenvalue equation for $\omega$ is found by setting the determinant of this linear system to be zero.  If $\omega$ has a negative imaginary part then \eqref{eq:stw} implies growth, and the onset of the Rayleigh-Taylor instability.

In the remainder of \S \ref{sec:modeling} we initially consider purely axisymmetric instabilities, first asymptotically for low rotation rates in \S \ref{sec:axisymchandra}--\ref{sec:critrate}, and then numerically for arbitrary rotation rates in \S \ref{sec:axisarb}.  We then consider asymmetric instabilities, firstly asymptotically for low rotation rates in \S \ref{sec:lrgwm}--\ref{sec:nocritrate}, and then numerically for arbitrary rotation rates in \S \ref{sec:marb} and \S \ref{sec:asymmarb}.

\subsection{Axisymmetric instability, $m=0$}
\label{sec:axisym}

In the first instance we consider purely axisymmetric motion: the special case $m=0$.  Setting $m=0$ in \eqref{eq:keq} shows that we sum over the zeros of $\mathcal{J}_1(k)$, which implies $k\in\mathbb{R}$.  

\subsubsection{Single mode, low rotation rate, gravity wave solutions: asymptotics}
\label{sec:axisymchandra}

Following \citet{miles64}, we initially consider a solution containing a single trial solution each in the upper and lower layers.  We further assume a low rotation rate such that $\alpha = \Omega^2 a/g \ll1$.  Using such an approximation Miles was able to explain the discrepancies between the theory of \citet{lamb32} and the experimental observations of \citet[][Fig.~12]{fultz62} and so we adopt this level of approximation for initial investigation.  Seeking an asymptotic expression for the eigenvalue equation for $\omega$, we take \eqref{eq:eigsol} for some single $n\in\mathbb{N}$.  By considering $\partial\Phi/\partial c_{1n}=0$ and $\partial\Phi/\partial c_{2n}=0$, and expanding in powers of $\alpha$ we find, after some significant manipulation, that an eigenvector of the solution is
\begin{equation}
\vc{c} \propto\left(1, -1-{\textstyle\frac{1}{6}}\coth(k_n\delta)\alpha + \mathcal{O}(\alpha^2)\right),
\end{equation}
where $\delta = d/a$, and the eigenvalue equation for $\omega$ is
\begin{multline} \label{eq:freq}
\omega^2 \sim g\Atw \frac{k_n}{a}\tanh(k_n\delta) + 2\Omega^2\left[1 + 2k_n\delta\textrm{csch}(2k_n\delta) \right. \\ \left.
- \frac{1}{24}k_n^2\Atw ^2\textrm{sech}^2(k_n\delta)\right] +\frac{g}{a}\mathcal{O}(\alpha^2). 
\end{multline}
We observe therefore that if $g\Atw <0$ then $\omega^2<0$ and interfacial perturbations will grow rather than oscillate -- the Rayleigh-Taylor instability.  The form of \eqref{eq:freq} suggests we may be able to suppress this growth to some extent by rotating the system, i.e., the second term in \eqref{eq:freq} may be used to compete with the first if it has the opposing sign.  However, it would be mistaken to suggest that \eqref{eq:freq} implies that given a sufficient rotation rate an unstable mode could be fully stabilized ($\omega^2>0$), as is concluded erroneously by \citet{sharma} in the context of particle laden Rayleigh-Taylor instability.  The expansion \eqref{eq:freq} is asymptotic and its validity breaks down when the second term is comparable to the first.  The correct approach is to consider an expansion when $\omega$, not $\Omega$, is small compared to $(a/g)^{1/2}$ (see \S \ref{sec:critrate}).

Whether the growth rate of a given wave mode is reduced or increased by rotation depends on the sign of the second term in \eqref{eq:freq}.  
Provided $|\Atw|/\delta\aleq8.72$ then there are no solutions for which the second term in \eqref{eq:freq} can be made negative, and so the effect of rotation is always to initially suppress a given wave mode.  (The threshold coefficient, $c\approx8.72$, is given by
\begin{displaymath}
c^2 = \frac{24}{\xi_0^2}\left[\xi_0\coth\xi_0 + \cosh^2\xi_0\right],
\end{displaymath}
where
\begin{displaymath}
\xi_0\left[\sinh(4\xi_0)-2\xi_0\right] = 2\left[\sinh(2\xi_0)+\xi_0\right]^2,
\end{displaymath}
giving $\xi_0\approx1.39$.)
However, if $|\Atw|/\delta\ageq8.72$, indicating a sufficiently strong stratification, or sufficiently shallow aspect ratio, then there may exist wave modes which are excited by rotating the system.  For example, $\Atw=-\frac{1}{2}$, $\delta = \frac{1}{18}$, $n=7$ gives $|\Atw|/\delta = 9>c$, $k_7 \approx 22.76$ and the second term of \eqref{eq:freq} is approximately $-0.14$, i.e., the seventh mode is excited rather than suppressed as the first six modes are.

Rather than considering the limit of low rotation rate, $\alpha\ll1$, we may substitute \eqref{eq:eigsol}  into \eqref{eq:Phi} with $m=0$ and take $\delta\to\infty$, which may be thought of as forcing a horizontal initial interface, rather than parabolic, to find
\begin{equation}
\omega^4 - 4\Omega^2\omega^2 - \omega_0^4=0,\quad\textnormal{where}\quad\omega_0^2 = g\Atw\frac{k_n}{a},
\end{equation}
the solution of which, selecting the physically appropriate branch by introducing the factor $\Atw/|\Atw|$, is Chandrasekhar's solution \citep[][eqs. 162, 163]{chandra} given by
\begin{equation} \label{eq:chandra}
\omega^2 = 2\Omega^2 + \frac{\Atw}{|\Atw|}\sqrt{4\Omega^4 + \omega_0^4},
\end{equation}
in the present notation.  We can expect that when we have large aspect ratio, $\delta$, and moderate values of $\alpha$, \eqref{eq:chandra} will be a better approximation to $\omega$ than the asymptotic expansion \eqref{eq:freq} since no small rotation rate approximation has been made in the case of \eqref{eq:chandra}.  (We note that the two solutions \eqref{eq:freq} and \eqref{eq:chandra} coincide, as they must, if $\delta\gg1$, $\Omega^2 a/g\ll1$.)  

\subsubsection{Single mode, low rotation rate, inertial wave solutions: asymptotics}
\label{sec:lriw}

We show the presence of inertial waves when $\omega^2\sim\mathcal{O}(\alpha)$.  We consider $\partial\Phi/\partial c_{1,n}=0$ and $\partial\Phi/\partial c_{2,n}=0$ for a single $n\in\mathbb{N}$, but specifically seek solutions for which $\omega^2$ does not have an order 1 contribution, but has a leading order contribution at $\mathcal{O}(\alpha)$.

In order to ensure that $\omega^2$ has no leading order contribution we find that we must satisfy
\begin{equation} \label{eq:iw1}
\sinh\left(\frac{2k_n\delta}{\sqrt{1-\mu^2}}\right) \sim \mathcal{O}(\alpha),
\end{equation}
which requires
\begin{equation} \label{eq:iw2}
\frac{\omega^2a}{g} \sim \frac{4\alpha}{1+[2k_n\delta_q]^2} + \mathcal{O}(\alpha^2),
\end{equation}
where $\delta_q = \delta/q\pi$, for $\pm q=1,2,\ldots$.  The frequencies associated with these wave modes depend upon whether $q$ is even or odd.  For $q$ odd
\begin{equation} \label{eq:qodd}
\frac{\omega^2a}{g}\sim\frac{4\alpha}{1+[2k_n\delta_q]^2}\left\{1\mp\frac{[2k_n\delta_q]^2}
{1+[2k_n\delta_q]^2}\frac{\alpha}{6\delta} + \mathcal{O}(\alpha^2)\right\},
\end{equation}
where the minus or plus sign is taken according as to whether the wave occurs mainly in the upper or lower fluid respectively.  The eigenvectors correspond to waves occurring either in predominantly the upper fluid, $\vc{c} = (1,\mathcal{O}(\alpha^2))$, or predominantly the lower fluid, $\vc{c} = (\mathcal{O}(\alpha^2),1)$.

For $q$ even
\begin{multline} \label{eq:milesinertia}
\frac{\omega^2 a}{g}\sim\frac{4\alpha}{1+[2k_n\delta_q]^2}\left\{
1- \frac{[2k_n\delta_q]^2}{
	\left(1+[2k_n\delta_q]^2\right)^2}\frac{1}{\Atw}
		\bigg[(4\delta_q)^2
		\right.\\\left.
		\pm\frac{1}{6}
			\left\{
				\left(1+[2k_n\delta_q]^2\right)
					\left(1+[2k_n\delta_q]^2-12(4\delta_q)^2\right)\Atw^2 \right.\right. \\ \left.\left. +36(4\delta_q)^4\right\}^{1/2}
					\bigg]\frac{\alpha}{\delta}+\mathcal{O}(\alpha^2)
		\right\}.\end{multline}
It is straightforward to show that when $\Atw =1$, $\delta_q$ is replaced by $\delta_q/2$,  and the minus sign is chosen in \eqref{eq:milesinertia} (corresponding to the flow taking place in the lower fluid) the solution in (4.13) \citet{miles64} is recovered.  The solutions Miles found correspond to the even $q$ solutions; hence $\delta_q$ must be replaced by $\delta_q/2$ above for comparison.   For even $q$ the associated eigenvector is
\begin{multline}
\vc{c} = \left(1,\frac{1}{6\left(1+\Atw\right)(4\delta_q)^2}\bigg\{\Atw\left(1+[2k_n\delta_q]^2-6(4\delta_q)^2\right)
\right.\\ \left.\left.
\mp\left[\Atw^2
\left(1+[2k_n\delta_q]^2\right)\left(1+[2k_n\delta_q]^2-12(4\delta_q)^2\right) \right.\right.\right. \\ \left.\left.\left. +36(4\delta_q)^4\right]^{1/2}\right\}+\mathcal{O}(\alpha)\right).
\end{multline}

The odd $q$ solutions have been missed in previous studies and, since the solutions are independent of $\Atw$, are present for all values of $\Atw$ including the special case $\Atw =1$.

\subsubsection{Single mode, critical rotation rate for stabilization}
\label{sec:critrate}

A critical rotation rate, $\Omega_c$, for which a single gravity wave mode is stable for $\Omega>\Omega_c$ and unstable for $\Omega<\Omega_c$ can be found by considering an asymptotic expansion of $\Phi$ as a series in $\omega^2 a/g$.  Near the stability threshold we are in a regime $\omega^2 a/g\ll1$ and thus an expansion to the first two terms of the series can be used to find the critical rotation rate.

We have that for $m=0$, $k_n$ is such that $\mathcal{J}_1(k_n)=0$ and so using the following results
\begin{multline} \label{eq:isbs}
\int_0^1 \frac{\mathcal{J}_0^2\left(k_n x\right)}{\mathcal{J}^2_0(k_n)}x\,\textrm{d}x = \frac{1}{2}, ~
\int_0^1 \frac{\mathcal{J}_0^2\left(k_n x\right)}{\mathcal{J}^2_0(k_n)}x^3\,\textrm{d}x =  \frac{1}{6},\\
\int_0^1 \mathcal{J}_0\left(k_n x\right)\mathcal{J}_1\left(k_n x\right)x^2\,\textrm{d}x = 0,
\end{multline}
we may show that if $\alpha = \alpha_0 +a \omega^2\alpha_1/g + \ldots$, to leading order the variational function $\Phi$ is proportional to
\begin{multline}
\frac{\omega^2a}{g}\left\{\left[\frac{1-\Atw}{\Atw}c_{1n}-\frac{1+\Atw}{\Atw}c_{2n}\right]^2  
\right. \\ \left.
+\frac{1-\Atw}{2\Atw}\left[\frac{k_n^2}{12}+\frac{\delta k_n^2}{\alpha_0}\right]c_{1n}^2 
-\frac{1+\Atw}{2\Atw}\left[\frac{k_n^2}{12}-\frac{\delta k_n^2}{\alpha_0}\right]c_{2n}^2\right\}.
\end{multline}
It follows that for non-trivial solutions of $\partial\Phi/\partial c_{1n}=0$ and $\partial\Phi/\partial c_{2n}=0$ we require to leading order
\begin{multline} \label{eq:eigeq}
\left\{\frac{1-\Atw}{\Atw} + \frac{1}{2}\left[\frac{k_n^2}{12}+\frac{\delta k_n^2}{\alpha_0}\right]\right\} \\ \times
\left\{\frac{1+\Atw}{\Atw} - \frac{1}{2}\left[\frac{k_n^2}{12}-\frac{\delta k_n^2}{\alpha_0}\right]\right\}
-\frac{1-\Atw^2}{\Atw^2}=0.
\end{multline}
At the instability threshold $\omega=0$ and hence $\alpha = \alpha_0$.  Thus, we may solve \eqref{eq:eigeq} for $\alpha_0=\alpha_c$, the critical value of $\alpha$ that yields $\omega=0$.  Hence, we find the critical rotation rate $\Omega_c$ to be given exactly by
\begin{multline} \label{eq:crit}
\frac{\Omega_c^2a}{g} = \frac{6\delta}{\Atw}\left(1 - \frac{k_n^2}{48}\right)^{-1} \\ \times\left[
\left\{1-\frac{k_n^2\Atw^2}{12}\left(1-\frac{k_n^2}{48}\right)\right\}^{1/2}-1\right].
\end{multline}
This result does not depend on exploiting a small rotation rate or other small external parameter and so is not asymptotic and is therefore true in general.  Since $\Omega_c\in\mathbb{R}$, \eqref{eq:crit} only applies for $-1\leqslant\Atw<0$, i.e., a critical rotation rate only exists if the fluid layers would be Rayleigh-Taylor unstable in a non-rotating regime, as might be anticipated on physical grounds.  Under this condition on $\Atw$, \eqref{eq:crit} can be shown to be a strictly monotonically increasing function in $k_n$, bounded such that $\alpha_c\in[0,12\delta)$.  

A key observation from \eqref{eq:crit} is that the monotonic dependence of $\alpha_c$ on $k_n$ means that for a given rotation rate all structures larger than the critical wavelength associated with $k_n$ are stabilised, whereas all structures smaller than the critical wavelength remain unstable.  This is in keeping with the physical arguments presented earlier in the introduction.

There exists a threshold rotation rate $\Omega = 4\delta$, where the hydrostatic interface intersects the lid and the base of the domain and, as a result, the assumed form of $\phi$ no longer satisfies the boundary conditions at $z=\pm d$.  So, although it follows from \eqref{eq:crit} that for a given radial wavenumber, $k_n$, there exists a critical rotation rate for stabilization, it is not guaranteed that this critical rotation rate is less than the threshold rotation rate $4\delta$. That is to say, although \eqref{eq:crit} implies that since there are no growing modes for $-1\leqslant\Atw<0$ when $\Omega_c^2a/g>12\delta$, suggesting all modes may therefore be made indefinitely stable, this absolute critical rotation rate cannot be attained before the model breaks down.  

In summary, \eqref{eq:crit} shows that for a given rotation rate there exists a critical wavelength, above which all axisymmetric modes are stable, but below which all short wavelength modes remain unstable.

\citet[][Chap.\,X \S 95]{chandra} considers the special case of a two-layer stratification of semi-infinite fluids with a horizontal interface and states that 
\begin{quote}
`\ldots it follows that in the present case rotation does not affect the instability or stability, as such, of a stratification \ldots'. 
\end{quote}
The critical rotation rate given in \eqref{eq:crit} shows that Chandrasekhar's (\citeyear{chandra}) result is a special case and not true in general for purely axisymmetric flows, supporting \citet{carnevale}.  The case of two semi-infinite fluids superposed is given by taking the limits $a\to\infty$, $d\to\infty$.  The assumption of a horizontal interface implies that these limits should be taken such that $\delta = d/a \to\infty$.  Taking the limit $\delta\to\infty$ in \eqref{eq:crit} shows that there is indeed no finite critical rotation rate to stabilize a given unstable mode as $\delta\to\infty$ since $\Omega_c\to\infty$.  However, as soon as $\delta<\infty$ there exists a finite critical rotation rate above which an unstable axisymmetric mode may be stabilized.

\subsubsection{\label{sec:axisarb} Single mode, arbitrary rotation rate solutions: numerics}

\begin{figure*} 
\begin{center}
\vspace*{4pt}
\setlength{\unitlength}{1pt}
\begin{picture}(218,212)(0,0)
\put(20,14){\includegraphics[width = 200pt]{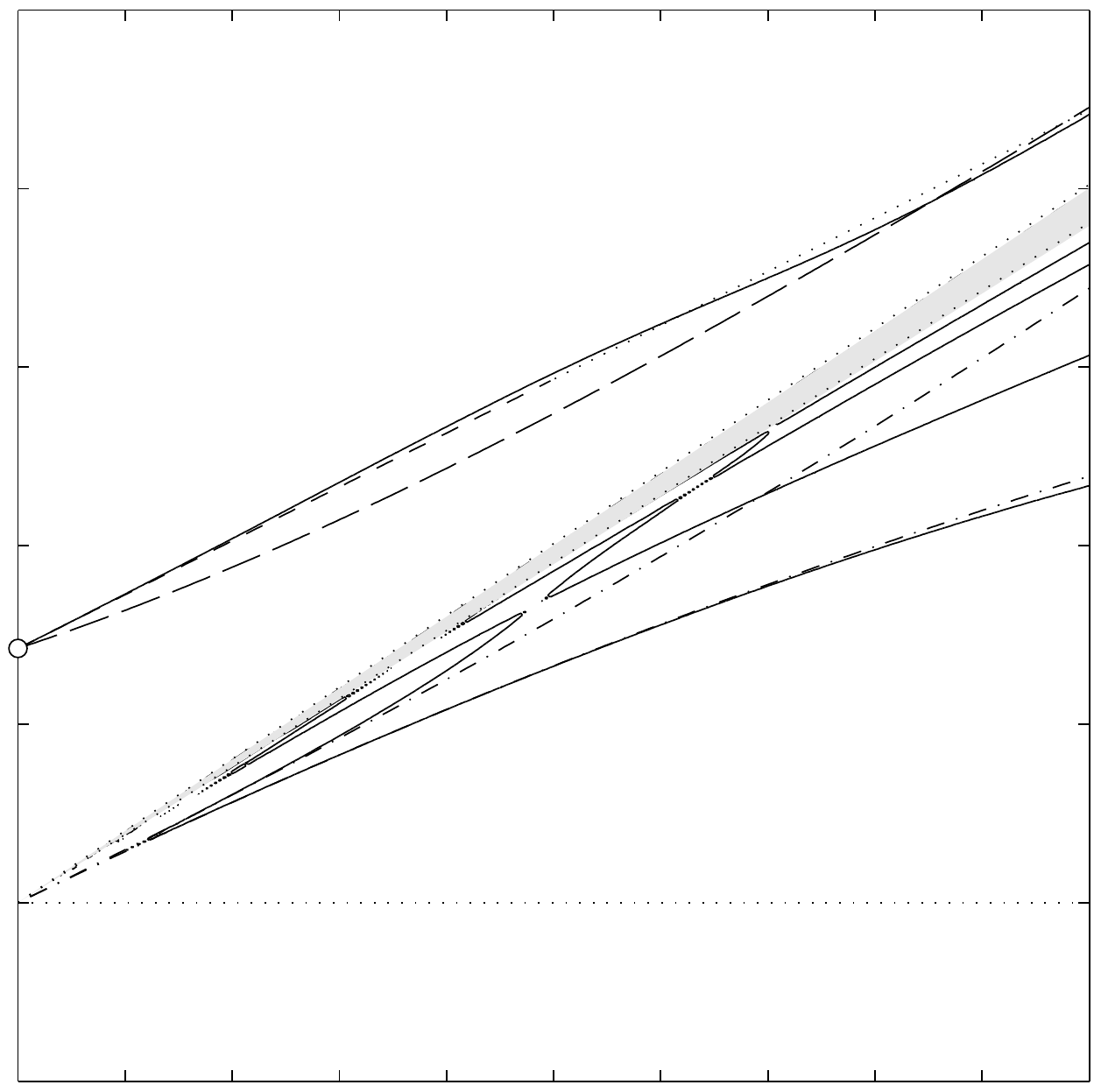}}
\put(2,102){\rotatebox{90}{$a\omega^2/g$}}
\put(118,-1){$\alpha$}
\put(30,194){(a) $\Atw=\frac{1}{2}$, $\delta = \frac{1}{4}$}
\put(21,8){{\small 0}}
\put(57,8){{\small 0.2}}
\put(95,8){{\small 0.4}}
\put(134,8){{\small 0.6}}
\put(173,8){{\small 0.8}}
\put(212,8){{\small 1.0}}
\put(13,13){{\small -1}}
\put(13,45){{\small \phantom{-}0}}
\put(13,77.5){{\small \phantom{-}1}}
\put(13,110){{\small \phantom{-}2}}
\put(13,143){{\small \phantom{-}3}}
\put(13,175){{\small \phantom{-}4}}
\put(13,208){{\small \phantom{-}5}}
\end{picture}
\hspace*{16pt}
\begin{picture}(218,212)(0,0)
\put(20,14){\includegraphics[width = 200pt]{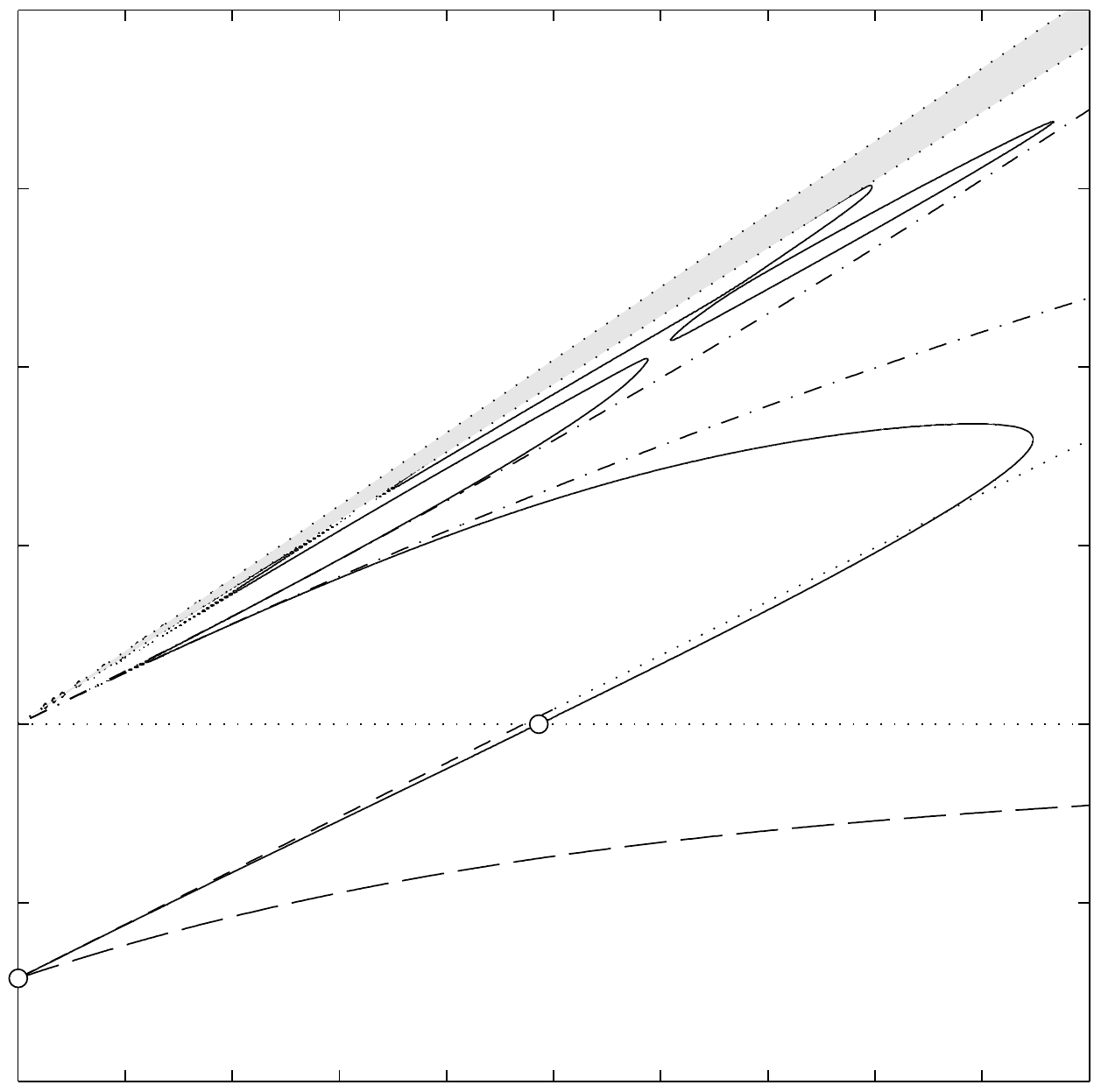}}
\put(2,102){\rotatebox{90}{$a\omega^2/g$}}
\put(118,-1){$\alpha$}
\put(30,194){(b) $\Atw=-\frac{1}{2}$, $\delta = \frac{1}{4}$}
\put(21,8){{\small 0}}
\put(57,8){{\small 0.2}}
\put(95,8){{\small 0.4}}
\put(134,8){{\small 0.6}}
\put(173,8){{\small 0.8}}
\put(212,8){{\small 1.0}}
\put(13,13){{\small -2}}
\put(13,45){{\small -1}}
\put(13,77.5){{\small \phantom{-}0}}
\put(13,110){{\small \phantom{-}1}}
\put(13,143){{\small \phantom{-}2}}
\put(13,175){{\small \phantom{-}3}}
\put(13,208){{\small \phantom{-}4}}
\end{picture}\\[12pt]
\begin{picture}(218,212)(0,0)  
\put(20,14){\includegraphics[width = 200pt]{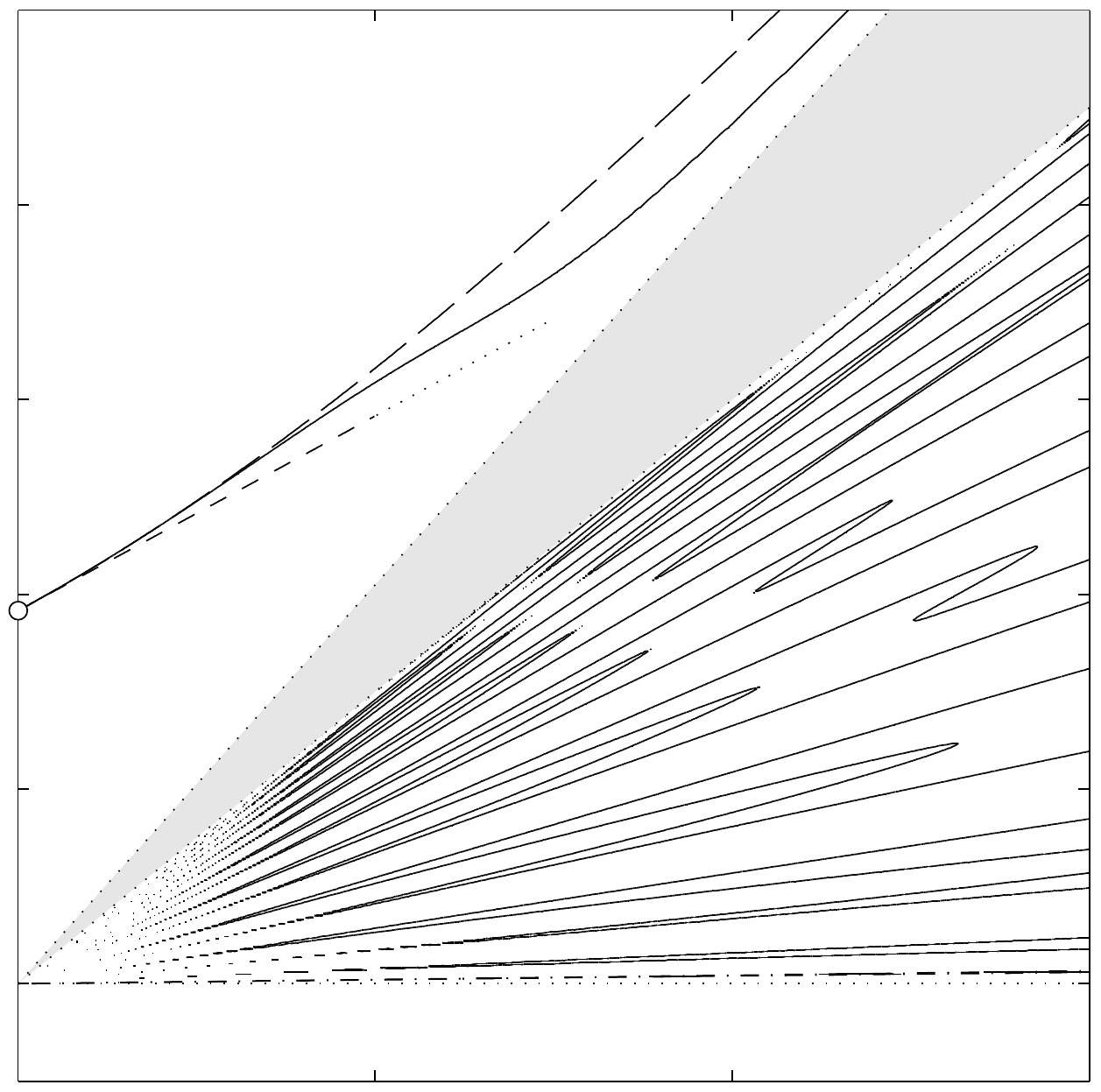}}
\put(2,102){\rotatebox{90}{$a\omega^2/g$}}
\put(118,-1){$\alpha$}
\put(32,194){(c) $\Atw=\frac{1}{2}$, $\delta = 4$}
\put(21,8){{\small 0}}
\put(83,8){{\small 0.5}}
\put(147,8){{\small 1.0}}
\put(212,8){{\small 1.5}}
\put(13,31){{\small \phantom{-}0}}
\put(13,66){{\small \phantom{-}1}}
\put(13,102){{\small \phantom{-}2}}
\put(13,137){{\small \phantom{-}3}}
\put(13,173){{\small \phantom{-}4}}
\put(13,208){{\small \phantom{-}5}}
\end{picture}
\hspace*{16pt}
\begin{picture}(218,212)(0,0)
\put(20,14){\includegraphics[width = 200pt]{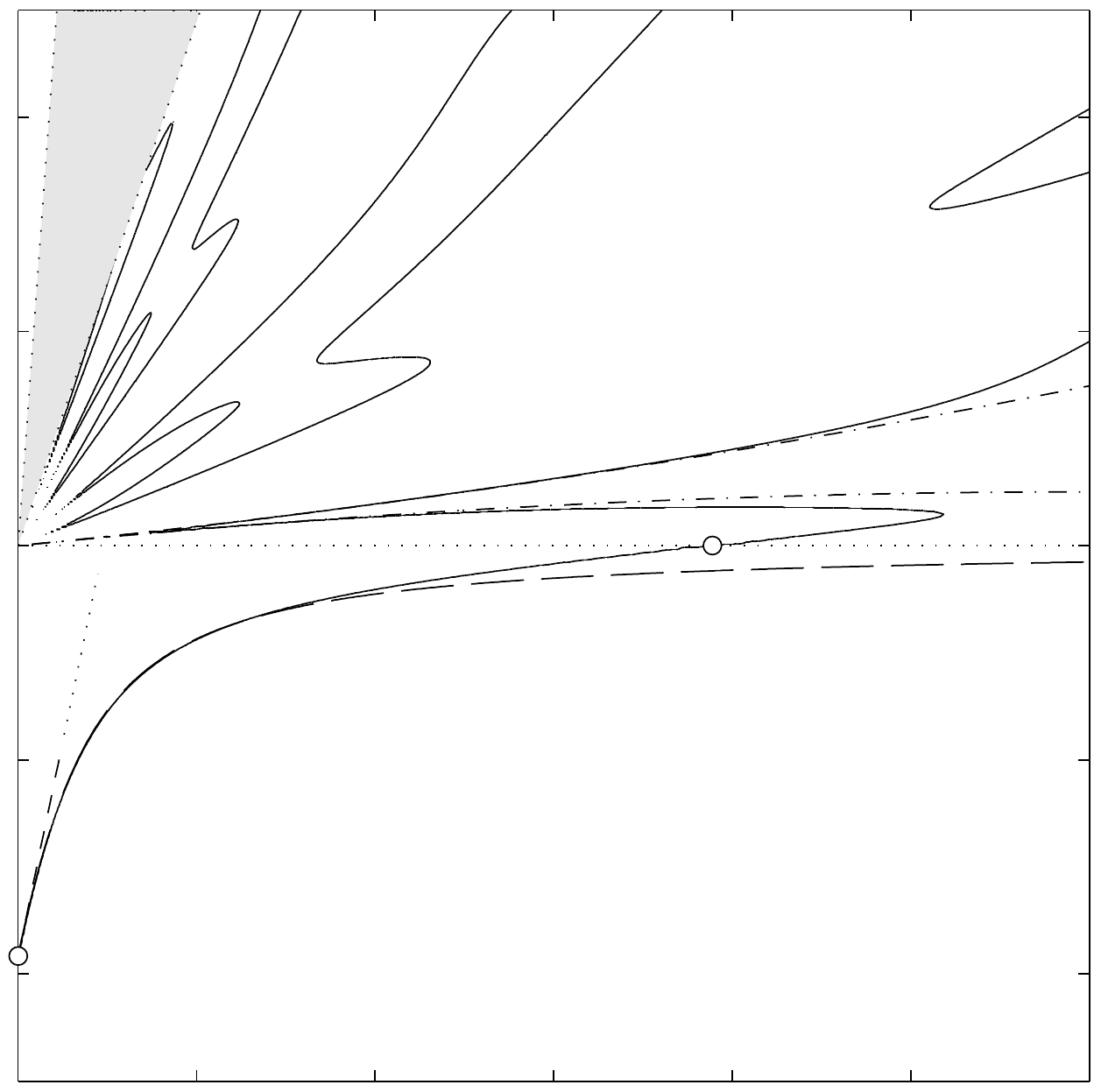}}
\put(2,102){\rotatebox{90}{$a\omega^2/g$}}
\put(118,-1){$\alpha$}
\put(34,194){(d) \hspace{90pt} $\Atw=-\frac{1}{2}$, $\delta = 4$}
\put(21,8){{\small 0}}
\put(53.5,8){{\small 2}}
\put(85.5,8){{\small 4}}
\put(118,8){{\small 6}}
\put(151,8){{\small 8}}
\put(180,8){{\small 10}}
\put(212,8){{\small 12}}
\put(13,32){{\small -2}}
\put(13,71){{\small -1}}
\put(13,110){{\small \phantom{-}0}}
\put(13,149){{\small \phantom{-}1}}
\put(13,188){{\small \phantom{-}2}}
\end{picture}
\end{center}
\caption{\label{fig:N1modes} Solutions of the eigenvalue problem, consistent with the assumptions of \S \ref{sec:growth}, describing the dispersion relation for Atwood numbers $\Atw =\pm\frac{1}{2}$, $\delta = \frac{1}{4},4$, $N=1$, $k = k_1 \approx 3.83$.   Solid lines are the exact solution calculated numerically.  The long-dashed lines correspond to Chandrasekhar's solution \eqref{eq:chandra}.  (a) Stable: $\Atw =\frac{1}{2}$, $\delta=\frac{1}{4}$.   The gravity wave solution coincides with the $\alpha=0$ axis at the value given by Taylor, indicated by a circle.  The asymptotic solution is shown dashed for $\alpha<0.5$ and continues dotted for larger values.  The first pair of inertial wave solutions \eqref{eq:qodd} corresponding to $q=1$ are shown (dot-dashed).  The greyed region contains an infinite number of possible inertial wave solutions corresponding to higher values of $q$.  (b) Unstable: $\Atw =-\frac{1}{2}$, $\delta=\frac{1}{4}$.  On $\alpha=0$ the unstable growth is predicted by Taylor's \citep{taylor50} result.  It can be seen that as the rotation rate $\alpha$ increases, one of the $q=1$ inertial wave solutions coalesces with the gravity wave solution.  The critical rotation rate is predicted by \eqref{eq:crit} and is given by $\alpha_c=0.49$.  (c) $\Atw=\frac{1}{2}$, $\delta = 4$.  With the increase in $\delta$ we see an improvement between the full solution and Chandrasekhar's solution, giving better agreement than the low rotation rate asymptotics \eqref{eq:freq}.  (d) $\Atw=-\frac{1}{2}$, $\delta = 4$.  There is excellent agreement with Chandrasekhar's solution for $\alpha<5$ compared with the low rotation rate asymptotics, but his solution remains in the unstable region as $\alpha\to\infty$, unlike the full solution.  The critical rotation rate, $\alpha_c = 7.78$, follows from \eqref{eq:crit}.  As in (b), one of the $q=1$ inertial wave solutions coalesces with the gravity wave solution. }
\end{figure*}
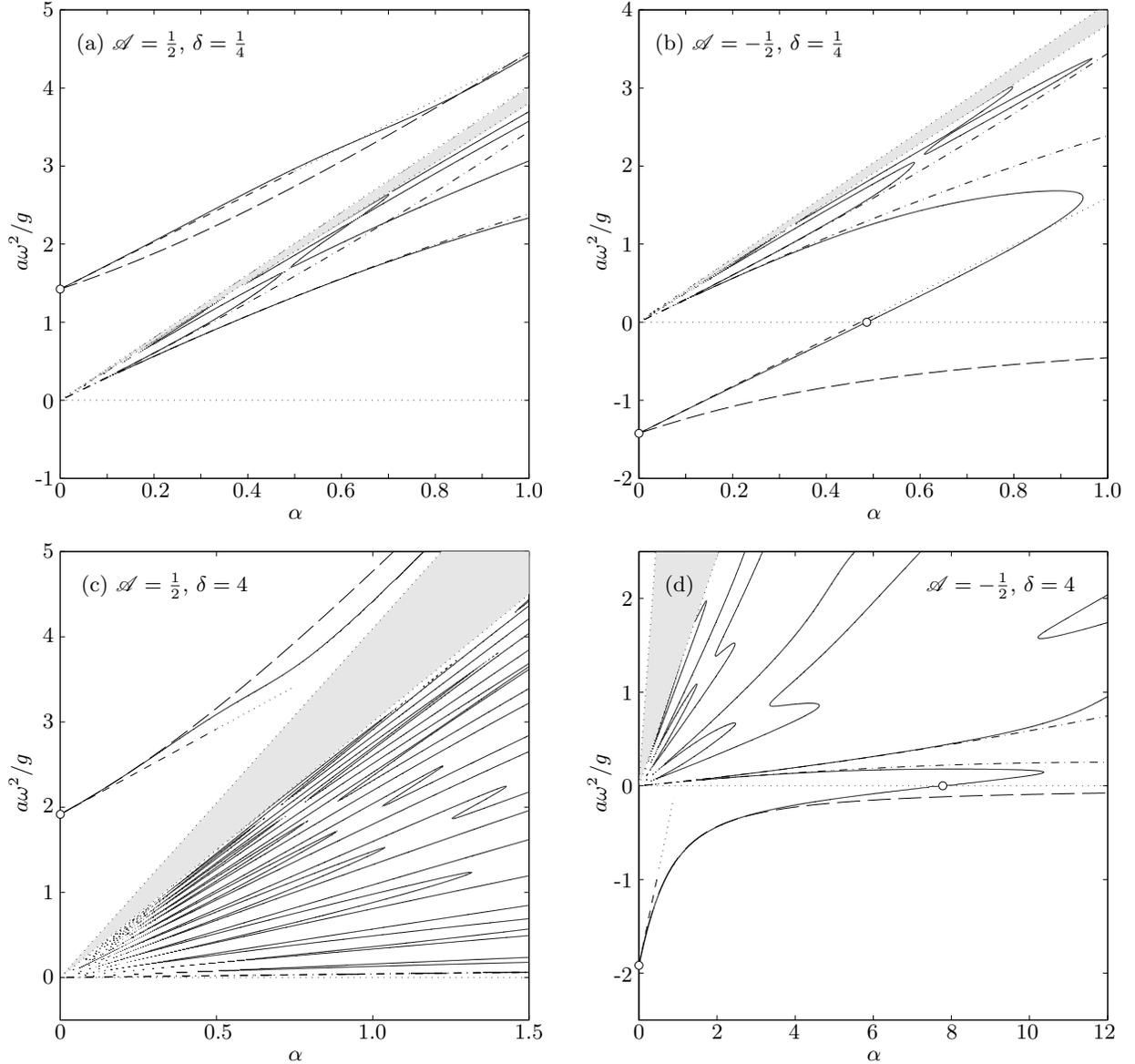

In order to obtain results at arbitrary rotation rate we proceed using a hybrid of analytical and numerical methods, whereby evaluation of integrals is carried out using Simpson's rule. 
For $N=1, n=1$ we construct the matrix of coefficients of $c_{jn}$ from the linear equations $\partial\Phi/\partial c_{jn}=0$ for $j=1,2$.  This yields a $2\times2$ matrix, $\mathsf{M}$, and the zeros of its determinant, corresponding to possible solutions, are calculated numerically and plotted in Fig.~\ref{fig:N1modes} for $\Atw  = \pm\frac{1}{2}$, $\delta = \frac{1}{4},4$.  The zero rotation rate solutions, as found by \citet{taylor50}, are indicated by white circles on the vertical axes.  Selecting $n=1$ gives $k=k_1$, the first zero of $\mathcal{J}_1$, and so we have $k\approx3.83$.

Inertial waves are present as a result of the rotation and it can be seen that these solutions all converge at the origin indicating that as the rotation rate tends to zero these waves are not supported, consistent with their definition.  The first pair of inertial wave solutions, corresponding to \eqref{eq:qodd} with $q=1$, are indicated by dot-dashed lines extending away from the origin.  The grayed-out regions contain an infinite number of inertial waves corresponding to the higher values of $q$.  Within this region the numerical contouring of $|\mathsf{M}|=0$ fails and so the region has been grayed-out.

In the stable cases, $\Atw=\frac{1}{2}$, shown in Fig.~\ref{fig:N1modes}a,\,c, the effect of the rotation on the gravity wave on the interface is only to increase its frequency, hence the comments of \citet{miles59} indicating that the effects of rotation are not especially interesting for axisymmetric waves on a single layer of fluid.  The asymptotic gravity wave solutions \eqref{eq:freq} are shown as the dashed lines extending away from the white circle on the vertical axes.  They are shown dashed for $\alpha<0.5$, after which we anticipate the approximations being less good and the solution is thereafter shown as dotted. 

In the unstable cases, $\Atw=-\frac{1}{2}$, shown in Fig.~\ref{fig:N1modes}b,\,d, the effect of rotation on the $k_1$ gravity wave at the interface is to change the sign of $\omega^2$ from negative (unstable -- Rayleigh-Taylor instability) to positive (stable -- standing wave solutions).  The rotation is able to completely stabilize the mode for $\alpha>\alpha_c$.  It can be seen that as the rotation rate is increased the gravity wave solution coalesces with the dominant inertial wave solution.   The predicted critical rotation rates are $\alpha_c\approx0.49$ for $\delta = \frac{1}{4}$ and $\alpha_c\approx7.78$ for $\delta = 4$.  It can be seen that for moderate values of $\alpha$ there is significant improvement in the agreement between the numerical solution and Chandrasekhar's \citep{chandra} solution for the larger value of $\delta$, as expected (see \S \ref{sec:axisymchandra}).  With the parameters used in Fig.~\ref{fig:N1modes}b, the asymptotic value of $\alpha_c$ calculated for large $N$ is within 3.4\% of that calculated using $N=1$ modes, as in \eqref{eq:crit}.

It can be shown that as a result of (\ref{eq:isbs}b), the key results of \S \ref{sec:axisym}, \eqref{eq:freq} and \eqref{eq:crit}, are independent of the $\Omega^2/(1-\mu^2)$ term in \eqref{eq:Phi}, the only term that has an explicit dependence on the profile $z_0$.  As the low rotation rate approximation \eqref{eq:freq} and the critical rotation rate \eqref{eq:crit} are independent of this term it follows that the unstable solution branch for $\omega$ can be well-approximated by neglecting this term.  Indeed, for low to moderate Atwood number ($\Atw\aleq\frac{1}{2}$) then
\begin{equation} \label{eq:approxPhi}
\Phi\propto\int_0^a\left.\left\{\omega^2\left(\hat\phi_2 - \hat\phi_1\right)^2 - g \Atw \pd{}{z}\left(\hat\phi_2^2 - \hat\phi_1^2\right)\right\}\right|_{z = z_0}r\,\textrm{d}r
\end{equation}
is a reasonable approximation to \eqref{eq:Phi}, with approximate $\mathcal{O}(\Atw^2)$ error.  The calculated critical rotation rate for the example considered in Fig.~\ref{fig:N1modes}b using \eqref{eq:approxPhi}, as opposed to \eqref{eq:Phi}, is $\alpha_c = 0.45$ compared to $\alpha_c=0.49$, an error of approximately 7.8\%.

\subsection{Non-axisymmetric instability, $m\ne0$}
\label{sec:asymm}

We now consider the more general case which includes non-axisymmetric modes.  Here, the right hand side of \eqref{eq:keq} can be non-zero, and so $\omega\in\mathbb{C}$, giving the possibility of both growth and precession of the instability.  As $\omega\in\mathbb{C}$ it follows that $k=k(\Omega,\omega)\in\mathbb{C}$ in general.  The fact that $k$ cannot be determined {\it a priori} for the whole solution space increases the difficulty of calculating solutions for the non-axisymmetric cases compared to the axisymmetric cases.

\subsubsection{Single mode, low rotation rate, gravity wave solutions: asymptotics}
\label{sec:lrgwm}

To find the corresponding low rotation rate asymptotics as in \S \ref{sec:axisym} we expand both $\omega$ and $k$ in terms of $\alpha$.  It follows from \eqref{eq:keq} for $\omega\sim\omega_0 + \omega_1\alpha^{1/2} + \omega_2\alpha + \ldots$ that
\begin{multline} \label{eq:kser}
\frac{k}{k_0}\sim 1 +\frac{2m}{k_0^2 - m^2} \left(\frac{\alpha g}{a\omega_0^2}\right)^{1/2}
-\frac{2m}{k_0^2-m^2}\left[\left(\frac{a\omega_1^2}{g}\right)^{1/2}
\right. \\ \left.
+\frac{m\left(k_0^2 + m^2\right)}{\left(k_0^2 - m^2\right)^2}\right]\left(\frac{\alpha g}{a\omega_0^2}\right)+\mathcal{O}(\alpha^{3/2}),
\end{multline}
where $k_0\in\mathbb{R}$ satisfies 
\begin{equation} \label{eq:k0}
k_0\mathcal{J}_{m+1}(k_0) = m\mathcal{J}_m(k_0).
\end{equation}
(Note that again there are a countable number of solutions $k_{0n}$ but for clarity we will use the notation $k_0$ and understand that it may not be the first zero of \eqref{eq:k0}.)  Substituting in and following a similar procedure to that in \S \ref{sec:axisym}, the first two terms for $\omega$ satisfy
\begin{subequations} \label{eq:om0om1}
\begin{equation} 
\frac{a\omega_0^2}{g} = \Atw k_0\tanh(k_0\delta),
\end{equation}
\begin{equation}
\sqrt{\frac{a}{g}}\omega_1=\frac{m}{k_0^2 - m^2}\left[1 + 2k_0\delta \textrm{csch}\left(2k_0\delta\right)\right].
\end{equation}
\end{subequations}
The leading order term $\omega_0$ is unchanged from \eqref{eq:freq}, noting the change in definition of $k_0$.  The $\omega_1$ term is not present in \eqref{eq:freq}, as a result of $m=0$ in the axisymmetric case.  However we note that $\omega_1\in\mathbb{R}$ and so this term can play no role in the growth or suppression of interfacial waves; it is merely contributing a modification to the precession velocity.  We also note that $\omega_1$ is independent of $\Atw$ and is therefore exactly the same as the first correction term found by \citet[][eq. (5.5)]{miles64}.

For comparison with the second term on the right hand side of \eqref{eq:freq} we now calculate $a(2\omega_0\omega_2+\omega_1^2)/g$ and find it to be
\begin{multline} \label{eq:om2}
2\left\{1 - \frac{2 m^2 k_0^2}{\left(k_0^2 - m^2\right)^3} + 2k_0\delta \textrm{csch}\left(2k_0\delta\right)
    \right. \\ \left.
     \times\left[1- \frac{m^2}{(k_0^2-m^2)^2}\left(\frac{k_0^2+m^2}{k_0^2-m^2}+2k_0\delta \textrm{coth}\left(2k_0\delta\right)\right)\right]
    \right. \\ \left.
    -    \frac{1}{8}k_0^2\Atw^2 \textrm{sech}\left(k_0\delta\right)^2\left[1+\frac{4}{k_0^2-m^2}
    \right.\right. \\ \left.\left. \times
    \left(\frac{m^2}{k_0^2}\cosh\left(k_0\delta\right)^2
     - k_0^2G(m,k_0)\right)\right]\right\},
\end{multline}
where we use \eqref{eq:isbs} and define
\begin{equation} \label{eq:Gdef}
G(m,k) = \int_0^1 \frac{\mathcal{J}_m(k x)^2}{\mathcal{J}_m(k)^2}x^3\,\textrm{d}x.
\end{equation}
Provided $k_0$ is a solution of \eqref{eq:k0} then in the limit $m\to0$, $G(m,k_0(m))\to\frac{1}{6}$ and we may recover the axisymmetric $m=0$ term in \eqref{eq:freq} from \eqref{eq:om2}.  The associated eigenvector  with the solution described by \eqref{eq:om0om1} and \eqref{eq:om2} is
\begin{multline}
\vc{c} = \left(1,-1 - \frac{k_0}{2}\,\textrm{coth}\,\left(k_0\delta\right)
\right. \\ \left. \times
\left[1+\frac{4}{k_0^2-m^2}\left(\frac{m^2}{k_0^2}
-k_0^2G(m,k_0)\right)\right]\alpha + \mathcal{O}(\alpha^2)\right),
\end{multline}
and we note that therefore to leading order the solution in the lower layer is growing and precessing in the opposite direction to the fluid in the upper layer, as might have been anticipated.

It follows from \eqref{eq:om2} that $\omega_2\in\mathbb{C}$ if $\omega_0\in\mathbb{C}$ and so may contribute to both precession and growth/decay.
Whether the growth rate of a wave mode is reduced or increased by a small amount of rotation, compared to its growth in a non-rotating system, is controlled by \eqref{eq:om2} too, since $\omega_1\in\mathbb{R}$.  

\subsubsection{Single mode, low rotation rate, inertial wave solutions: asymptotics}

As with the axisymmetric case, for $\omega^2$ to have a leading order contribution of $\mathcal{O}(\alpha)$ we require \eqref{eq:iw1} and hence \eqref{eq:iw2} to be satisfied.  Writing $\omega\sim \omega_1\alpha^{1/2} + \omega_2\alpha+\ldots$ and $k\sim k_0 + k_1\alpha^{1/2} + k_2\alpha + \ldots$ we have that
\begin{equation} \label{eq:iwm1}
\frac{\omega_1^2a}{g} = \frac{4}{1+[2k_0\delta_q]^2}\quad\textnormal{for}\quad \delta_q \equiv \frac{\delta}{q\pi},\quad\textnormal{and}\quad q\in\mathbb{N}.
\end{equation}
The leading order balance of \eqref{eq:keq} is therefore
\begin{equation} \label{eq:iwm2}
\mathcal{J}_{m+1}(k_0) = \frac{m}{k_0}\left(1+\frac{2}{\omega_1}\right)\mathcal{J}_m(k_0).
\end{equation}
Combining \eqref{eq:iwm1} and \eqref{eq:iwm2} we have that for a given $m\ne0$ and $\delta_q$, $k_0$ must satisfy
\begin{equation}
1+[2k_0\delta_q]^2 = \left(1-\frac{k_0}{m}\frac{\mathcal{J}_{m+1}(k_0)}{\mathcal{J}_m(k_0)}\right)^2.
\end{equation}
The solutions fall into two categories according as to whether $q$ is odd or even, as before.  For $q$ odd
\begin{multline} \label{eq:m_podd}
\frac{\omega^2 a}{g}\sim\frac{4\alpha}{1+[2k_0\delta_q]^2}
\\ \times
\left\{1\mp\frac{\alpha[2k_0\delta_q]^2}{2\delta}
\left(1+4\left[\delta_q^2m^2 - G(m,k_0)\right]\right) \right.\\\left.
\times\left[\left(1+4\delta_q^2m^2\right)\left(1+[2k_0\delta_q]^2\right) - \frac{8m\delta_q^2}{\omega_1}\right]^{-1}\right\}.
\end{multline}
The expression for $q$-even is lengthy and so here we note only the solutions for extreme values of $\delta$, specifically for $q$ even, $m\ne0$, then for $\delta \ll 1$
\begin{multline}
\frac{\omega^2a}{g}\sim
 4\alpha\Bigg\{1\pm\frac{2\delta_q^2\alpha}{\delta}\left(k_0^2\left[1-4G(m,k_0)\right]+4m\right)
\\ + \mathcal{O}(\alpha^2)\Bigg\} ,
 \end{multline}
 and for $\delta \gg 1$
\begin{equation}
\frac{\omega^2a}{g}\sim4\alpha\left\{\frac{1}{[2k_0\delta_q]^2}\pm\frac{\alpha}{\delta}\frac{m}{\delta_qk_0^3}+ \mathcal{O}(\alpha^2)\right\}.
\end{equation}

A further, higher order, solution exists, provided $m\ne0$, for $k\sim k_0 + \mathcal{O}(\alpha)$ where $\mathcal{J}_m(k_0)=0$ and
\begin{multline}
\frac{\omega^2a}{g}\sim\left(\frac{2m}{k_0^2\delta}\right)^2\alpha^3\Bigg\{1 - \frac{\alpha}{2\delta}\Bigg[4G^+(m,k_0)
\\
 - 1 + \frac{4}{k_0^2}\left(1\pm\frac{2}
{\Atw}\right)\Bigg] + \mathcal{O}(\alpha^2)\Bigg\},
\end{multline}
where
\begin{equation}
G^+(m,k) = \int_0^1\frac{\mathcal{J}_m^2(k x)}{\mathcal{J}_{m+1}^2(k x)}x^3\,\textrm{d}x.
\end{equation}

\subsubsection{Single mode, critical rotation rate for stabilization}
\label{sec:nocritrate}

In \S \ref{sec:critrate} it was shown that for $\delta<\infty$ there exists a critical rotation rate, $\Omega_c$, above which an axisymmetric wave mode
can be stabilized for a given unstable Atwood number.  Here we show that such a critical rotation rate does not exist in the case $m\ne0$.

For $m\ne0$ and $\Omega\sim\Omega_0\left[1 + (\Omega_1/\Omega_0)\omega + \mathcal{O}(\omega^2)\right]$, \eqref{eq:keq} implies that
\begin{equation}
\frac{k}{k_0}\sim 1 - \frac{\omega}{2m\Omega_0} + \mathcal{O}(\omega^2),\quad\textnormal{where}\quad \mathcal{J}_m(k_0)=0,
\end{equation}
noting that $m\ne0$ changes the definition of $k_0$ from the axisymmetric definition $\mathcal{J}_{m+1}(k_0)=0$, to $\mathcal{J}_{m}(k_0)=0$.  The eigenvalue equation for $\Omega$ becomes
\begin{equation} \label{eq:nocrit}
\frac{1-\Atw^2}{\Atw^2}\left[a^2\,m\,\Omega_0\,\mathcal{J}_{m+1}^2(k_0)\right]^2\omega^2 + \mathcal{O}\left(\omega^3\right)=0.
\end{equation}
It can be seen that there is no non-zero critical rotation rate, $\Omega_0$, that can force the leading order term in \eqref{eq:nocrit} to be zero.  Therefore, unlike the axisymmetric $m=0$ case, there does not exist a critical rotation rate that can be used to stabilize a given wave mode.  However, a given wave mode may still be suppressed (or indeed excited) by rotation, but a change of stability cannot occur.

\subsubsection{Single mode, arbitrary rotation rate solutions: numerics}
\label{sec:marb}

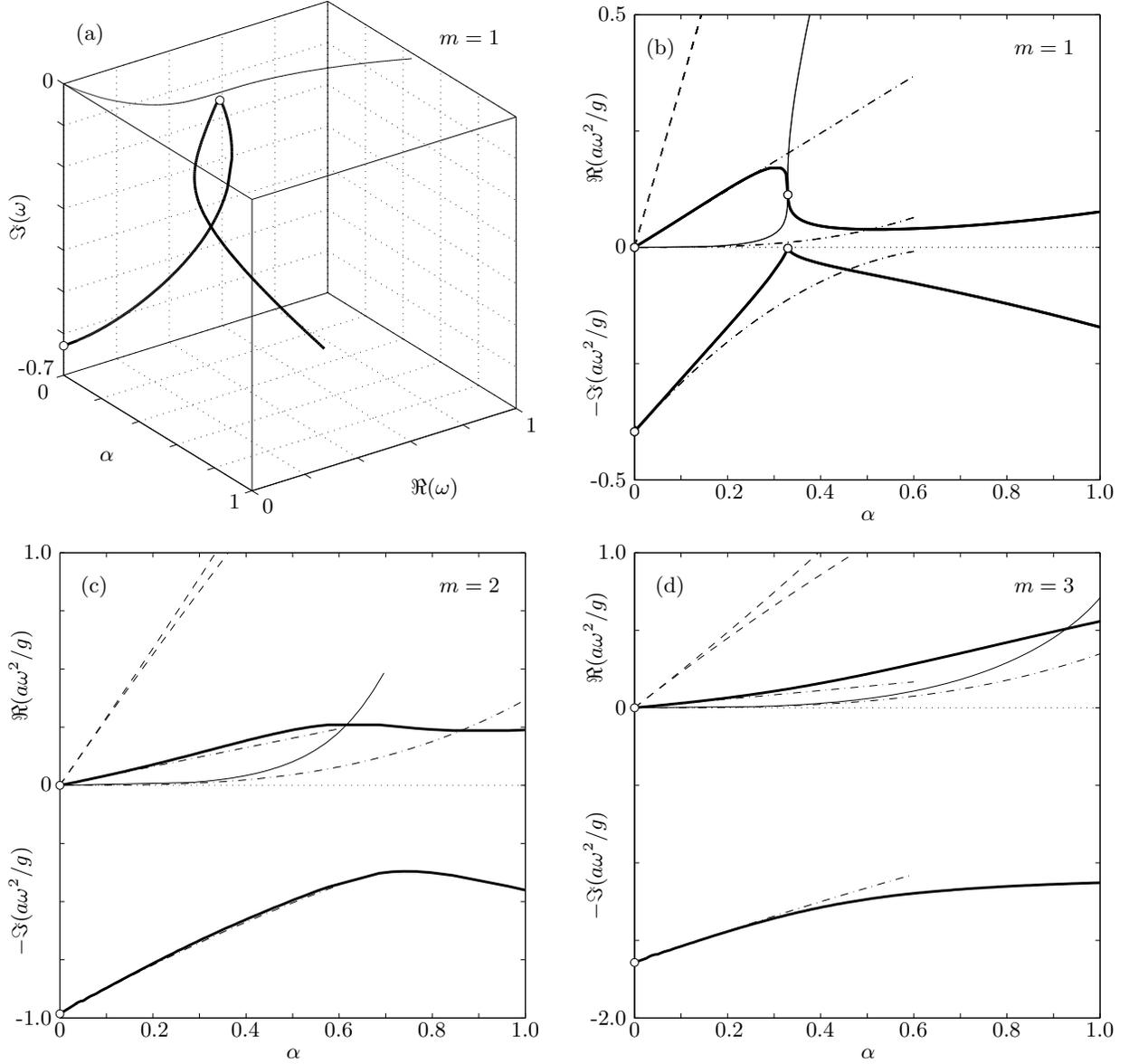
\begin{figure*}
\begin{center}
\vspace*{4pt}
\setlength{\unitlength}{1pt}
\begin{picture}(218,212)(0,0)   %
\put(20,7){\includegraphics[width = 200pt]{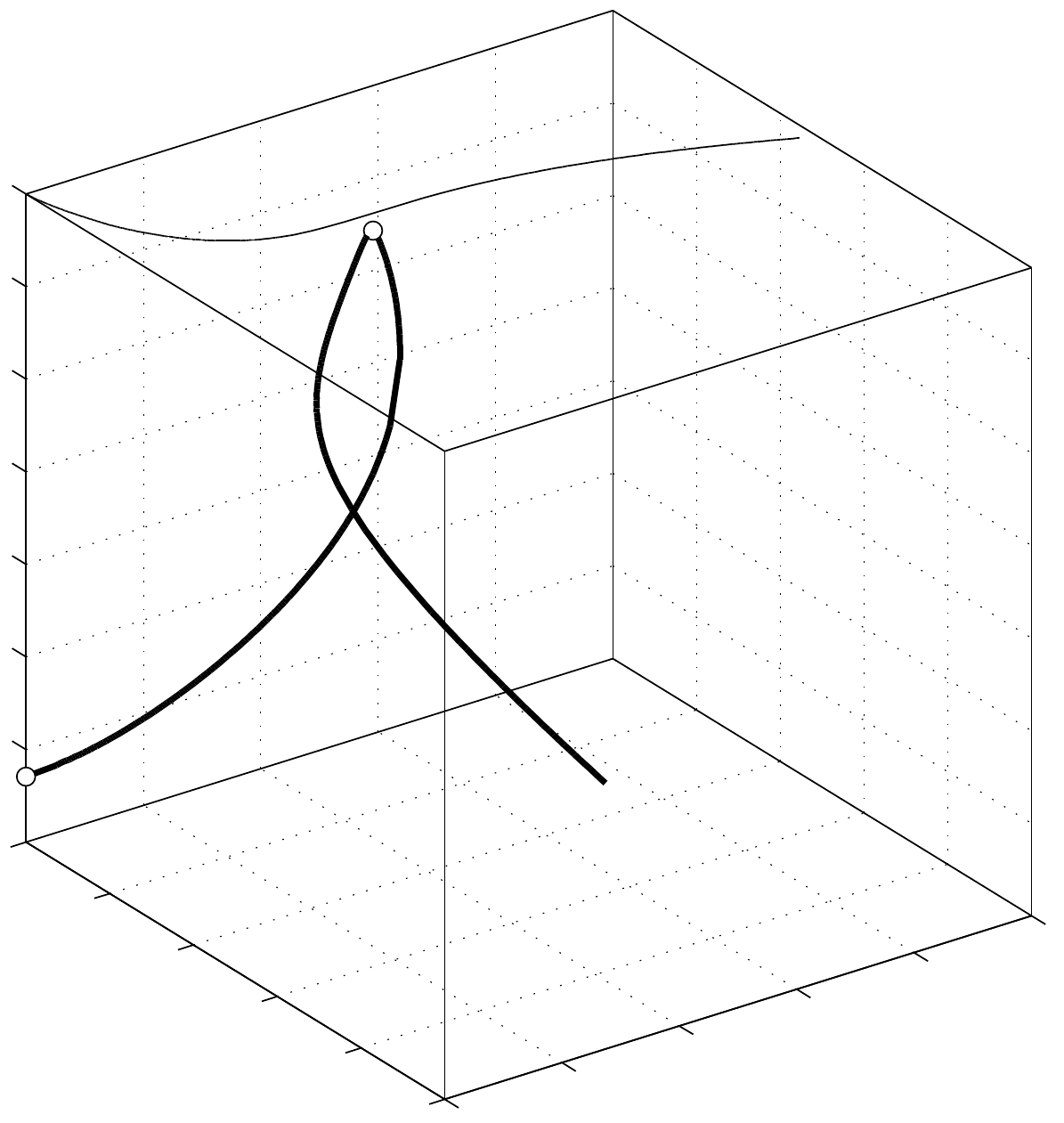}}
\put(2,116){\rotatebox{90}{$\Im(\omega)$}}
\put(40,24){$\alpha$}
\put(170,10){$\Re(\omega)$}
\put(30,200){(a)}
\put(108,2){{\small 0}}
\put(218,36){{\small 1}}
\put(14,50){{\small 0}}
\put(94,4){{\small 1}}
\put(16,180){{\small 0}}
\put(6,60){{\small -0.7}}
\put(182,198){$m=1$}
\end{picture}
\hspace*{16pt}
\begin{picture}(218,212)(0,0)   %
\put(20,14){\includegraphics[width = 200pt]{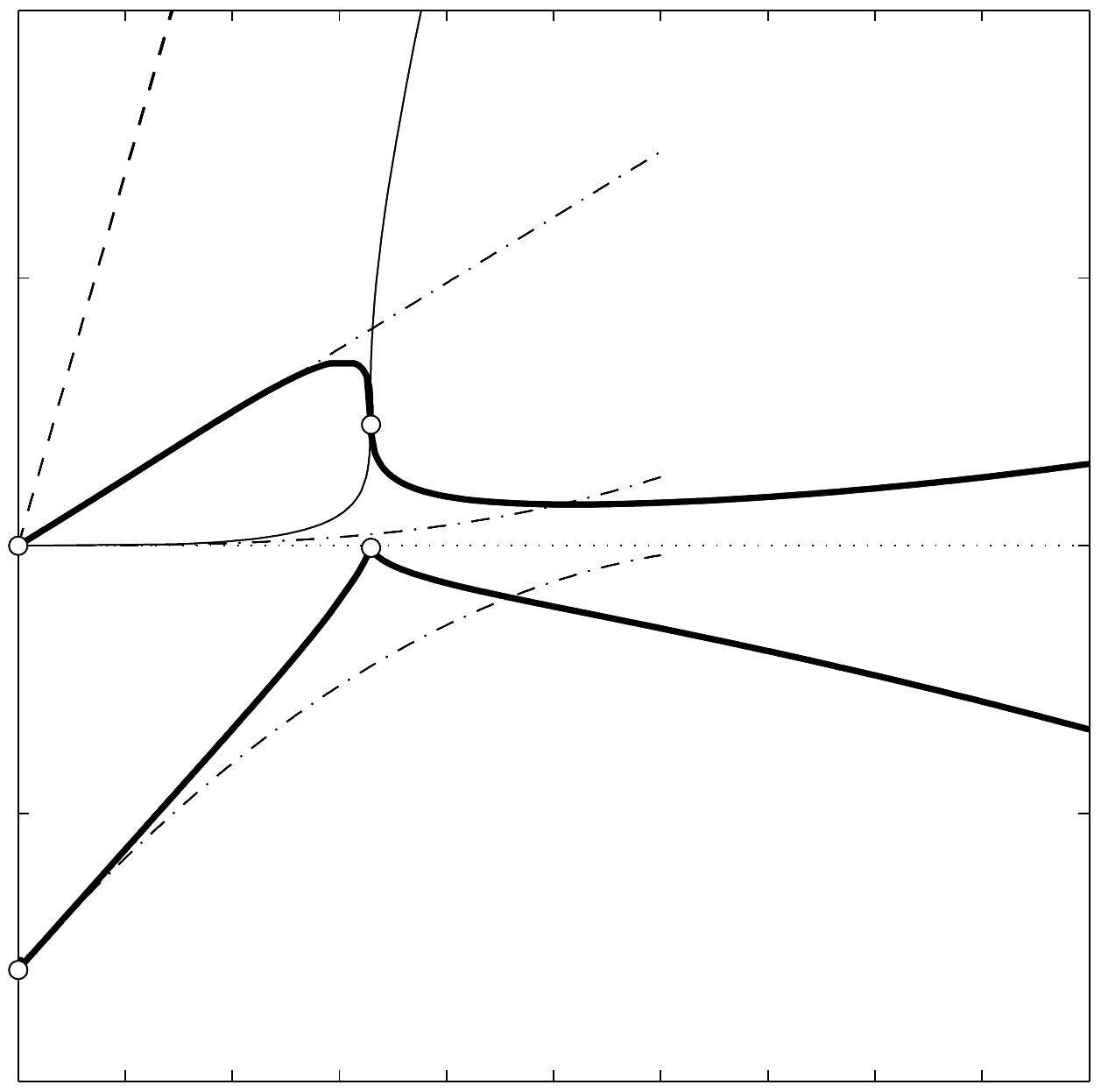}}
\put(2,140){\rotatebox{90}{$\Re(a\omega^2/g)$}}
\put(2,42){\rotatebox{90}{$-\Im(a\omega^2/g)$}}
\put(118,-1){$\alpha$}
\put(28,194){(b)}
\put(182,194){$m=1$}
\put(21,8){{\small 0}}
\put(57,8){{\small 0.2}}
\put(95,8){{\small 0.4}}
\put(134,8){{\small 0.6}}
\put(173,8){{\small 0.8}}
\put(212,8){{\small 1.0}}
\put(5,13){{\small -0.5}}
\put(5,111){{\small \phantom{-0.}0}}
\put(5,208){{\small \phantom{-}0.5}}
\end{picture}\\[12pt]
\begin{picture}(218,212)(0,0)   %
\put(20,14){\includegraphics[width = 200pt]{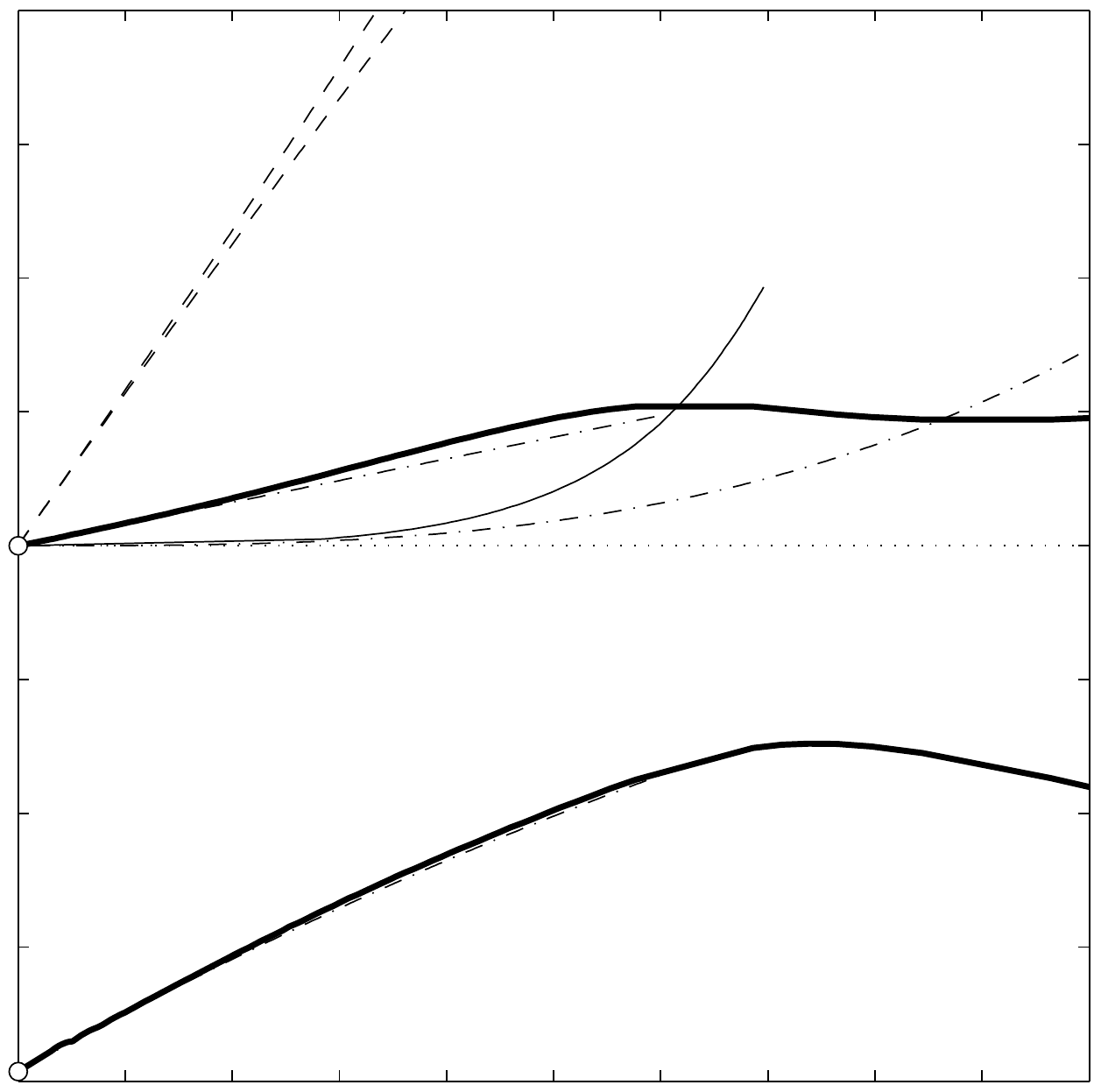}}
\put(2,140){\rotatebox{90}{$\Re(a\omega^2/g)$}}
\put(2,42){\rotatebox{90}{$-\Im(a\omega^2/g)$}}
\put(118,-1){$\alpha$}
\put(32,194){(c)}
\put(182,194){$m=2$}
\put(21,8){{\small 0}}
\put(57,8){{\small 0.2}}
\put(95,8){{\small 0.4}}
\put(134,8){{\small 0.6}}
\put(173,8){{\small 0.8}}
\put(212,8){{\small 1.0}}
\put(5,13){{\small -1.0}}
\put(5,111){{\small \phantom{-0.}0}}
\put(5,208){{\small \phantom{-}1.0}}
\end{picture}
\hspace*{16pt}
\begin{picture}(218,212)(0,0)   %
\put(20,14){\includegraphics[width = 200pt]{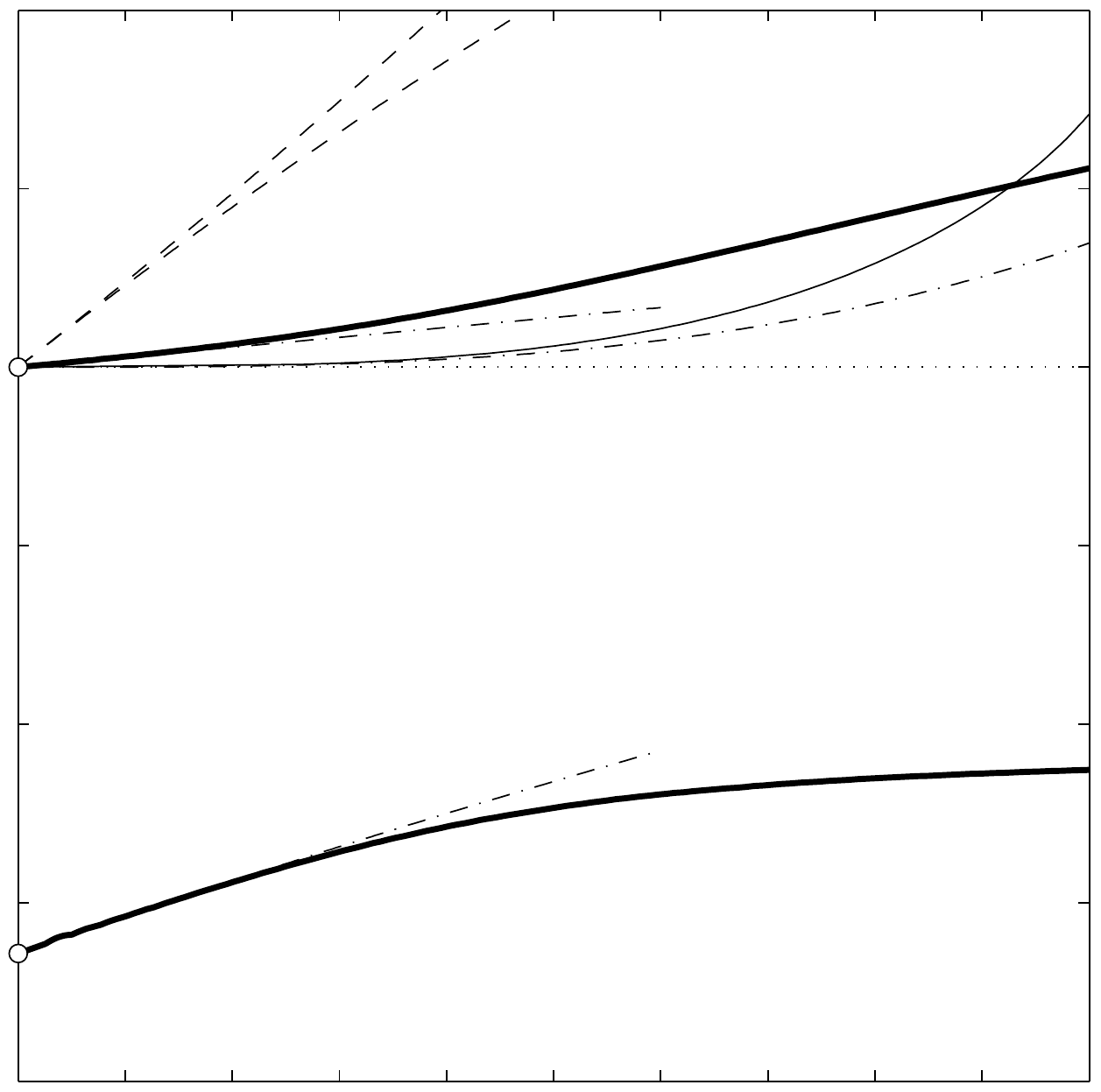}}
\put(2,156){\rotatebox{90}{$\Re(a\omega^2/g)$}}
\put(2,57){\rotatebox{90}{$-\Im(a\omega^2/g)$}}
\put(118,-1){$\alpha$}
\put(32,194){(d)}
\put(182,194){$m=3$}
\put(21,8){{\small 0}}
\put(57,8){{\small 0.2}}
\put(95,8){{\small 0.4}}
\put(134,8){{\small 0.6}}
\put(173,8){{\small 0.8}}
\put(212,8){{\small 1.0}}
\put(5,13){{\small -2.0}}
\put(5,143){{\small \phantom{-0.}0}}
\put(5,208){{\small \phantom{-}1.0}}
\end{picture}
\end{center}
\caption{\label{fig:m1-4modes} (a) The constructed solution of $|\mathsf{M}|=0$ for $\omega\in\mathbb{C}$ and $\Atw=-\frac{1}{2}$, $\delta=\frac{1}{4}$, $N=1$, $n=1$, $m=1$ (solid lines).  (b)--(d) Are projections of the solution squared, for comparison with Fig.~\ref{fig:N1modes}b.  Bold solutions have non-zero imaginary component.  It can be seen that unstable wave modes are not stabilized by increasing the rotation rate, but are suppressed initially.  Asymptotic gravity wave approximations \eqref{eq:om0om1}, \eqref{eq:om2} to the solution are shown dot-dashed.  Numerical inertial wave solutions have not been plotted for clarity, but the first asymptotic solutions for inertial waves, with $q=1$ \eqref{eq:m_podd}, are shown dashed.}
\end{figure*}

The solutions of the eigenvalue problem are calculated numerically for $N=1$, $n=1$, $\delta=\frac{1}{4}$, $\Atw=-\frac{1}{2}$, and $m = 1, 2, 3$ (see Fig.~\ref{fig:N1modes}b for comparison with the axisymmetric case, $m=0$).  

The numerical solution was  calculated by evaluating the determinant of $\mathsf{M}$ for a given $\alpha$ over a plane $\omega\in\mathbb{C}$ (numerical integration was carried out using Simpson's rule).  The zeros of the real part of $|\mathsf{M}|$ were contoured and intersections with the zero contour of the imaginary part of $|\mathsf{M}|$ were found.  The solution was constructed by then allowing $\alpha$ to vary over the range $[0, \alpha_T]$ (see Fig.~\ref{fig:m1-4modes}a).  Figs~\ref{fig:m1-4modes}b--d are projections of the three-dimensional solution to allow comparison with Fig.~\ref{fig:N1modes}b.  The positive vertical axis shows a projection of $\Re(\omega)^2a/g$ and the negative vertical axis shows a projection of $-\Im(\omega)^2a/g$ so that the plots coincide with the axisymmetric case when $\omega\in\mathbb{R}$ or $\omega\in\textrm{i}\mathbb{R}$.  It can be seen that for $m\ne0$ the dominant gravity wave solution is not able to cross from the unstable lower half of the domain into the stable upper half, unlike the $m=0$ solution shown in Fig.~\ref{fig:N1modes}b.

\subsubsection{\label{sec:asymmarb} Multiple mode, arbitrary rotation rate solutions}

\begin{figure}
\begin{center}
\vspace*{4pt}
\setlength{\unitlength}{1pt}
\begin{picture}(218,212)(0,0)
\put(20,7){\includegraphics[width = 200pt]{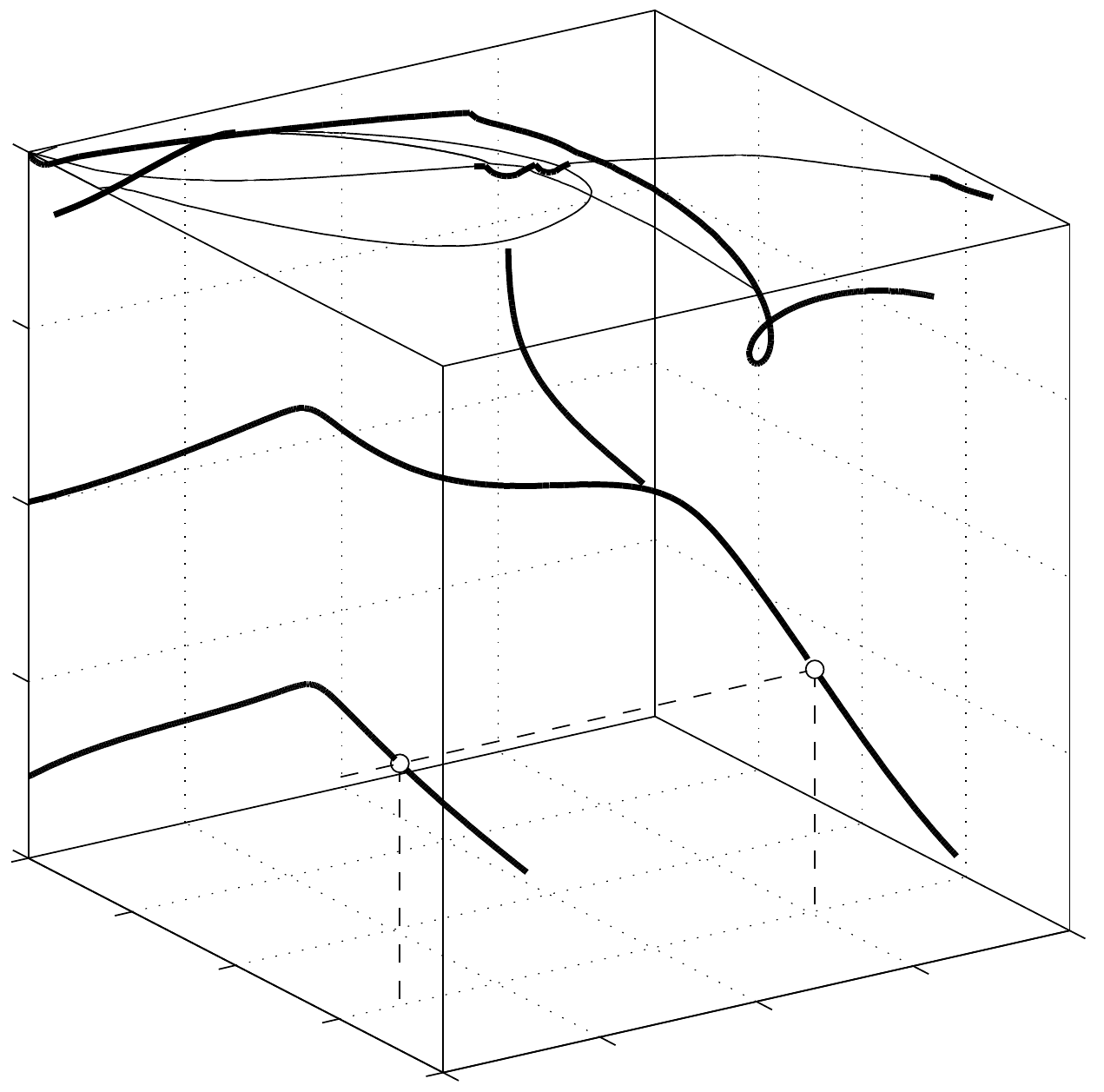}}
\put(4,100){\rotatebox{90}{$\Im(\omega)$}}
\put(48,18){$\alpha$}
\put(160,8){$\Re(\omega)$}
\put(25,188){(a)}
\put(105,3){{\small 0}} 
\put(212,26){{\small 2}}
\put(18,41){{\small 0}}
\put(90,4){{\small 2}}
\put(18,172){{\small 0}}
\put(15,50){{\small -2}}
\end{picture}\\[12pt]
\begin{picture}(218,212)(0,0) 
\put(20,14){\includegraphics[width = 200pt]{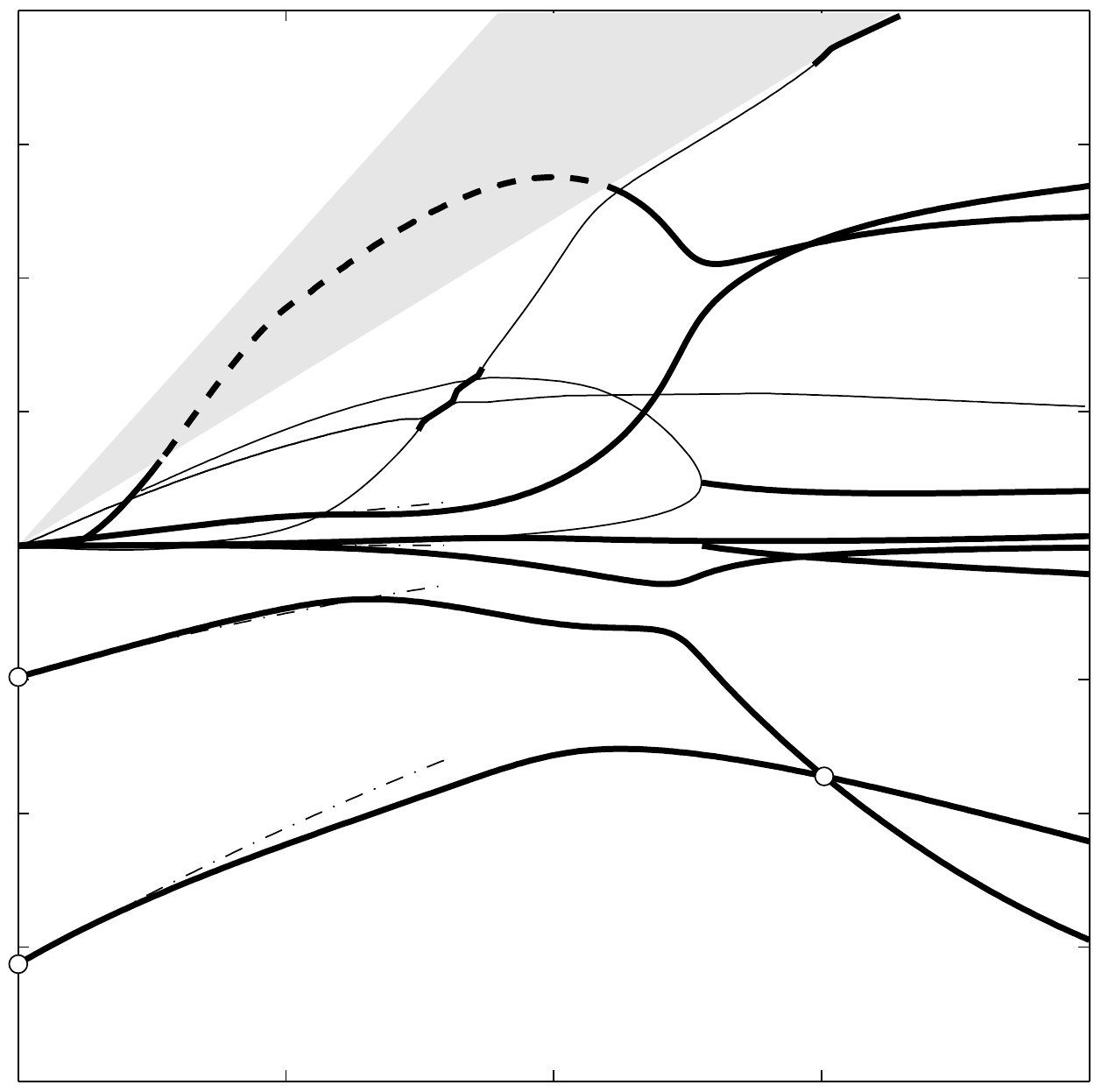}}
\put(2,140){\rotatebox{90}{$\Re(a\omega^2/g)$}}
\put(2,42){\rotatebox{90}{$-\Im(a\omega^2/g)$}}
\put(118,-3){$\alpha$}
\put(32,194){(b)}
\put(184,192){$\begin{array}{rcl} m&\!\!=\!\!&2 \\ N&\!\!=\!\!&2\end{array}$}
\put(21,8){{\small 0}}
\put(67,8){{\small 0.5}}
\put(115,8){{\small 1.0}}
\put(163,8){{\small 1.5}}
\put(212,8){{\small 2.0}}
\put(12,13){{\small -4}}
\put(12,111){{\small \phantom{-}0}}
\put(12,208){{\small \phantom{-}4}}
\end{picture}
\end{center}
\caption{\label{fig:m2N2modes} Gravity wave solutions of $|\mathsf{M}|=0$ for $\Atw=-\frac{1}{2}$, $\delta = \frac{1}{4}$, $m=2$ for $N=2$ and $n=1,2$, i.e., $k_{01}\approx3.054$ and $k_{02}\approx6.706$.  Bold solutions have non-zero imaginary component.  (a) Three-dimensional representation of the solution: the most unstable branches cross at $\alpha\approx1.505$, where $\omega_1 \approx 1.514 -1.313\textrm{i}$ and $\omega_2 \approx 0.189-1.313\textrm{i}$ indicated by circles.  (b) The projected solutions for comparison with Fig.~\ref{fig:N1modes}b.  Although $\alpha_T = 1$, the $\alpha$ axis has been extended to show the possibility of rotation causing some modes to become more unstable than others.}
\end{figure}

Fig.~\ref{fig:m2N2modes} shows the possible wavemodes for $\Atw=-\frac{1}{2}$, $\delta = \frac{1}{4}$, $m=2$, $N=2$, and $n=1,2$.  As the rotation rate is increased the unstable gravity wave modes are seen to be suppressed, though the suppression is greater for the more unstable $n=2$ mode.  The plot shows that suppressing a higher wavemode to such an extent that it becomes more stable than a lower wavemode is possible since the solution's projections cross (at $\alpha\approx1.505$, $\Im(\omega)\approx-1.313$, shown as circles, though in this case the crossing occurs for $\alpha>\alpha_T$ where the solution is not strictly valid).  Comparing Fig.~\ref{fig:m1-4modes}c with Fig.~\ref{fig:m2N2modes}b, it can be seen that the addition of a single extra mode significantly increases the number of possible modes of behavior.

\subsection{\label{sec:theory_summary} Summary of key results}

In \S \ref{sec:growth} the approach developed by \citet{miles64} to model surface waves on a rotating body of water was generalised to the two-layer case, allowing for either a stable (positive Atwood number) or an unstable (negative Atwood number) initial stratification.  The dispersion relation for axisymmetric perturbations at low rotation rates was derived in \S \ref{sec:axisym}, \eqref{eq:freq} and it shows that gravitationally unstable perturbations may be made less unstable by rotating the system.  This suggests that at least partial suppression of the Rayleigh-Taylor instability may be achieved through rotation of the system, though we note that \eqref{eq:freq} is only valid in the limit $a\Omega^2/g \ll 1$.  In \S \ref{sec:critrate} an exact result,  \eqref{eq:crit}, was found for the critical rotation rate required to completely stabilise an otherwise gravitationally unstable axisymmetric wave mode.  This critical rotation rate depends on the aspect ratio of the system which is the reason an exchange of stability was not found in the model of \citet{chandra}.  \eqref{eq:crit} indicates that a rotation rate $\alpha_c=12\delta$ is required to stabilise all axisymmetric wave modes, but the model solutions \eqref{eq:eigsol} are invalid for $\alpha>4\delta$.  Evaluating \eqref{eq:crit} at $\alpha = 4\delta$ suggests that modes in the range $0<k^2<6(3-|\Atw|)|\Atw|^{-1}$ may be stabilised, i.e., wave modes with wavelengths $\lambda$ in the range
\begin{displaymath}
\frac{\lambda}{a} \gtrsim 2\pi \left\{\frac{|\Atw|}{6\left(3-|\Atw|\right)}\right\}^{1/2} \sim 1.48|\Atw|^{1/2} + \mathcal{O}(\Atw^2),
\end{displaymath} 
though this calculation does not account for the influence of one wave mode upon another as the summation, in \eqref{eq:modesum}, has been ignored.

In \S \ref{sec:asymm} the dispersion relation for asymmetric wave modes was derived \eqref{eq:kser}--\eqref{eq:Gdef}.  This dispersion relation includes axisymmetric perturbations, $m=0$, as a special case.  In the asymmetric case, $m\ne0$, it was shown that the wavenumber cannot be determined {\it a priori}, it depends on both the rotation rate, $\Omega$, and the mode frequency, $\omega$.  The dispersion relation reveals, as might be anticipated, that the mode frequency contains both real and imaginary parts in general.  Hence, the developing instability is characterised by both a growth and a precession of a given wave mode.  It was also shown, \S \ref{sec:nocritrate}, that a general critical rotation rate to stabilise an asymmetric mode does not exist, unlike the axisymmetric case.

\section{\label{sec:mag}Rayleigh-Taylor Instability in paramagnetic and diamagnetic fluids}

For the purposes of comparison with our experiments \citep{scirep} (see also Supplementary Information), we now consider the rotating Rayleigh-Taylor instability induced between two magnetically susceptible fluids in the presence of a  gradient magnetic field, and how this compares with the classical case considered in \S \ref{sec:modeling}.  The rotating Euler equation, including the magnetic body force (see Supplementary Information) is 
\begin{multline} \label{eq:mag1}
\rho\Dv{\vc{u}}{t} = -\nabla p + \rho\,\vc{g} - \rho\,\vc{\Omega}\times(\vc{\Omega}\times\vc{x}) \\ - 2\rho\,\vc{\Omega}\times\vc{u} + \mu_0 M \nabla H.
\end{multline}
Here $H$ is the magnitude of the applied gradient magnetic field $\vc{H}$ and $M$ is the magnitude of the magnetization $\vc{M}$ of the fluid.  The introduction of the magnetic body force modifies \eqref{eq:pdef2L}, but only appears at order 1 as a hydrostatic background effect and so does not modify the governing equation for the generalized potential $\phi$.  (See Supplementary Information for a derivation of this result).  Hence, \eqref{eq:potgov2L} and \eqref{eq:bcs1} are unchanged by including magnetic effects.  

There exists a technique for separating mixtures of materials on the basis of their density known as `sink-float separation' \citep{rosenweigannrev} which is useful to consider at this point.  In this process the density of an object is measured by observing the `apparent density' of a surrounding ferrofluid in a magnetic field when the object is neutrally buoyant.  The apparent density of the ferrofluid, $\rho'$ is defined as the density of an equivalent non-magnetic fluid that is subject to a gravitational force equal to the combined magnetic and gravitational force on the ferrofluid.  That is to say (taking the non-rotating, static form of \eqref{eq:mag1}), $\rho'\vc{g} = \rho\vc{g} + \mu_0 M \nabla H$.  Hence, for a constant magnetic field with the gradient aligned vertically
\begin{equation} 
\rho' = \rho - \frac{\mu_0 M}{g}\dv{H}{z}.
\end{equation}
For para- and diamagnetic fluids, $|\chi|\ll1$, and we may write
\begin{equation} \label{eq:sf1}
\rho' \sim \rho\left\{1 - \frac{1}{2g\mu_0} \frac{\chi}{\rho}\dv{B^2}{z}\left[1 + O(\chi)\right]\right\},
\end{equation}
\citep[see (39),][]{rosenweigannrev}.  (Here $B$ is the magnitude of the magnetic induction $\vc{B}$, related to $\vc{M}$ and $\vc{H}$ by $\vc{B} = \mu_0\left(\vc{M} + \vc{H}\right)$.)  Equivalently we can define a modified gravity such that $\rho\vc{g}' = \rho\vc{g} + \mu_0 M\nabla H$.  Then, in this case
\begin{equation} \label{eq:modg2}
g' = g - \frac{\mu_0 M}{\rho}\dv{H}{z} \sim g\left\{1 - \frac{1}{2g\mu_0}\frac{\chi}{\rho}\dv{B^2}{z}\left[1+O(\chi)\right]\right\}.
\end{equation}

We see that the correction factors in curly braces in \eqref{eq:sf1} and \eqref{eq:modg2} are identical.
Hence, with regard to calculating the vertical force only, it is equivalent to consider either the magnetic field to be modifying the gravitational field, or the apparent density field.
However, the concept of effective density is not fully compatible with the rotating case as different effective densities must be employed in the radial and vertical directions.  We therefore proceed by considering an effective gravity, $g'$, acting on each fluid in the magnetic field, which is applicable to both radial and vertical directions.

Continuity of normal stress across the perturbed boundary, $z = z_0(r) + \epsilon \textrm{e}^{\textrm{i}(\omega t + m \theta)} \zeta(r)$, is satisfied at order 1 by the hydrostatic initial conditions,
and at order $\epsilon$ by
\begin{equation} \label{eq:pcontmag}
\textrm{i}\,\omega\mu^2\hat\zeta = \frac{2\Omega^2}{g'}\left(1 - \frac{1}{\mu^2}\right)\left(\frac{1+\Atw }{\Atw }\hat\phi_2 -\frac{1-\Atw }{\Atw }\hat\phi_1\right),
\end{equation}
(where $\mu$ is as defined in \S\,\ref{sec:modeling} and not to be confused with magnetic permeability) on $z=z_0$, where 
\begin{equation} \label{eq:modg1}
g' = g\left[1 - \frac{1}{2g\mu_0}\left(\frac{\chi_2 - \chi_1}{\rho_2 - \rho_1}\right)\left.\pd{B^2}{z}\right|_{z = z_0(r)}\right],
\end{equation}
is a modified, and spatially varying effective gravitational acceleration.  Note that \eqref{eq:pcontmag} has exactly the form of \eqref{eq:pcont2} with the only difference being the modified gravity.  In \eqref{eq:modg1}, $\chi_j$ is the magnetic susceptibility, defined by $\vc{M} = \chi \vc{H}$, of each fluid layer.  The form of \eqref{eq:modg1} is a two-layer generalization of a modified gravity in the single-layer sink-float applications \eqref{eq:modg2}.

The expression for the modified gravity, $g'$, in \eqref{eq:modg1} indicates that if there exists a region in the magnetic field where $\partial B^2/\partial z<0$, and we have an appropriate choice of $\chi$ and $\rho$ in each layer, then, depending on the magnitude of $\partial B^2/\partial z$, we may be able to reverse the sign of $g'$.  This implies the following experimental method is possible: a gravitationally stable stratification may be prepared and placed in the magnetic field; then, if there is sufficient downward magnetic attraction of the upper layer and magnetic repulsion of the lower layer, the stablizing effect of the gravitational field may be overcome.  By choosing fluids such that $\chi_1>\chi_2$ and $\rho_2>\rho_1$ then the second term on the right hand side of \eqref{eq:modg1} is negative.  If the system can be placed in a region of the magnetic field where $|\partial B^2/\partial z| > 2g\mu_0(\rho_2-\rho_1)/|\chi_2 - \chi_1|$, then the right hand side of \eqref{eq:modg1} is negative, the stabilizing effect of the gravitational field is weaker than the destabilizing effect of the magnetic field and a Rayleigh-Taylor-like instability may be initiated.

The discussion above motivates us to ask whether the use of a magnet can closely replicate the onset of the classical Rayleigh-Taylor instability, particularly in a rotating regime where the number of technical difficulties involved experimentally increases for standard barrier-removal methods.  In the static non-rotating sink-float separation technique discussed in \citet{rosenweigannrev}, there was an equivalence between an `apparent density' and a modified gravity.  However, in the rotating case, we observe from \eqref{eq:interface} that, while manipulating the fluid densities in each layer does not change the parabolic hydrostatic interface, %
as soon as magnetically susceptible fluids are placed in a magnetic field the parabolic  interface is changed in general.   Nevertheless, %
we can compare the growth rates of modes of instability using the full magnetic expressions derived above, with the classical expressions derived in \S \ref{sec:growth} using an apparent Atwood number approximation and show that the supported modes and predicted growth rates may be extremely closely matched.

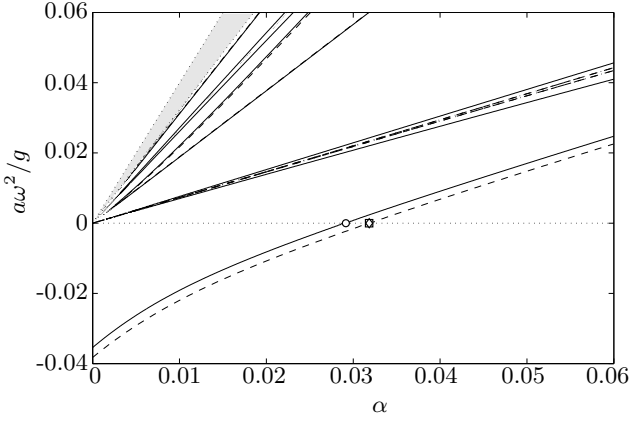
\begin{figure}
\begin{center}
\setlength{\unitlength}{1pt}
\begin{picture}(200,136)(-10,0)
\put(0,6){\includegraphics[width = 200pt]{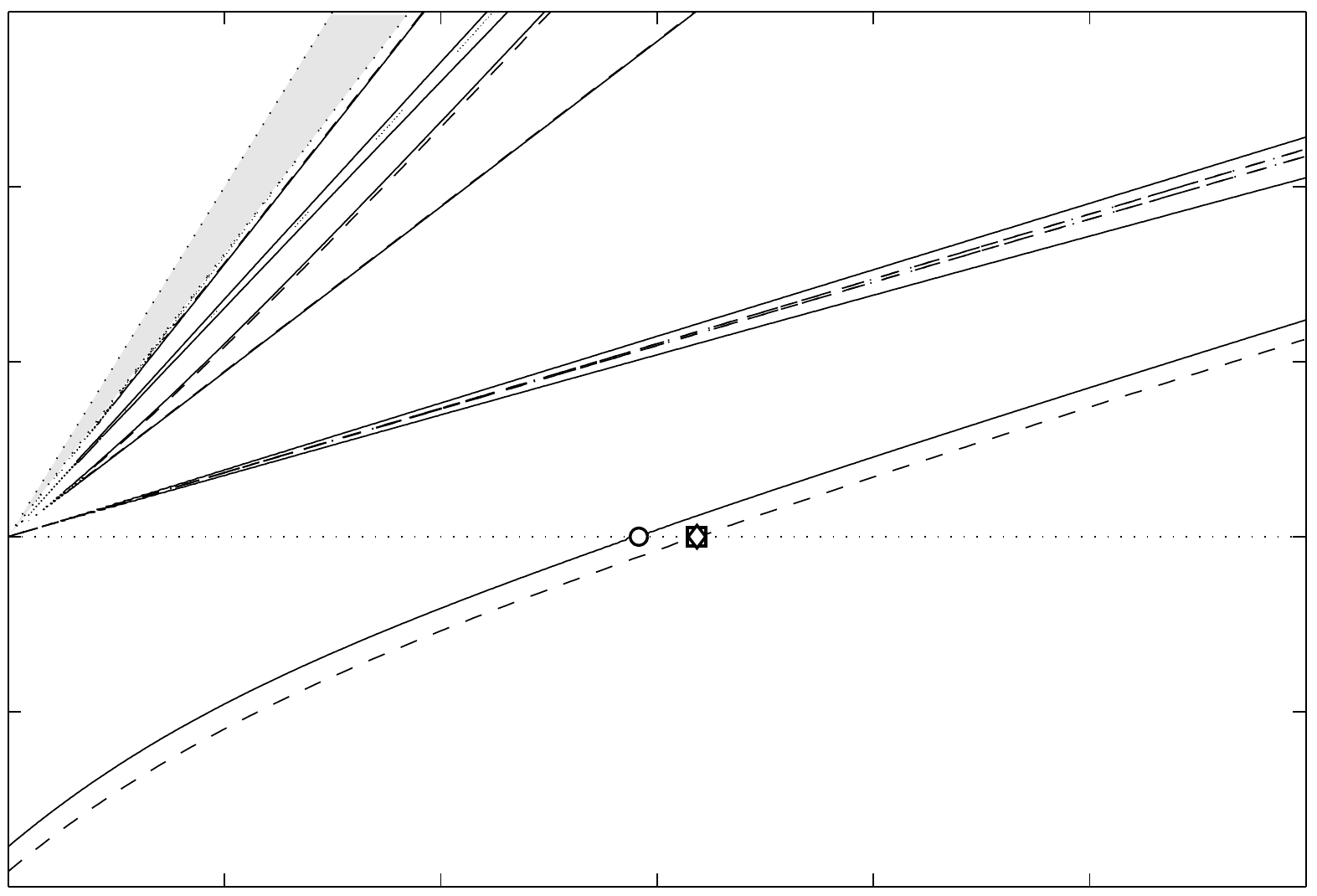}}
\put(0,0){\small 0}
\put(26,0){\small 0.01}
\put(58,0){\small 0.02}
\put(91,0){\small 0.03}
\put(123,0){\small 0.04}
\put(156,0){\small 0.05}
\put(188,0){\small 0.06}
\put(107,-11){$\alpha$}
\put(-20, 5){\small -0.04}
\put(-20, 32){\small -0.02}
\put(-20, 58){\small \phantom{-0.0}0}
\put(-20, 85){\small \phantom{-}0.02}
\put(-20, 111){\small \phantom{-}0.04}
\put(-20, 138){\small \phantom{-}0.06}
\put(-32,64){\rotatebox{90}{$a\omega^2/g$}}
\end{picture}
\end{center}
\caption{\label{fig:magcompare} Supported eigenvalues of the fundamental mode of instability for: (solid lines) the full magnetic model using the magnetic field and experimental parameters used in \S \ref{sec:exp}; (dashed lines) the classical model using an `apparent Atwood number'.  Not shown are solutions using a uniform gradient magnetic field which are indistinguishable at this scale from the apparent Atwood number classical solution.  The predicted critical rotation rate for stabilization using the classical model (white square) is approximately $4.3\%$ too large compared to the solution of the full magnetic problem (white circle).  The error between the classical model, and the critical rotation rate from the magnetic model with a uniform gradient field (white diamond) is less than $0.02\%$.}
\end{figure}

Proceeding as in \S \ref{sec:growth} with the modified stress continuity condition \eqref{eq:pcontmag} and the unchanged kinematic condition \eqref{eq:kin1}, gives the magnetic counterpart of \eqref{eq:Phi} as
\begin{multline} \label{eq:Phimag}
\Phi \propto \int_0^a\left\{\omega^2\left(\frac{g}{g'}\right)\left[\frac{1+\Atw }{\Atw }\hat\phi_2 - \frac{1-\Atw }{\Atw }\hat\phi_1\right]^2
\right.\\ \left.
+\left[\left(\frac{gz_0'}{\Omega^2r}\right)\frac{\Omega^2}{1-\mu^2}\left(r\pd{}{r} + 2\mu m\right) - g\pd{}{z}\right]  \right. \\ \left.
\left[\frac{1+\Atw }{\Atw }\hat\phi_2^2 - \frac{1-\Atw }{\Atw }\hat\phi_1^2\right] \right\}\Bigg|_{z=z_0(r)}r\textrm{d}r.
\end{multline}
In the case $B=0$, $g' = g$ and $z_0$ is as given in \eqref{eq:interface}, and so \eqref{eq:Phi} is recovered.

We can therefore seek to achieve experimentally an effective, ideal, unstable Atwood number, $\Atwid<0$, in the following manner.  We define an apparent Atwood number, $\Atwapp$, via  $g'\Atwlab = g\Atwapp$, where $\Atwlab>0$ represents a gravitationally stable stratification.  This implies
\begin{equation}
\Atwapp=\Atwlab\left[1 - \frac{1}{2g\mu_0}\left(\frac{\chi_2 - \chi_1}{\rho_2 - \rho_1}\right)\pd{B^2}{z}\right].
\end{equation}
With the appropriate choices for the magnetic susceptibility and fluid density in each layer, and suitable position in the magnetic field, we can therefore closely approximate $\Atwid$ with $\Atwapp$. The experiment may be prepared using standard techniques, as $\Atwlab$ is a gravitationally stable stratification that is realisable in a laboratory.  The stable stratification may be spun-up into solid body rotation before applying the magnetic field, changing the effective Atwood number from $\Atwlab$ to $\Atwapp \approx \Atwid$.  The relationship between the magnetically-induced instability and the gravitationally-induced `classical' instability is evident in the limit of small Atwood numbers, where in the classical case for Atwood number $|\Atwid|\ll1$ we have approximately
\begin{multline} \label{eq:Phiasym}
\Phi \propto \int_0^a\left\{\frac{\omega^2}{\Atwid}\left(\hat\phi_2 - \hat\phi_1\right)^2
+\left[\left(\frac{gz_0'}{\Omega^2r}\right)\frac{\Omega^2}{1-\mu^2}
\right.\right. \\ \left.\left. \times
\left(r\pd{}{r} + 2\mu m\right) - g\pd{}{z}\right]\left(\hat\phi_2^2 -\hat\phi_1^2\right) \right\}\Bigg|_{z=z_0}r\textrm{d}r,
\end{multline}
and in the magnetic case where $|\Atwlab|\ll1$, we have approximately
\begin{multline} \label{eq:Phimagasym}
\Phi \propto \int_0^a\left\{\omega^2\left(\frac{g}{g'\Atwlab}\right)\left(\hat\phi_2 - \hat\phi_1\right)^2
+\left[\left(\frac{gz_0'}{\Omega^2r}\right)\frac{\Omega^2}{1-\mu^2}
\right.\right. \\ \left.\left. \times
\left(r\pd{}{r} + 2\mu m\right) - g\pd{}{z}\right]\left(\hat\phi_2^2 - \hat\phi_1^2\right) \right\}\Bigg|_{z=z_0}r\textrm{d}r.
\end{multline}
So provided that the effect of the differences between $z_0$ in the classical case \eqref{eq:interface}, and $z_0$ in the magnetic case (see Supplementary Information) %
on the structure of $\Phi$ are small, the magnetically induced instability, governed by \eqref{eq:Phimagasym}, can be expected to accurately represent the classical gravitationally induced instability, governed by \eqref{eq:Phiasym}, where the ideal Atwood number $\Atwid$ is replaced by an apparent Atwood number $\Atwapp = g'\Atwlab/g$.   The difference between the two interface profiles %
depends on the specific profile of the magnetic field imposed.  In the case of a uniform gradient magnetic field, aligned vertically, and zero rotation, %
the two profiles coincide and the structure of $\Phi$ is identical -- a magnetically induced instability is exactly equivalent to a gravitationally induced instability.

We now consider the supported modes for rotating Rayleigh-Taylor instability in an ideal system, compared to systems with magnetically-induced instabilities.  Fig.~\ref{fig:magcompare} is a comparison between the supported frequencies of oscillation for the fundamental mode of instability for: the magnetically induced instability from the magnetic field used in \citep{scirep} and \S \ref{sec:exp} (solid); the magnetically induced instability from a magnetic field with uniform vertical gradient and no radial gradient (using the centerline of the magnetic field in \citep{scirep} and \S \ref{sec:exp}); and the classical solution, as derived in \S \ref{sec:modeling} using an `apparent Atwood number' approximation (dashed).

The solution using a uniform gradient field is not shown as it is visually indistinguishable, at the scale shown, from the classical `ideal $\Atw$' solution.  The two curves coincide at the limit of no rotation, as they must, and differ by less than $0.02\%$ at the critical rotation rate $\alpha_c=0.032$.  The magnetic field used in the calculations is the field used in \citep{scirep} (see Fig.~3 therein).

If an approximately uniform gradient magnetic field, such as those described in the sink-float separation technique \citep{rosenweigannrev}, can be employed to induce the instability there is an excellent agreement between the predicted growth rates between the magnetically induced flow and a classical flow using an apparent Atwood number approximation.  The agreement is better for low rotation rates, with the two methods exactly coinciding at zero rotation rate.  Using the magnetic field of a solenoid magnet, that has some radial gradient in strength, to induce the instability, yields good quantitative agreement between the two methods.  A magnetically induced realisation is therefore a reasonable experimental method for approximating a Rayleigh-Taylor instability and useful initial test of the critical rotation rate predicted in \eqref{eq:crit}.

\section{\label{sec:exp}Experiments}

In \cite{scirep} we described a series of three experiments that used a superconducting solenoidal magnet
to induce a light paramagnetic fluid to impinge into a dense diamagnetic layer triggering an instability.  The first series investigated the reduction in growth rate of the Rayleigh-Taylor instability due to rotation; in the second series we measured the effect of rotation on the size and form of the structures in the instability as it developed.  The third series of experiments investigated the effect of fluid viscosity on the observed structures.  In each experiment the two layers comprised a paramagnetic manganese chloride solution (upper layer), and a denser diamagnetic sodium chloride solution (lower layer).  In the third series of experiments glycerol was added to both layers when the effects of viscosity were investigated.  Further details are in \citep{scirep} (see also Supplementary Information).

\subsection{\label{sec:structure}Structure of the instability}

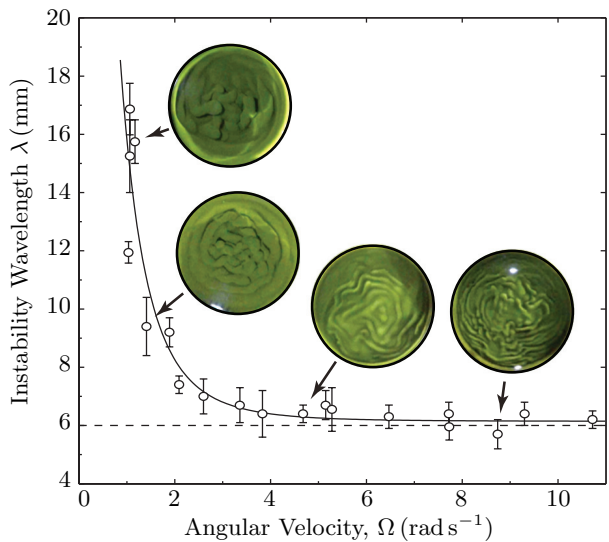
\begin{figure}
\begin{center}
\setlength{\unitlength}{1pt}
\begin{picture}(200,177)(0,0)
\put(10,0){\includegraphics[width = 200pt]{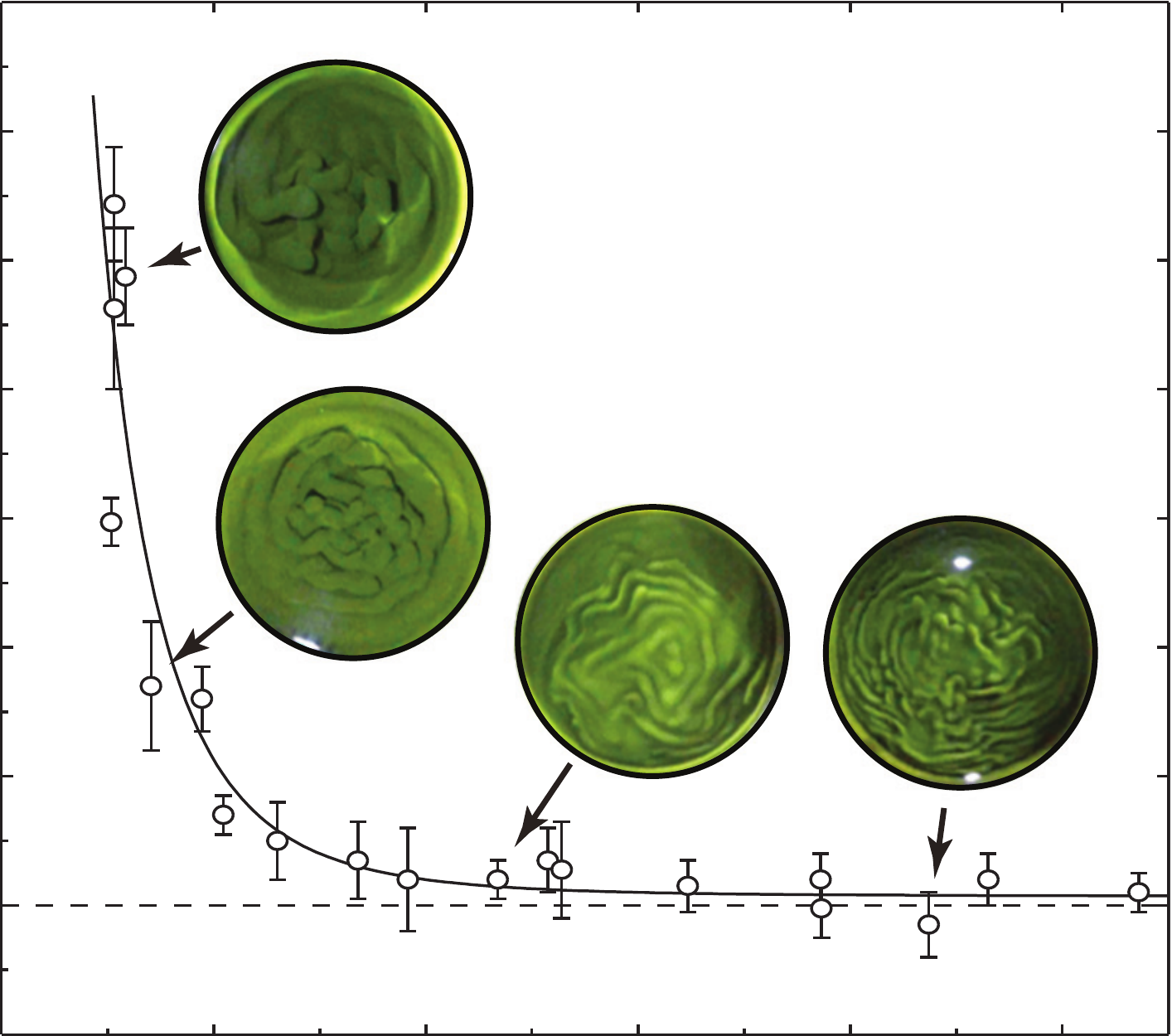}}
\put(-2,-1){{\small \phantom{0}4}}
\put(-2,21){{\small \phantom{0}6}}
\put(-2,43){{\small \phantom{0}8}}
\put(-2,65){{\small 10}}
\put(-2,87){{\small 12}}
\put(-2,108){{\small 14}}
\put(-2,130){{\small 16}}
\put(-2,152){{\small 18}}
\put(-2,174){{\small 20}} 
\put(10,-8){{\small 0}}
\put(44,-8){{\small 2}}
\put(80,-8){{\small 4}}
\put(117,-8){{\small 6}}
\put(152.5,-8){{\small 8}}
\put(187,-8){{\small 10}}
\put(50,-19){Angular Velocity, $\Omega$\,(rad\,s$^{-1}$)}
\put(-16,30){\rotatebox{90}{Instability Wavelength $\lambda$\,(mm)}}
\end{picture}\\[15pt]
\caption{\label{fig:wavelengths} The dominant scales of perturbation after the onset of instability.  Error bars are associated with goodness-of-fit from the autocorrelation algorithm.  It can be seen that as the rotation rate is increased, the observed scale of motion asymptotes to approximately 6\,mm for the experimental parameters chosen.  The solid line is the empirical best fit $\lambda/a = 0.12 + \left[0.017/\left(\alpha + 0.023\right)\right]^{3.5}$, where $a$ is the tank radius and $\alpha=\Omega^2 a/g$.}
\end{center}
\end{figure}

In the second series of experiments the effect that the rotation had on the structure of the developing instability could be seen.  The inset images to Fig.~\ref{fig:wavelengths} (green discs) show the development of the Rayleigh-Taylor instability at the interface between the two fluids for different rotation rates (additional images taken from the experiments are shown in Fig.~5a \citep{scirep}). 
At early times ($t\sim 0.5$--$1.0$\,s) a perturbation to the interface can be seen which exhibits a dominant length scale.  Structures reminiscent of snake-like convection rolls \citep[e.g.,][]{rossby} can be observed.  It is apparent that with an increase in rotation rate, the observed instability decreases in length scale.  At the lower rotation rates the paths followed by the initial disturbance structures have significant radial deviation, meandering in towards the center of the tank and back out to the side walls again.  At the lowest rotation rates the instability is more cellular than serpentine.  As the rotation rate is increased the cellular initial perturbation is no longer observed and a more serpentine-like structure appears.  With increasing rotation rate the width of these structures decreases.  It can also be observed that the amount of radial meandering decreases too.  It can be seen that \citep[][Fig.~5]{scirep}, for increasing rotation rates, the instability develops radially first with the azimuthal perturbations becoming more pronounced as time evolves.  By the time $t\approx3.0$\,s it is difficult to distinguish which structures arose due to a radial or azimuthal perturbation.  The key observation is that the observed length scale of the structures is smaller for greater rotation rates.

Fig.~\ref{fig:wavelengths} shows that as the rotation rate increases the observed length scale of the instability decreases, and asymptotes to an approximately constant, finite length, over the range of parameters explored.  For the parameters chosen the finite length is approximately 6\,mm.   From \S\,\ref{sec:modeling}, we anticipate that rotation inhibits the formation of large structures, at least in the radial direction, and the faster the rotation the more inhibited the larger structures are in general.  However, unlike the inviscid analysis presented in \S \ref{sec:modeling}, the fluids are viscous and it has been shown \citep[e.g.,][]{chandra} that the viscosity of the fluid inhibits the formation of small structures in classical non-rotating Rayleigh-Taylor instability.  We therefore interpret the 6\,mm asymptote of observed lengthscale with increasing rotation rate as being the result of two competing effects: inhibition of large structures due to rotation and inhibition of small structures due to fluid viscosity.

\subsection{Suppression of the growth of the instability}
\label{sec:expres}

\begin{figure*}
\begin{center}
\vspace*{20pt}
\begin{picture}(460,172)(-6,-15)
\put(10,0){\includegraphics[width = 460pt]{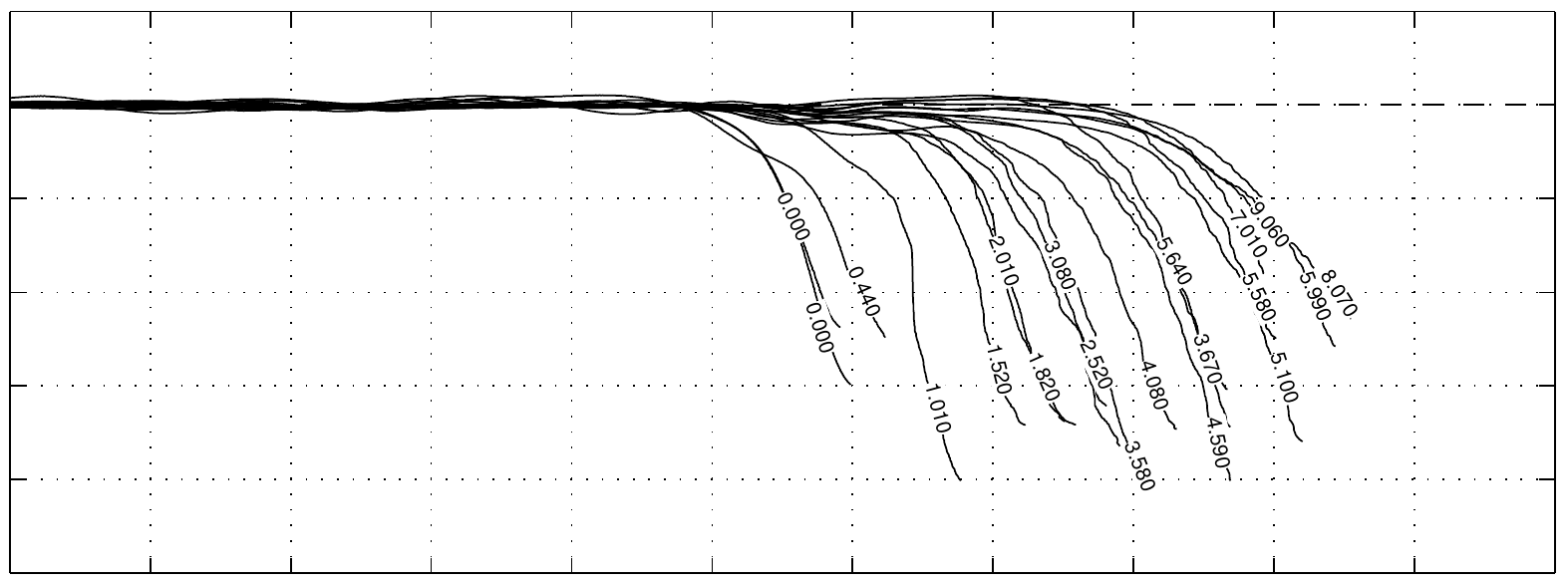}}
\put(-2,0){{\small -25}}
\put(-2,28){{\small -20}}
\put(-2,55){{\small -15}}
\put(-2,83){{\small \phantom{-}{10}}}
\put(-2,110){{\small \phantom{-1}{5}}}
\put(-2,138){{\small \phantom{-1}{0}}}
\put(-2,165){{\small \phantom{-1}{5}}}
\put(6,-6){{\small -12}}
\put(46,-6){{\small -10}}
\put(90,-6){{\small -8}}
\put(131,-6){{\small -6}}
\put(172,-6){{\small -4}}
\put(214,-6){{\small -2}}
\put(258,-6){{\small 0}}
\put(299,-6){{\small 2}}
\put(340,-6){{\small 4}}
\put(381,-6){{\small 6}}
\put(422.5,-6){{\small 8}}
\put(461,-6){{\small 10}}
\put(218,-19){Time, $t$\,(s)}
\put(-12,20){\rotatebox{90}{Relative interface position (mm)}}
\end{picture}
\caption{\label{fig:onsettimea} The position of the interface, relative to the tank, for 20 different rotation rates (rad\,s$^{-1}$).  There is a trend that the slower the rotation rate of the experiment the sooner the instability appears to grow, and the faster it appears to grow.}
\end{center}
\end{figure*}
\begin{figure}
\begin{center}
\begin{picture}(200,200)(-2,0)
\put(10,0){\includegraphics[width = 200pt]{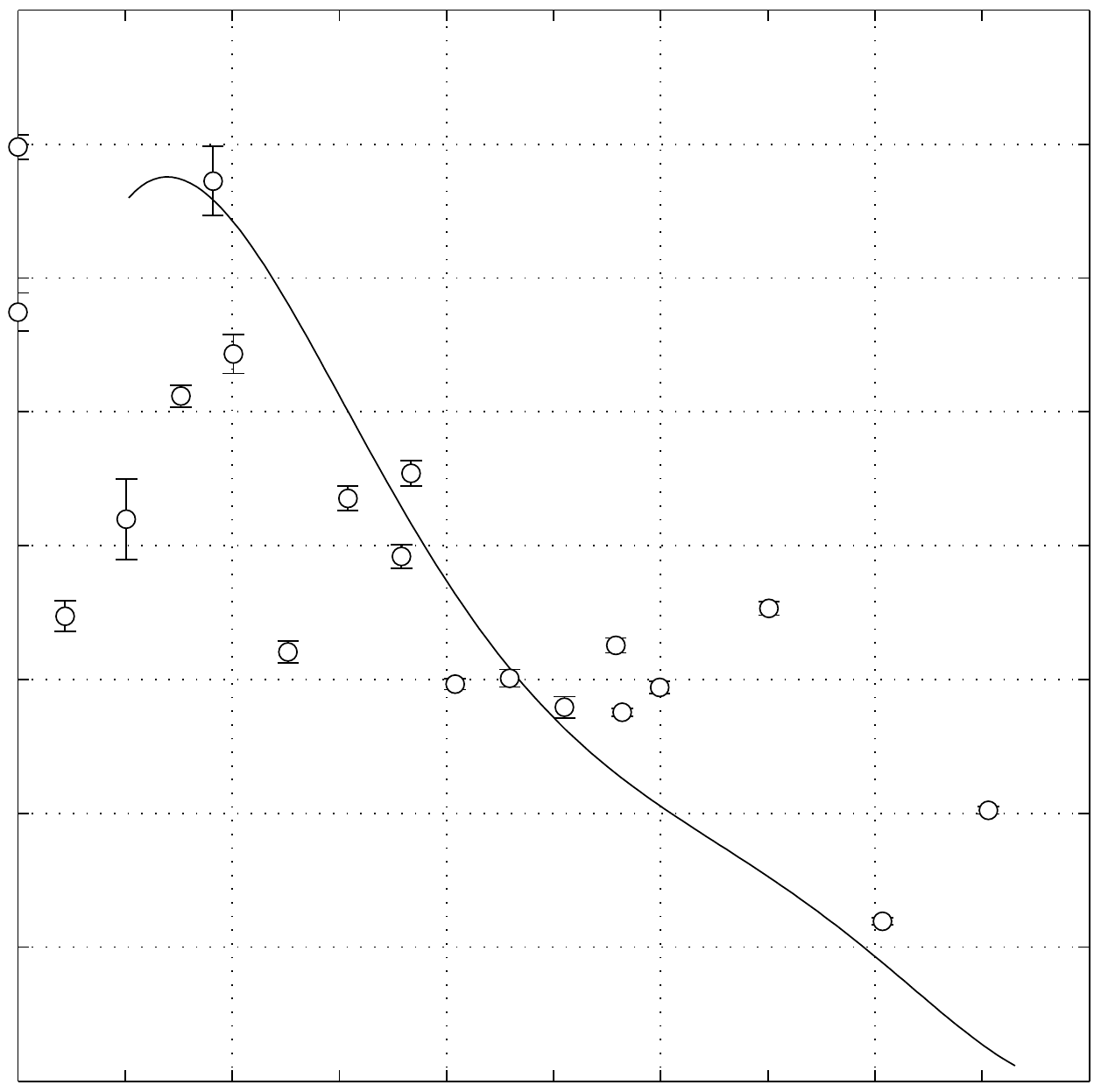}}
\put(-1,0){{\small 0.4}}
\put(-1,24){{\small 0.6}}
\put(-1,49){{\small 0.8}}
\put(-1,73){{\small 1.0}}
\put(-1,97){{\small 1.2}}
\put(-1,121){{\small 1.4}}
\put(-1,146){{\small 1.6}}
\put(-1,170){{\small 1.8}}
\put(-1,194){{\small 2.0}}
\put(11,-6){{\small 0}}
\put(49,-6){{\small 2}}
\put(88,-6){{\small 4}}
\put(127,-6){{\small 6}}
\put(166,-6){{\small 8}}
\put(203,-6){{\small 10}}
\put(51,-18){Angular Velocity, $\Omega$\,(rad\,s$^{-1}$)}
\put(-15,48){\rotatebox{90}{Growth rate, $-\Im(\omega)$\,(s$^{-1}$)}}
\end{picture}\\[20pt]
\caption{\label{fig:onsettime} 
 (c) Exponential growth rates of the form $\exp\{\textrm{i}\,\omega t\}$ for the interface profiles shown in (a) (points, with 95\% confidence interval associated with the best-fittting), the solid line is the estimated growth rate based on an assumption of axisymmetric perturbations at the dominant wavelength of growth.  The Reynolds numbers attained in the experiments range from $\mathcal{O}(100)$ for the slowest rotating experiments with the fastest growth and largest observed length scales down to $\mathcal{O}(20)$ for the most rapidly rotating experiments with the slowest growth and smallest observed length scales.}
\end{center}
\end{figure}

Fig.~\ref{fig:onsettimea} is a plot of the interface position relative to the tank against time for 20 different rotation rates in the range $\Omega\in[0,9.06]$\,rad\,s$^{-1}$.  The specific rotation rate is labelled on each curve.  The plot shows a trend towards slower growth rates with increasing angular velocity.  In Fig.~\ref{fig:onsettime} an exponential curve of the form $\exp\{-\Im(\omega) t\}$ has been best-fitted to the interface profiles in Fig.~\ref{fig:onsettimea} and the growth rate $-\Im(\omega)$ plotted.  The error bars represent a 95\% confidence interval for the fit to the profile only.  A trend can again be observed with the experiments with the highest rotation rates having the lowest growth rates.  The solid line in Fig.~\ref{fig:onsettime} is the prediction of the growth rate, calculated as in \S \ref{sec:axisarb}, of the dominant observed wavelength.  Here a dominant axisymmetric mode has been assumed with $k_n$ in the range $1 \leqslant n\leqslant 15$ where $\lambda\approx 2\pi a/k$ and the dominant wavelength has been calculated using an empirical relationship $\lambda/a = 0.12 + \left[0.017/\left(\alpha + 0.023\right)\right]^{3.5}$ (see \S \ref{sec:structure}, Fig.~\ref{fig:wavelengths}) for the experimentally obtained $\lambda$.  

\begin{figure}
\begin{center}
\setlength{\unitlength}{1pt}
\begin{picture}(200,157)(-2,0)
\put(10,15){\includegraphics[width = 200pt]{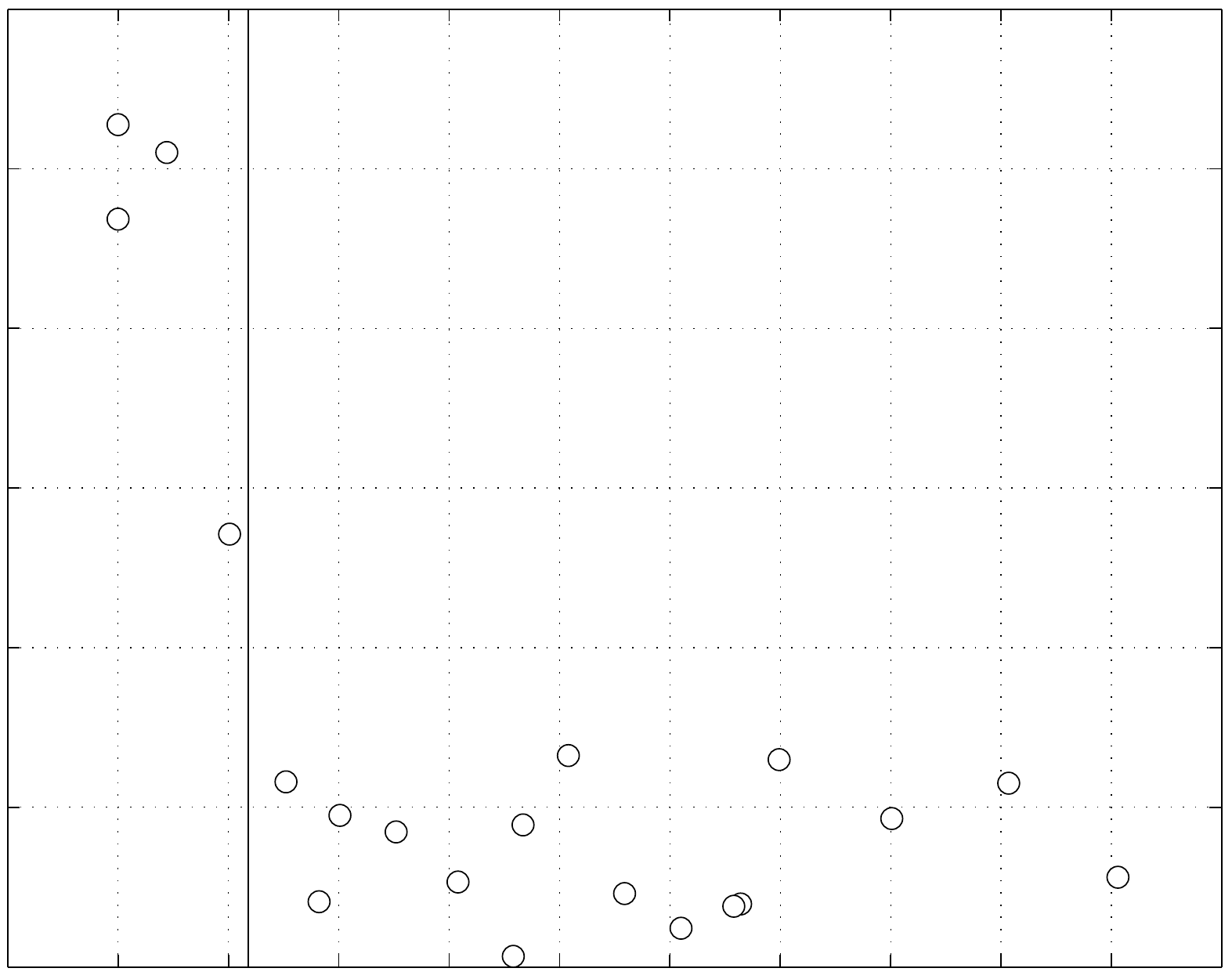}}
\put(29,8){{\small 0}}
\put(64,8){{\small 2}}
\put(99,8){{\small 4}}
\put(133,8){{\small 6}}
\put(168,8){{\small 8}}
\put(203,8){{\small 10}}
\put(51,-4){Angular Velocity, $\Omega$\,(rad\,s$^{-1}$)}
\put(-2,15){{\small \phantom{0.}0}}
\put(-2,41){{\small 0.1}}
\put(-2,66){{\small 0.2}}
\put(-2,92){{\small 0.3}}
\put(-2,117){{\small 0.4}}
\put(-2,143){{\small 0.5}}
\put(-2,168){{\small 0.6}}
\put(-16,34){\rotatebox{90}{Proportional magnitude of $c_1$}}
\end{picture}
\end{center}
\caption{\label{fig:modecomposition} The proportional mode-one contribution to the interfacial profile, when represented as a sum of Bessel functions $\mathcal{J}_0(k_n r/a)$, at time $t=T$, against rotation rate.   The vertical solid line is the critical rotation rate, $\Omega=\Omega_c\approx 1.18$\,rad\,s$^{-1}$, above which the linear theory of section \S \ref{sec:critrate} predicts suppression of the mode-one instability, but below which predicts that the mode remains unstable.}
\end{figure}

We can show from \eqref{eq:crit} that for $0<-\Atw\ll1$ we have
\begin{subequations}  \label{eq:asymcrit}
\begin{equation}
\alpha_c \sim \frac{k^2}{4}\delta|\Atw| + \mathcal{O}(\Atw^2)
\end{equation}
and therefore
\begin{equation}
\Omega_c\approx 1.92 \frac{\sqrt{dg|\Atw|}}{a},
\end{equation}
\end{subequations}
for the largest mode.  Hence for the chosen experimental parameters ($d = 3.93\times10^{-2}$\,m, $a = 4.50\times10^{-2}$\,m, $g = 9.81$\,m\,s$^{-2}$) and given that the instability is observed experimentally to occur where the apparent Atwood number lies in the range $-2\times10^{-3}\lesssim\Atw<0$ (see \citep{scirep} Fig.~3), then both \eqref{eq:crit} and \eqref{eq:asymcrit} indicate that the largest mode is stabilised at $\Omega_c\approx1.18$\,rad\,s$^{-1}$.  Fig.~\ref{fig:modecomposition} shows the proportional contribution of mode-one, the largest mode, to the overall interface profile of the developing instability at the time at which the instability has an amplitude of $0.05 d$.  The experimental images were stretched to remove the effects of rotation such that $r\mapsto r$, $z\mapsto z - \Omega^2(r^2 - a^2/2)/2g$ and the initial approximately parabolic profile was rendered horizontal.  Contouring the image then yielded a fit of the interface of the form $z = \zeta(r)$ and the coefficients $c_n$ were found such that $\zeta(r) = \sum c_n \mathcal{J}_0(k_nr/a)$ where $k_n$ are the zeros of $\mathcal{J}_1$ as in \S \ref{sec:modeling}.  The proportional contribution of mode-one was then calculated as $|c_1|/\sum{|c_n|}$, as plotted in Fig.~\ref{fig:modecomposition}.  The solid vertical line is the critical rotation rate, $\Omega_c\approx1.18$\,rad\,s$^{-1}$.  As can be seen, for rotation rates above the critical rotation rate the mode-one instability appears suppressed, but for rotation rates below the critical rotation rate the proportional contribution of mode-one is significantly greater.  The suppression of the mode-one instability can be readily observed by eye in \citep{scirep} Fig.~1.

\section{\label{sec:conc}Discussion and Conclusions}

We have considered theoretically the effects of rotation upon the classical Rayleigh-Taylor instability and compared this theory with experimental results obtained using linearly magnetizable fluids.  The dispersion relation for interfacial disturbances at low rotation rates \eqref{eq:freq} suggests that axisymmetric modes of a developing Rayleigh-Taylor instability may have their rate of growth inhibited by rotation.  Indeed, if the critical rotation rate for the mode is below the threshold $2(gd)^{1/2}/a$, then \eqref{eq:crit} indicates that the mode may be stabilised indefinitely.  Rotation was also seen in some cases to be able to slow the growth of asymmetric modes.  Our observations from the experiments are broadly inline with our theoretical predictions, specifically that by rotating the system we inhibit the growth of large wave modes and suppress the growth rate of the instability (see e.g., Fig.~\ref{fig:quali}).  If the dominant wavelength of instability is obtained from experiment, the theoretically predicted growth rates compare well with our observations.  We have derived a critical rotation rate for stabilizing axisymmetric modes and observed experimentally that the magnitude of the largest fundamental mode is significantly suppressed above the calculated critical rotation rate (see Fig.~\ref{fig:modecomposition}).

We can understand our observations in the following qualitative manner: a rotating fluid is known to organise itself into coherent vertical structures aligned with the axis of rotation, so-called `Taylor columns' \citep{taylor23}, whereas a perturbation to an unstable two-layer density stratification will lead to baroclinic generation of vorticity at the interface, tending to break-up any vertical structures.  Hence the system under investigation undergoes competition between the stabilising effect of the rotation, that is organising the flow into vertical structures and preventing the layers passing each other, and the destabilising effect of the denser fluid overlying the lighter fluid that generates an overturning motion at the interface.  With increased rotation rate the ability of the fluid layers to move radially, with opposite sense to each other, in order to rearrange themselves into a more stable configuration, is increasingly prohibited by the Taylor-Proudman theorem \citep[see][]{proudman, taylor17}.  The radial movement is therefore reduced and the observed structures that materialize as the instability develops are smaller in scale.

The experimentally observed structures are also affected by the viscosity of the fluid layers.  As is known, viscous diffusion suppresses the growth of small-scale structures in non-rotating Rayleigh-Taylor flows and we have have observed the same system response in the rotating regimes too.  Higher viscosity in the fluid layers leads to larger observed structures in the developing instability.  We therefore also have competition between rotation suppressing the larger structures and viscosity suppressing the smaller structures, leading to an observed instability structure that depends on both the rotation rate and the fluid viscosity.  The length scale of the observed dominant wavelength of the instability remains an open question.   We can say that the structure of the Rayleigh-Taylor instability is significantly altered by rotating the system, with the scale of the instability decreasing with increasing rotation rate to a limit controlled by the fluid viscosity.

A key question is whether the instability could be stabilized indefinitely given a sufficiently high rotation rate.  Equivalently we could ask whether it is possible to find a rotation rate rapid enough to completely suppress the growth of the instability for any desired length of time?  It seems reasonable to suggest that it would be possible to rotate the system quickly enough that the large-scale structures are suppressed by rotation, and any remaining small-scale structures are suppressed by viscosity.  However, noting that the theory presented in \S \ref{sec:modeling} is limited to a maximum rotation rate $\alpha<4\delta$ and the experiments presented are necessarily rotation rate limited too, the evidence presented in Fig.~\ref{fig:onsettime} indicates that while the instability can be suppressed, it cannot be suppressed indefinitely, at least not in the configurations considered.  

MMS acknowledges funding from EPSRC under grant number EP/K5035-4X/1, RJAH acknowledges support from EPSRC Fellowship EP/I004599/1.  We thank L.~Eaves, P.~Linden, E.~Hall and M.~Swift for useful discussions, and T.~Wright and O.~Larkin for technical support.

\end{document}


\title{The Rotating Rayleigh-Taylor Instability: Supplementary Information} 

\author{M.~M.~Scase$^1$, K.~A.~Baldwin$^2$ \& R.~J.~A.~Hill$^3$}
\email{matthew.scase@nottingham.ac.uk}
\affiliation{$^1$School of Mathematical Sciences, University of Nottingham, Nottingham NG7 2RD, UK\\
$^2$Faculty of Engineering, University of Nottingham, Nottingham NG7 2RD, UK\\
$^3$School of Physics and Astronomy, University of Nottingham, Nottingham NG7 2RD, UK}
\date{February 29, 2016} 
\maketitle

\section{Equivalence between initial conditions and system acceleration for non-rotating systems}

Taylor [42] considered a two-layer non-rotating system: an upper layer, of density $\rho_1$, overlying a lower layer, of density $\rho_2$, with an initially horizontal hydrostatic interface perpendicular to the direction of gravity. It was shown that a stable two-layer stratification supported stable interfacial standing waves with a temporal dependence of the form $\exp\left\{\textrm{i}\,\omega t\right\}$, where the frequency, $\omega$, satisfied $\omega^2 \propto g \Atw$ (the constant of proportionality depending on the wavenumber of the wave).  Taylor showed that by subjecting the system to a constant acceleration vertically downwards, at a rate greater than that of gravity, the system became unstable.  The inertia of the denser lower layer is greater than that of the lighter upper layer and so the two layers are induced to swap position.  It was shown that if the system was subjected to a vertical acceleration $g_1$ (using Taylor's notation) then $\omega^2\propto(g+g_1)\Atw$ [42].  It can be seen therefore that for $g_1<-g$, $\omega^2<0$, i.e., $\omega$ is imaginary, and perturbations to the interface grow rather than oscillate as standing waves.  This growth represents the onset of the Rayleigh-Taylor instability.  Exactly the same growth can be seen if, instead of subjecting the system to a bulk acceleration $g_1$, the sign of the Atwood number, $\Atw$, is changed by inverting the initial stable stratification.  There is therefore an equivalence between accelerating a stable initial stratification rapidly vertically downwards ($g_1<-g$) and creating an unstable initial stratification ($\Atw <0$). 

This equivalence between acceleration of the system and unstable initial stratification does not exist when the system is rotating.  Independent of the density field, the isobaric surfaces have identical paraboloidal profiles, as is shown by (2).  This equation also shows that if the direction of acceleration is instantaneously inverted, the isobaric surfaces are simultaneously inverted, becoming convex (as can be seen by changing the sign of $g$ in (3)).  However, the physical interface between the two fluids cannot instantaneously invert, and so the fluid arrangement is no-longer hydrostatic and an instability develops from non-hydrostatic conditions.  The equivalence of changing the sign of $\Atw$ and $g$ is lost except in the special case when the interface is horizontal, i.e., the non-rotating $\Omega=0$ case, when inverting the isobaric surfaces in the vertical direction has no effect.

\section{\label{sec:mag}Rayleigh-Taylor Instability in paramagnetic and diamagnetic fluids}

In the absence of magnetic and viscous effects, the stress tensor associated with the short-range molecular forces [see e.g., 2] depends only upon the thermodynamic pressure, $\tn{\sigma} = -p\,\tn{I}$.  The stress tensor must be modified however in the presence of a magnetic field and the additional terms [see 34, (4.29)], $\tn{\sigma}\,_m$ are given by
\begin{equation} \label{eq:ms1}
\tn{\sigma}\,_m = -\left\{\int_0^H\mu_0\left.\pd{\left(\nu M\right)}{\nu}\right|_{H,T}\,\textrm{d}H + \frac{\mu_0}{2}H^2\right\}\tn{I} + \vc{B}\vc{H},
\end{equation}
where $H = |\vc{H}|$, $M = |\vc{M}|$, $T$ is the temperature, $\nu=\rho^{-1}$ is the specific volume, and the dyadic product is taken in the final term. 

The magnetic induction, $\vc{B}$ is defined by $\vc{B} = \mu\vc{H}$, where $\mu$ is the magnetic permeability [see e.g., 19].  In a vacuum $\vc{B} = \mu_0 \vc{H}$, where $\mu_0$ is the magnetic permeability of free-space, and takes a value $\mu_0 = 4\pi\times10^{-7}$\,N\,A$^{-2}$.   The magnetization per unit volume, $\vc{M}$, is defined via $\vc{B} = \mu_0(\vc{H} + \vc{M})$.

We assume that the fluids are linearly magnetizable, consistent with the choice of diamagnetic and paramagnetic fluids used in [1] and discussed in \S IV.  The fluids may therefore be characterized by a constant magnetic volume susceptibility, $\chi$ (nondimensional), such that $\vc{M} = \chi\vc{H}$.  For the experiments in [1], typical orders of magnitude of $\chi$ were $10^{-5}$ or smaller.  It follows that we may take $\mu = \mu_0(1+\chi)\approx\mu_0$, and $\vc{H}$, $\vc{B}$, and $\vc{M}$ are collinear.

We further assume that the magnetization per unit mass, $\nu M$, is independent of density, so that the integral in \eqref{eq:ms1} vanishes [34].  The implication of this final assumption is that for an isothermal flow, the `magnetostrictive pressure'
\begin{equation}
p_s \equiv \mu_0 \int_0^H \nu\left.\pd{M}{\nu}\right|_{H,T}\,\textrm{d}H
\end{equation}
exactly balances the `fluid-magnetic pressure'  [as defined in 34, (4.36b)], $p_m$, as
\begin{multline}
p_m \equiv \mu_0\int_0^H M\,\textrm{d}H  \\ = \int_0^H \mu_0 \left[\pd{(\nu M)}{\nu} - \nu\pd{M}{\nu}\right]_{H,T}\,\textrm{d}H = 0 - p_s.
\end{multline} 
Equivalently, the `composite pressure' [34], $p^* \equiv p + p_s + p_m$, is identical to the thermodynamic pressure, $p$.
The full stress tensor for the flow may therefore be written as
\begin{equation} \label{eq:ms2}
\tn{\sigma} = -\left\{p + \frac{\mu_0}{2}H^2\right\}\tn{I} + \vc{B}\vc{H}.
\end{equation}
The contribution to the total body force on a fluid element due to magnetic effects is given by $\vc{f}_m = \nabla\cdot\tn{\sigma}\,_m$, hence
\begin{equation} \label{eq:fm0}
\vc{f}_m = -\frac{\mu_0}{2}\nabla H^2 + \left(\vc{B}\cdot\nabla\right)\vc{H} + \left(\nabla\cdot\vc{B}\right)\vc{H}.
\end{equation}
Maxwell's equations state that the magnetic induction is solenoidal, and so the third term on the right hand side of \eqref{eq:fm0} is zero.  For the present analysis, we assume magnetostatic conditions.  Furthermore, we model an idealised fluid system in which there are no free electric currents, $\vc{J}_\textrm{free}$, in the fluids, which is equivalent to assuming that the fluids have negligible conductivity.  In this case, $\nabla\times\vc{H}=\vc{J}_\textrm{free}=\vc{0}$.
Using the identity $\frac{1}{2}\nabla H^2 = (\vc{H}\cdot\nabla)\vc{H} + \vc{H}\times(\nabla\times\vc{H})$, we may rewrite the magnetic body force as
\begin{equation} \label{eq:fm}
\vc{f}_m = -\frac{\mu_0}{2}\nabla H^2 + \frac{\mu}{2} \nabla H^2 \equiv \mu_0 M\nabla H \equiv \frac{\mu_0\chi}{\mu^2}B\nabla B.
\end{equation}
For constant $|\chi|\ll1$, $\mu_0\mu^{-2}\sim\mu_0^{-1}[1+O(\chi)]$ and so the final expression may be well-approximated by 
\begin{equation}
\vc{f}_m \approx \frac{\chi}{\mu_0}B\nabla B = \nabla\left[\frac{\chi B^2}{2\mu_0}\right].
\end{equation}
We see that the body force due to the magnetic effects is conservative and may be absorbed into a modified pressure.  There is therefore an analogy with buoyancy driven flow in a two-layer fluid system.  A wave may be supported at an interface between two fluid layers with uniform, but contrasting, densities.  No fluid element in the bulk of either layer experiences a `buoyancy force' upon itself as it is surrounded by fluid entirely of its own density.  However, the effect of the difference in density is felt throughout both fluids via the continuity of normal stress across the interface, and hence a baroclinically driven flow may be realised and sustained.  Similarly here, the body force in each fluid layer due to the gradient magnetic field may be removed by consideration of a suitable background pressure, but a flow may still be driven due to the difference in magnetic susceptibility of the two fluids and the stress continuity condition at their interface.

For the purposes of calculating the continuity of stress across the fluid-fluid interface we now consider the appropriate generalization of Bernoulli's equation to an isothermal rotating magnetic system in the absence of viscosity.  This will enable the calculation of the profile of the interface between the two fluid layers once spun up into a hydrostatic state.  
The rotating Euler equation, including the magnetic body force \eqref{eq:fm} is
\begin{multline} \label{eq:mag1}
\rho\Dv{\vc{u}}{t} = -\nabla p + \rho\,\vc{g} - \rho\,\vc{\Omega}\times(\vc{\Omega}\times\vc{x}) \\ - 2\rho\,\vc{\Omega}\times\vc{u} + \mu_0 M \nabla H.
\end{multline}
It follows that in a hydrostatic system with $\vc{\Omega} = \Omega \vc{e}_z$ that
\begin{equation}
\vc{0} = -\nabla \left[ p + \rho g z - \frac{\rho\,\Omega^2r^2}{2} - \mu_0 \int^H M\,\textrm{d}H \right],
\end{equation}
[cf. 34].  For the case of linearly magnetizable fluids considered here, $M=\chi H$ and so the final term is taken to be $\frac{1}{2}\mu_0\chi H^2$.  Hence we have that
\begin{equation} \label{eq:bn1}
p + \rho g z - \frac{\rho\,\Omega^2r^2}{2} - \frac{\mu_0\chi}{2}H^2 = \textrm{const}.
\end{equation}

Using the same notation as in Fig.~2 and taking $\vc{n}$ to be a unit normal to the interface directed from fluid 2 into fluid 1, then \eqref{eq:ms2} implies that
\begin{multline} \label{eq:sg1}
\tn{\sigma}\cdot\vc{n}\Big|^+_- = -\left[p_2 - p_1 + \frac{\mu_0}{2}(H_2^2 - H_1^2)\right]\vc{n} \\ + \left(\vc{B}\cdot\vc{n}\right)\left[\vc{H}_2 - \vc{H}_1\right],
\end{multline}
where subscript 1 denotes evaluation just above the interface, and subscript 2 denotes evaluation just below the interface.  Continuity of $\vc{B}\cdot\vc{n}$ has been used in \eqref{eq:sg1} which follows as a result of $\vc{B}$ being solenoidal (or equivalently from Gauss's Law, $\oint_S \vc{B}\cdot\textrm{d}\vc{S}=0$).  The tangential component of $\vc{H}$ is continuous across the interface [see 34, (5.17), (5.18)], giving $\vc{H}_2 - \vc{H}_1 = (H_{2n} - H_{1n})\vc{n}$ and $H_2^2 - H_1^2 = H_{2n}^2 - H_{1n}^2$, where $H_{jn}\equiv\vc{H}_j\cdot\vc{n}$.  Writing $B_n \equiv \vc{B}\cdot\vc{n}$, \eqref{eq:sg1} is therefore
\begin{multline} \label{eq:sg2}
\tn{\sigma}\cdot\vc{n}\Big|^+_- = -\Big[p_2 - p_1 + \frac{\mu_0}{2}(H_{2n} + H_{1n})(H_{2n} - H_{1n})  \\ - B_n(H_{2n}-H_{1n})\Big]\vc{n}.
\end{multline}
It follows from $\vc{B} = \mu_0(\vc{H} + \vc{M})$ that $H_n = B_n/\mu_0 - M_n$. Substituting this into \eqref{eq:sg2} and simplifying gives
\begin{equation} \label{eq:sg3}
\tn{\sigma}\cdot\vc{n}\Big|^+_- = -\left[p_2 - p_1 - \frac{\mu_0}{2}(M_{2n}^2 - M_{1n}^2)\right]\vc{n}.
\end{equation}
The terms involving $M_n^2$ represent a traction at the interface and are referred to as the `magnetic normal traction' [34].  

To find the profile of the hydrostatic interface between the two fluid layers, we enforce continuity of normal stress across the fluid interface, given by
\begin{equation}
\tn{\sigma}:\vc{n}\vc{n}\Big|_-^+ = 0\quad\Rightarrow\quad p_2 - p_1 + \frac{\mu_0}{2}(M_{2n}^2 - M_{1n}^2) = 0.
\end{equation}
Substituting in from the generalized Bernoulli equation, \eqref{eq:bn1}, we have therefore that
\begin{multline} \label{eq:if1}
(\rho_2 - \rho_1)gz - (\rho_2 - \rho_1)\frac{\Omega^2 r^2}{2} - \frac{\mu_0}{2}(\chi_2H_2^2 - \chi_1H_1^2) \\ + \frac{\mu_0}{2}(M_{2n}^2 - M_{1n}^2) = \textrm{const}.
\end{multline}
Equation \eqref{eq:if1} represents a generalization of the two ferrofluid case considered in Rosenweig [35, (42)] for a rotating system, but restricted to linearly magnetizable fluids.

We may approximate \eqref{eq:if1} under the assumption that $|\chi_j|\ll1$ and see that the final term, representing the magnetic normal traction is an order of magnitude smaller in $\chi$ than the third term, representing the magnetic body force.  Hence, we neglect the magnetic normal traction in this case and approximate $\mu_0 H_j^2\sim B^2\mu_0^{-1}(1 + O(\chi_j))$ to write the hydrostatic interface profile, $z_0(r)$, as
\begin{equation} \label{eq:if2}
z_0(r) - \frac{\Omega^2 r^2}{2g} - \frac{\chi_2- \chi_1}{\rho_2-\rho_1}\frac{B(r,z_0(r))^2}{2g\mu_0}  = \textrm{const}.
\end{equation}
As $B$ varies vertically in general, \eqref{eq:if2} is an implicit relation for the profile of the interface between two hydrostatic rotating fluid layers.  In the absence of the magnetic field, or in the case that the two fluids have identical magnetic susceptibility, the parabolic profile in (3) is recovered.  The profiles in \eqref{eq:if2} and (3) also coincide in the no-rotation limit if the magnetic field has no radial variation, with both profiles being a horizontal plane, $z_0 = \textrm{const.}$  Equation \eqref{eq:if2} expresses the fact that the sum of the gravitational potential, $U_g$, centrifugal potential, $U_c$, and magnetic potential, $U_m$, of a fluid element of volume $\textrm{d}V$, of susceptibility and density $\chi_2$ and $\rho_2$ respectively immersed in a fluid of susceptibility and density $\chi_1$ and $\rho_1$ respectively, is constant on the surface $z_0(r)$, where
\begin{displaymath}
U_g = \left(\rho_2 - \rho_1\right)gz\,\textrm{d}V, \quad
U_c = {\textstyle \frac{1}{2}}\left(\rho_2 - \rho_1\right)r^2\Omega^2 \,\textrm{d}V,
\end{displaymath}
\begin{displaymath}
U_m = -\left(\chi_2 - \chi_1\right)B^2\,\textrm{d}V/(2\mu_0).
\end{displaymath}

The magnetic body force, $\vc{f}_m$, may be absorbed into a modified pressure term, $p - \chi B^2/(2\mu_0)$.  The approach taken in \S II.1 to investigate the Rayleigh-Taylor instability therefore carries over to the magnetic case almost entirely unchanged. The equation of motion for each fluid layer may be written
\begin{equation}
\Dv{\vc{u}_j}{t} = -\frac{1}{\rho_j}\nabla \left[p_j - \frac{\chi_j B^2}{2\mu_0}\right]  + \vc{g} - \vc{\Omega}\times\left(\vc{\Omega}\times\vc{x}\right) - 2\vc{\Omega}\times\vc{u}_j,
\end{equation}
with (5) unchanged and (6) modified to account for the background magnetic induction as
\begin{multline} \label{eq:magpress}
p_j = p_0 - \rho_j\left\{gz - \frac{\Omega^2r^2}{2} - \frac{B^2}{2\mu_0}\frac{\chi_j}{\rho_j}\right\} \\ - \epsilon\rho_j\left\{\pd{\phi_j}{t} + \frac{1}{4\Omega^2}\pd{^3\phi_j}{t^3}\right\}.
\end{multline}
The analysis of \S III then follows.

\section{\label{sec:exp}Experiments}
 
\begin{figure}
\begin{center}
\setlength{\unitlength}{1pt}
\begin{picture}(200,256)(-4,0)
\put(0,0){\includegraphics[width = 200pt]{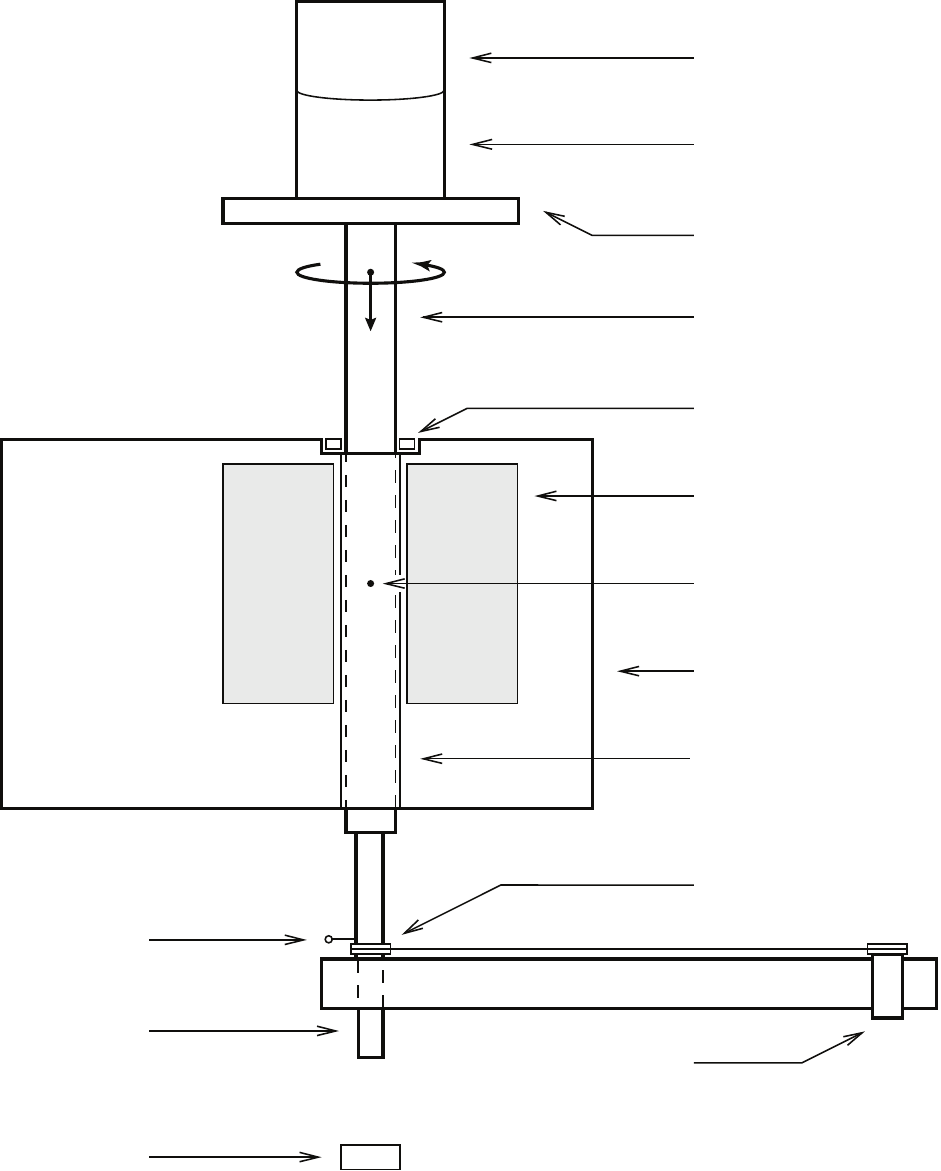}}
\put(150,234){Light layer}
\put(150,215){Dense layer}
\put(150,197){Stage}
\put(150,178){Copper cylinder}
\put(150,160){PTFE bearing}
\put(150,141){Solenoid}
\put(150,122){Solenoid center}
\put(150,104){Magnet}
\put(150,85){Magnet bore}
\put(150,59){Slip bearing}
\put(-18,47){Holding pin}
\put(119.5,20.5){Motor}
\put(-16,27){Drive shaft}
\put(-23,1){PTFE buffer}
\end{picture}
\end{center}
\caption{Experimental arrangement (not to scale).  A gravitationally stable two-layer stratification was spun-up above the magnet.  By removing the holding pin the system was lowered slowly ($\sim 10$\,mm\,s$^{-1}$) into the magnetic field under its own weight.  The motor rotated the drive shaft about a vertical axis and the slip ring allowed the stage to descend, whilst being simultaneously rotated.  Arrows indicate the direction of rotation and descent of the stage.  In the first series of experiments a polytetrafluoroethylene (PTFE) buffer halted the tank's descent at a prescribed height so that thereafter the magnetic field acting on the fluids was steady.}
\label{fig:exprig}
\end{figure}

In our previous experiments [1] a weak upward magnetic force acted on a diamagnetic lower layer of NaCl aqueous solution due to the interaction of the imposed strong magnetic field with the orbital motion of the electrons in the water molecules [see, e.g., 17]. The downward magnetic force on the paramagnetic upper layer of MnCl$_2$ aqueous solution was due to the interaction of the magnetic field with the spins of the Mn$^{2+}$ ions [see, e.g., 17] which was stronger than the upward magnetic force on the water molecules in this layer.   
As discussed in \S III, these magnetic forces are body forces: the force per unit volume is given to a close approximation by $\vc{f}_m = \chi B \nabla B/\mu_0$, in which $\chi$ is a positive quantity for the paramagnetic fluid and a negative quantity for the diamagnetic fluid. Note that, in contrast to a ferrofluid, the magnetic susceptibility of the fluids in the experiments was sufficiently small (the magnitude of the SI volume susceptibility $|\chi|\sim10^{-6}$--$10^{-5}$) that the magnetic field generated by the magnetisation of the fluid could be considered insignificant compared to the applied field (approximately $1$\,T).

In modeling the two-fluid system in \S \ref{sec:mag}, we assumed the conductivity to be zero. In our experiments, since we used ionic solutions, the conductivity was finite but small compared to a fluid metal or plasma for example. We measured the conductivity to be $\sigma \sim 1$--$4$\,S\,m$^{-1}$. The effect of this small conductivity on the fluids was confined to introducing a weak, viscous-like damping of the fluids, as we now discuss. Firstly, magnetic fields generated by electromagnetic induction were negligible compared to the externally imposed magnetic field. The magnetic Reynolds number $\textrm{Re}_\textrm{m} = \mu_0 U L \sigma$ gives a measure of the strength of the magnetic field generated this way. For the largest length scale, $L$, and velocity scale, $U$, in our experiments, $\textrm{Re}_\textrm{m} \sim 10^{-8}$--$10^{-7}$. Secondly, the Lorentz force, which is the force generated by current-carrying ions moving through the magnetic field, is weak compared to the Coriolis force. This may be shown by considering the Elsasser number, a measure of the relative importance of the Lorentz force to the Coriolis force. The Elsasser number for our experiments, $\textrm{El} = \sigma B^2/(\rho\Omega)$, was of the order of $10^{-2}$ for rotation rates $\Omega > 1$\,rad\,s$^{-1}$. We sought to verify that the conductivity of the fluids played a negligible role in the experiments by performing a separate series of experiments using more massive zinc sulphate ions in the lower layer, replacing the sodium chloride ions. This was carried out in such a way that the density of the lower layer remained the same, but the conductivity was reduced by a factor of approximately 35\%; we observed no difference between the experimental runs using the two different solutions in the lower layer.

In some cases the relative magnitude of the Lorentz forces and viscous forces became comparable.  The square of the Hartmann number, $B^2L^2\sigma/(\rho\nu)$, a measure of the relative strength of the Lorentz body force per unit mass to the viscous diffusion term in the Navier-Stokes equation was of the order of 1 for the largest length scales ($L\sim10^{-1}$\,m) and lowest viscosities ($\nu\sim10^{-6}$\,m$^2$\,s$^{-1}$) we measured.  It is known that in some cases the Lorentz force can act similarly to a viscous force in that it acts to weakly dampen motion that deviates from solid body rotation [see, e.g., 34].  We therefore investigated the effect of changing the fluid viscosities by the addition of equal amounts glycerol to each layer, see \S \ref{sec:viscosity}, in order to elucidate the possible side-effects of the magnet in this limit.

\subsection{Experimental method}

The experimental tank consisted of a 110\,mm$\times$110\,mm$\times$135\,mm perspex box containing a cylindrical inner tank.  For the first series of experiments the inner cylindrical tank had internal radius 45\,mm and height 130\,mm; the tank's axis of symmetry was aligned with the axis of rotation.  A volume of 250\,ml colourless brine (NaCl$_\textnormal{(aq)}$ 0.43\,mol\,l$^{-1}$) formed the lower, dense diamagnetic layer of the stratification.  The layer had density $\rho_2 = 1012.9\pm1.2$\,kg\,m$^{-3}$.  The layer's magnetic susceptibility was $\chi_2 = -9.03\times10^{-6}$ (SI units), making the layer weakly repelled by the magnetic field.  A lighter, paramagnetic layer of manganese chloride solution (MnCl$_{2\textnormal{(aq)}}$, 0.06\,mol\,l$^{-1}$) was then floated on top of this dense layer using standard floatation boat methods.   The lighter upper layer was dyed for visualisation using water-tracing dye (Cole-Parmer 00295-18).  The lighter layer, with density $\rho_1=998.2\pm0.5$\,kg\,m$^{-3}$, was filled to a volume slightly greater than 250\,ml. This layer's magnetic susceptibility was $3.37\times10^{-6}$ (SI units), making the upper layer weakly attracted by the magnetic field.  The solutions were prepared from distilled deionized water that was allowed to come up to the ambient temperature of the laboratory before use.  The preparation of the stratification took approximately 20\,minutes to complete.  The magnetic susceptibility was measured using a Guoy balance.  The conductivity of the lower layer was measured to be in the range $\sigma\approx3$\,--\,$4$\,S\,m$^{-1}$ and the conductivity of the upper layer was measured to be in the range $\sigma\approx1.5$\,--\,$2$\,S\,m$^{-1}$.  The temperature of the fluids was in the range $T=22\pm2^\circ$C over the run of experiments, with a kinematic viscosity of approximately $\nu = 1.0\times10^{-6}$\,m$^2$\,s$^{-1}$.  The aspect ratio of this series experiments was $\delta\approx0.87$.  

The upper layer was floated on top of the lower layer and then a clear perspex tank lid was submerged into the upper layer, avoiding trapping air bubbles.  The lid was lowered into a position such that, between the tank base and lid, the fluid layers were of equal volume and therefore equal height, approximately 39\,mm.  The cavity between the inner cylinder and the square tank was then filled with distilled water so that when viewed square-on, lensing effects were minimised.  This gravitationally stable two-layer stratification was then placed on the experimental stage (see schematic in Fig.~\ref{fig:exprig}).  With the holding pin in place the tank was spun-up such that both layers were in solid body rotation, which, depending on the chosen angular velocity, typically took between 20\,minutes and 1\,hour.  This cautious spin-up time, comparable to the viscous spin-up timescale $d^2/\nu$, and long compared to the Ekman spin-up time, $\textrm{Ek}^{-1/2}/\Omega$, where the Ekman number $\textrm{Ek} = \nu/(\Omega d^2)$, was due to the presence of the interface modifying the Ekman pumping with behavior comparable to that of a free surface [see e.g., 4, 27].  The angular velocity was progressively ramped up to the final, target rotation rate.  Typical ramping rates were approximately $2\times10^{-3}$\,rad\,s$^{-2}$.  A series of control experiments with interfacial tracer particles were used to determine the spin-up times for the various rotation rates employed.  

The experimental stage was mounted on a copper cylinder (see figure~\ref{fig:exprig}).  After the holding pin was removed the cylinder descended smoothly through the magnet's bore under its own weight.  To maintain a straight vertical descent and prevent precession a PTFE bearing was placed at the top of the magnet bore. The constant speed of descent was governed by the strength of the induced eddy currents in the copper cylinder.  Descent speeds were approximately 9--11\,mm\,s$^{-1}$.  A weak dependence of the descent speed on rotation rate, owing to friction with the slip bearing, was corrected for by the addition of small brass weights to the top of the outer experimental tank.  The copper cylinder was attached at its base to a drive shaft of narrower diameter that had a key shape cross-section.  The drive shaft passed through a slip-bearing with a key-hole shape cut through its center.  An off-axis motor was used to drive the slip bearing around such that the drive shaft dropped smoothly through the bearing whilst simultaneously rotating.  

The experiments were lit using an array of white LEDs and were recorded using a high speed digital video camera ($1920\times1080$\,px, $60$--$240$\,Hz).  Inevitably both the process of floating on the upper layer and the spin up lead to some diffusion of the interface.  The thickness of the diffuse layer was measured as being of the order of 2\,mm at the time the holding pin was released.  

\begin{figure}
\begin{center}
\setlength{\unitlength}{1pt}
\begin{picture}(210,206)(-10,0)
\put(10,0){\includegraphics[width = 200pt]{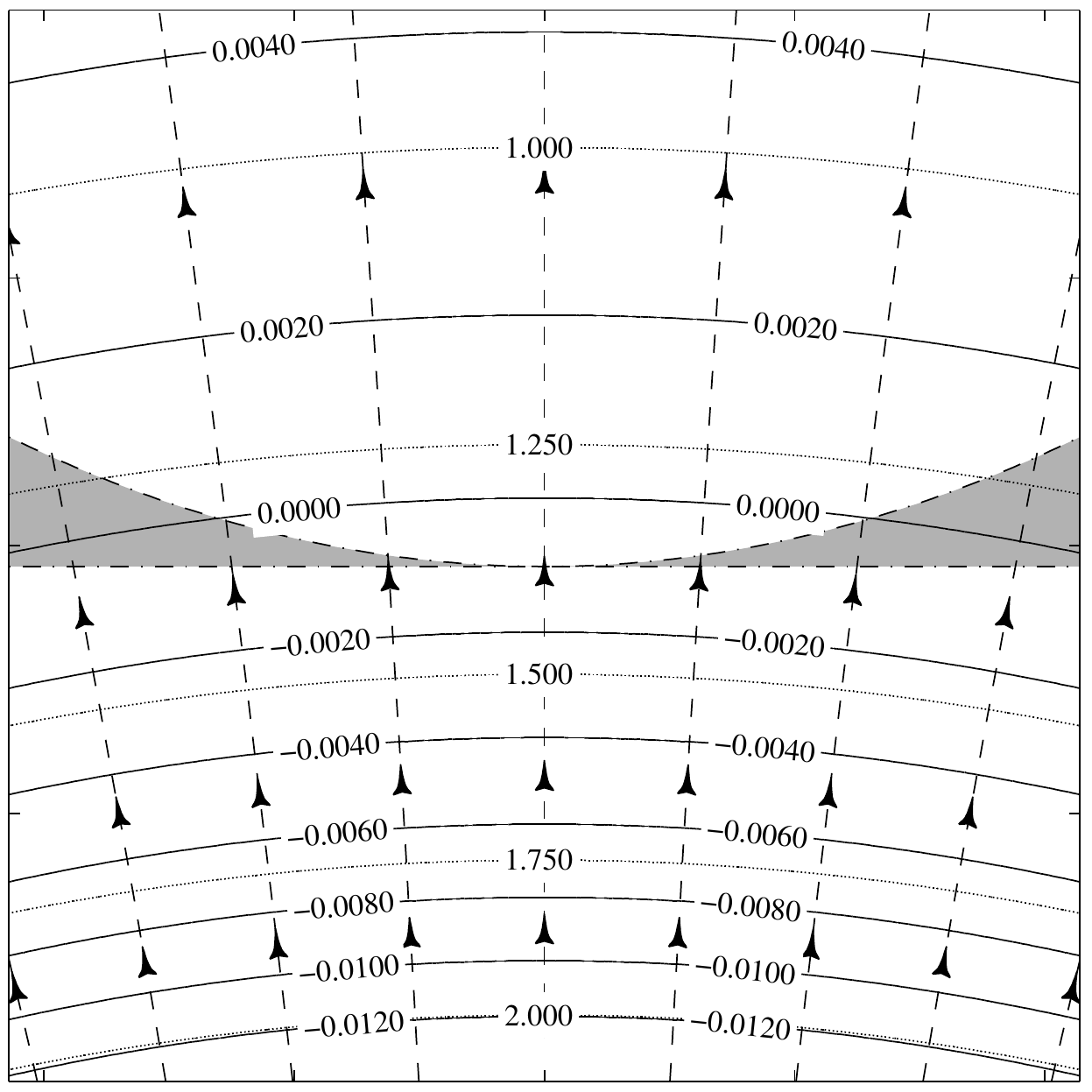}}
\put(-4,0){{\small 250}}
\put(-4,49){{\small 275}}
\put(-4,98){{\small 300}}
\put(-4,147){{\small 325}}
\put(-4,196){{\small 350}} 
\put(11,-6){{\small -50}}
\put(58,-6){{\small -25}}
\put(104,-6){{\small 0}}
\put(151,-6){{\small 25}}
\put(197,-6){{\small 50}}
\put(13,-17){Horizontal position $x$\,(mm) relative to bore axis}
\put(-18,8){\rotatebox{90}{Vertical position above solenoid center $z$\,(mm)}}
\end{picture}\\[20pt]
\end{center}
\caption{Contours of the apparent Atwood number (solid) and magnetic field strength (dotted, measured in Tesla) for the system.  The aspect ratio of the image is 1:1.  The dashed lines are the magnetic field lines with arrows indicating the direction of the magnetic field, $\vc{B}$.  The grey region shows the range of curvatures of the initial interface for rotation rates in the range $\Omega\in[0, 9.06]$\,rad\,s$^{-1}$.}
\label{fig:effdens}
\end{figure}

Based on the concentration of MnCl$_2$ in the upper layer, NaCl in the lower layer, and the strength of the magnetic field (dotted contours, Fig.~\ref{fig:effdens}, it was possible to calculate the `apparent Atwood number', between the two fluids {\it a priori} and hence anticipate the height at which we might expect the Rayleigh-Taylor instability to begin to develop.  Fig.~\ref{fig:effdens} shows the contours of apparent Atwood number for typical experimental parameters.  As a result of the spatial variation of the magnetic field strength, the contours of apparent Atwood number are curved in the opposite direction (convex) to the curve of the interface between the fluid layers (concave). 
This means that as the tank was lowered into the magnetic field, the Atwood number of fluid at the interface on the axis of the tank became negative before fluid at the interface further from axis, i.e., the change of stability process does not quite occur uniformly across the interface.  This undesirable effect was more pronounced the higher the rotation rate was.  Though the region of the interface at the center of the tank becomes unstable slightly before the regions at the edge, there is no clear indication of this in our observations: the instability, to a good approximation, is initiated across the whole extent of the tank simultaneously.  For the purposes of comparison, the range of parabolic interface profiles used in the experiments is shown in Fig.~\ref{fig:effdens} in gray.  The interface has a maximum deflection from horizontal on the order of 6\,mm, whereas the lines of equipotential deflect from the horizontal on the order of 2.5\,mm across the width of the experiment.  For further details of the experimental method see [1]

\begin{table*}
\begin{tabular}{c|ccccc}
Experiment & Rotation rate & Front velocity & Length scale$^*$  & Reynolds number  & Rossby number  \\
& $\Omega$\,(rad\,s$^{-1}$) & $U$\,(mm\,s$^{-1}$) & $L$\,(mm) & $UL/\nu$ & $U/(L\Omega)$\\
\hline
1 & 0.00 & 9.3 & 21.9 & 205 &  $\infty$\\ 
2 & 0.00 & 11.4 & 21.9 & 250 &  $\infty$\\ 
3 & 0.44 & 7.0 & 19.8 & 140 & 0.81\\ 
4 & 1.01 & 13.1 & 13.9 & 183 & 0.94\\ 
5 & 1.52 & 11.3 & 9.6 & 109 & 0.77\\ 
6 & 1.82 & 10.0 & 8.0 &  80 & 0.69\\ 
7 & 2.01 & 10.4 & 7.3 &  76 & 0.71\\ 
8 & 2.52 & 7.6 & 6.2 &  47 & 0.49\\ 
9 & 3.08 & 9.6 & 5.6 &  54 & 0.56\\ 
10 & 3.58 & 9.5 & 5.4 &  52 & 0.49\\ 
11 & 3.67 & 8.8 & 5.4 &  47 & 0.44\\ 
12 & 4.08 & 8.1 & 5.3 &  43 & 0.37\\ 
13 & 4.59 & 8.9 & 5.3 &  47 & 0.36\\ 
14 & 5.10 & 7.8 & 5.3 &  41 & 0.29\\ 
15 & 5.58 & 6.9 & 5.3 &  36 & 0.23\\ 
16 & 5.64 & 7.8 & 5.3 &  41 & 0.26\\ 
17 & 5.99 & 6.7 & 5.2 &  35 & 0.21\\ 
18 & 7.01 & 6.2 & 5.2 &  32 & 0.17\\ 
19 & 8.07 & 4.1 & 5.2 &  22 & 0.10\\ 
20 & 9.06 & 4.1 & 5.2 &  21 & 0.09\\ 
\end{tabular}
\caption{\label{tab:exppars} Measurements and key nondimensional quantities for the 20 experimental runs in \S IV.2.  Front velocity is the approximately constant velocity attained after the initial growth of the instability, length scale is the dominant length scale of the instability.  $^*$The length scales were not directly measured, but estimated using the fit shown in Fig.~7, justified in \S IV.1.}
\end{table*}

The measured velocity and estimated wavelength for each experiment shown in Fig.~9 is tabluated in Table \ref{tab:exppars}, together with the attained Reynolds number and Rossby number in each experiment.

\subsection{Structure of the instability}
\label{sec:structure}

In the second series of experiments the effect that the rotation had on the structure of the developing instability was observed.  For this series, a slightly wider inner tank was used and the PTFE buffer was removed.  The inner tank was formed by wrapping transparent cellulose acetate around a perspex circular disc that formed the base.  The internal radius was 53.5\,mm.  A volume of 300\,ml of dyed brine (NaCl$_\textnormal{(aq)}$, 25\,g\,l$^{-1}$, $\rho = 1.014$\,g\,cm$^{-3}$) was filled to a depth of approximately 33\,mm.  As before, the brine was prepared from distilled deionized water that was allowed to come up to the temperature of the laboratory before use.  The lower layer was dyed with red and blue water-tracing dyes (Cole-Parmer 00295-18 and 00295-16) to make the layer opaque.  To this, a small amount of fluorescein sodium salt (Sigma-Aldrich F6377) was added so that when the layer was illuminated most of the visual signal was the fluorescence of the interface.  This dye technique has previously been used by Dalziel [7], Davies-Wykes and Dalziel [8].  An upper, lighter layer of fluid was then floated on top.  The upper layer was made paramagnetic by adding a manganese salt (MnCl$_2\textnormal{(aq)}$, 12\,g\,l$^{-1}$, $\rho=1.008$\,g\,cm$^{-3}$) but no dye, and was therefore transparent.  The upper layer was filled to a depth slightly greater than 33\,mm.  The lid of the tank was formed by inserting another perspex disc into the upper layer, ensuring that no air bubbles were trapped.  The disc was positioned such that the layer volumes, and hence depths were equal at approximately 33\,mm each.  This procedure gave an aspect ratio for the experiment of $\delta \approx 0.62$.  As with the first series of experiments, floating the upper layer on and inserting the lid took approximately 20\,minutes to complete.  The cavity between the outer box and the inner cylinder was filled with clear distilled water. The layers had magnetic susceptibilities of $\chi_1=3.37\times10^{-6}$ (SI units) in the upper layer and $\chi_2 = -9.03\times10^{-6}$ (SI units) in the lower layer.  

The camera was focussed on a plan view of the interface of the two-layer system by use of a mirror  positioned above the experiment and angled at 45$^\circ$ to the vertical.  The camera's position was fixed throughout the experiment.  Prior to analysis the rotation of the experimental images was removed digitally, as if the camera had rotated with the tank.  

\subsection{Effects of fluid viscosity}
\label{sec:viscosity}

The effect of the viscosity is to suppress growth of small-scale structures and therefore select a preferred wavelength for growth.  Since the fluid is both rotating and viscous we might anticipate that above a certain critical rotation rate large structure are inhibited by the rotation and small structures are inhibited by viscosity.  Our observation is that this inhibition selects a preferred wavelength.
With reference to Fig.~7, for $\Omega\aleq 4$\,rad\,s$^{-1}$, structures larger than 6\,mm appear, but above this critical rotation rate, structures on this scale are inhibited.

To test the hypothesis that the observed length scale above the critical rotation rate depends crucially upon the fluid viscosity, a third set of experiments were conducted keeping the rotation rate fixed ($\Omega = 7.8\pm0.1$\,rad\,s$^{-1}$), but varying the fluid viscosity.  The viscosity of the fluid was altered by the addition of glycerol to each layer.  The amount added to each layer was such that the viscosity of the two layers was equal for a given experiment.  The viscosity was varied from $\mu_w = 1.00\times10^{-2}$\,g\,cm$^{-1}$\,s$^{-1}$, the viscosity of water at room temperature, to $\mu = 26.73\times10^{-2}$\,g\,cm$^{-1}$\,s$^{-1}$.  The density difference and magnetic susceptibility difference between the two layers were maintained at their previous values.  All other parameters from the second series of experiments were maintained too, including the dimensions of the experimental arrangement.

\begin{figure}
\begin{center}
\setlength{\unitlength}{1pt}
\begin{picture}(200,160)(0,0)
\put(10,0){\includegraphics[width = 200pt]{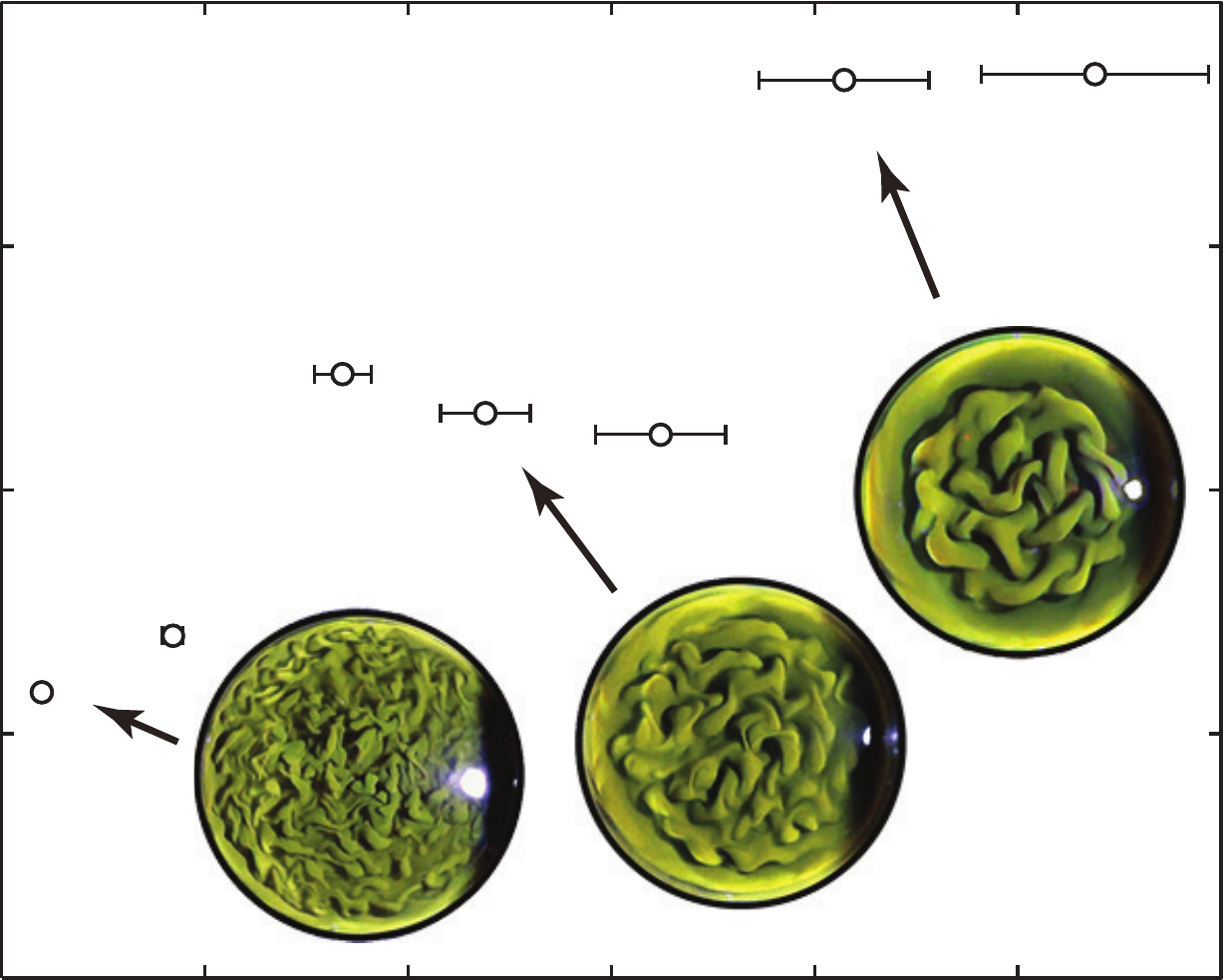}}
\put(-1,-2){{\small \phantom{1}0}}
\put(-1,38){{\small \phantom{1}5}}
\put(-1,77){{\small 10}}
\put(-1,117){{\small 15}}
\put(-1,157){{\small 20}}
\put(10,-8){{\small 0}}
\put(41,-8){{\small 5}}
\put(72,-8){{\small 10}}
\put(105,-8){{\small 15}}
\put(138,-8){{\small 20}}
\put(171.5,-8){{\small 25}}
\put(204,-8){{\small 30}}
\put(50,-19){Viscosity, $\mu$\,($\times10^{-2}$\,g\,cm$^{-1}$\,s$^{-1}$)}
\put(-16,20){\rotatebox{90}{Instability Wavelength $\lambda$\,(mm)}}
\end{picture}\\[15pt]
\caption{\label{fig:viscwavelengths} The dominant scales of perturbation after the onset of instability.  Error bars are associated with goodness-of-fit from the autocorrelation algorithm.  It can be seen that as the viscosity is increased the dominant lengthscale of the instability increases.  An increasing wall effect is observed as the viscosity of the fluid layers is increased.  The data suggests that the limit as $\mu\to0$ is non-zero indicating possible magnetically induced apparent viscosity in the fluid.}
\end{center}
\end{figure}

The variation of the observed length scale of instability with viscosity is shown in Fig.~\ref{fig:viscwavelengths}.  The data points are the white circles.  
It is apparent that as the viscosity of the two layers is increased the observed length scale increases.  In the most viscous case shown the observed length scale is approximately 18\,mm compared to the 6\,mm length scale observed in the least viscous case.  It can also be seen that in the most viscous case there appears to be a strong wall effect (Fig.~\ref{fig:viscwavelengths} insert).  The rotation rate chosen for the experiments shown in Fig.~\ref{fig:viscwavelengths}, $\Omega = 7.8\pm0.1$\,rad\,s$^{-1}$, was selected as Fig.~7 indicates that this puts the experiments in the regime where the observed wavelength has asymptoted to $6$\,mm.  In the absence of viscosity it might have been anticipated that the wavelength data would asymptote to 0\,mm.  This series of experiments, varying the fluid viscosity, was to designed to investigate and explain the observed finite asymptote ($\approx 6$\,mm for $\Omega \ageq 4$\,rad\,s$^{-1}$) in Fig.~7.  It can be seen from Fig.~\ref{fig:viscwavelengths} that if the viscosity of the fluid layers had been increased, then, at least at $\Omega = 7.8$\,rad\,s$^{-1}$, the observed wavelength would have increased, and the horizontal asymptote in Fig.~7 would have been raised.  Fig.~\ref{fig:viscwavelengths} also suggests that were it possible to reduce the fluid viscosity to zero, the observed wavelength of instability may not have tended to zero.  One explanation for this is that at this low viscous limit the relative importance of viscous diffusion becomes small compared to the effect of the Lorentz force, i.e., the Hartmann number becomes large, and we may have observed an effective increase in viscosity due to the magnetic field.